\documentclass[paper,10pt,nofootinbib,notitlepage,showpacs,showkeys]{revtex4}
\usepackage{graphicx}
\usepackage{amsmath}
\usepackage{amssymb}
\usepackage{dcolumn}
\usepackage{bm}
\usepackage{color}
\usepackage{dsfont}

\newcommand \beq{\begin{eqnarray}}
\newcommand \eeq{\end{eqnarray}}

\newcommand{\rt}[1]{{}}
\renewcommand{\k}{{\bf k}}
\newcommand{\q}{{\bf q}}
\newcommand{\p}{{\bf p}}
\newcommand{\rb}{{\bf r}}

\begin{document}
\allowdisplaybreaks

\title{Broken phase effective potential\\ in the two-loop $\Phi$-derivable approximation\\ and nature of the phase transition in a scalar theory}

\author{Gergely Mark{\'o}}
\email{smarkovics@hotmail.com}
\affiliation{Department of Atomic Physics, E{\"o}tv{\"o}s University, H-1117 Budapest, Hungary.}

\author{Urko Reinosa}
\email{reinosa@cpht.polytechnique.fr}
\affiliation{Centre de Physique Th{\'e}orique, Ecole Polytechnique, CNRS, 91128 Palaiseau Cedex, France.}

\author{Zsolt Sz{\'e}p}
\email{szepzs@achilles.elte.hu}
\affiliation{Statistical and Biological Physics Research Group of the Hungarian Academy of Sciences, H-1117 Budapest, Hungary.}

\date{\today}

\begin{abstract}
We study the phase transition of a real scalar $\varphi^4$ theory in the two-loop $\Phi$-derivable approximation using the imaginary time formalism, extending our previous (analytical) discussion of the Hartree approximation. We combine Fast Fourier Transform algorithms and accelerated Matsubara sums in order to achieve a high accuracy. Our results confirm and complete earlier ones obtained in the real time formalism \cite{Arrizabalaga:2006hj} but which were less accurate due to the integration in Minkowski space and the discretization of the spectral density function. We also provide a complete and explicit discussion of the renormalization of the two-loop $\Phi$-derivable approximation at finite temperature, both in the symmetric and in the broken phase, which was already used in the real-time approach, but never published. Our main result is that the two-loop $\Phi$-derivable approximation suffices to cure the problem of the Hartree approximation regarding the order of the transition: the transition is of the second order type, as expected on general grounds. The corresponding critical exponents are, however, of the mean-field type. Using a ``RG-improved'' version of the approximation, motivated by our renormalization procedure, we find that the exponents are modified. In particular, the exponent $\delta$, which relates the field expectation value $\bar\phi$ to an external field $h$, changes from $3$ to $5$, getting then closer to its expected value $4.789$, obtained from accurate numerical estimates \cite{Pelissetto:2000ek}.
\end{abstract}


\maketitle 

\section{Introduction}
It is well known that conventional perturbation theory fails to describe the second order phase transition of a $\varphi^4$ scalar model with $Z_2$ symmetry in four dimensions, and more generally any system involving bosonic degrees of freedom, because the perturbative expansion is plagued with infrared divergences \cite{Zinn-Justin:2002}. This great sensitivity to the infrared corresponds physically to the fact that a system of bosons becomes three dimensional near a second order or a weakly first order phase transition. Several resummation procedures have been put forward in order to cure the breakdown of the perturbative expansion and thus to provide a correct description of the phase transition in a given model, including the order of the transition and, if the transition is of the second order, the corresponding critical exponents. The quality of the description of the phase transition depends usually on the level of approximation considered within these methods.

The ring (daisy) resummation \cite{Kapusta:2006pm} was proposed in order to cure the infrared divergences of a massless theory through the resummation of the leading order thermal effects which lead to the generation of a thermal mass. This method produces in the effective potential a term which becomes cubic in the field at the temperature where the quadratic term  vanishes (this is observed already at the one-loop level of the original perturbation theory) and as a result the phase transition turns out to be of the first-order type \cite{Takahashi:1985vx,Carrington:1991hz}. However, it was argued in \cite{Arnold:1992rz} that one cannot rely on the ring-improved perturbation theory in order to distinguish between a first- and a second-order phase transition, because its loop expansion parameter $\lambda T/m_\textnormal{eff}$ becomes of $\mathcal{O}(1)$ at the nontrivial minimum of the potential, where the effective mass $m_\textnormal{eff}$ is $\mathcal{O}(\lambda T)$.

An even larger class of perturbative diagrams is resummed in the superdaisy \cite{Dolan:1973qd} or self-consistent Hartree-Fock resummation scheme which, as the daisy resummation, is a local resummation scheme in that it results in a momentum independent self-energy. Numerical studies and the use of the high temperature expansion revealed that this resummation also fails to reproduce the true nature of the phase transition \cite{Espinosa:1992gq,AmelinoCamelia:1992nc}. Recently, the same conclusion was obtained analytically \cite{Reinosa:2011ut}. 

It was shown in \cite{Buchmuller:1993bq} that in the $SU(2)$ Higgs model the phase transition in the Higgs-Goldstone sector turns into second order when going beyond the super-daisy resummation by including the scalar bubble diagram in the propagator with which the one-loop effective potential is calculated. There are other indications in the literature that with the inclusion of the setting-sun diagram in the effective action the phase transition turns into a second order one \cite{Inagaki:1997gi,Ogure:1998je,Chiku:2000eu,Smet:2001un,Farias:2008fs}. Recently the type of the temperature phase transition in the $\varphi^4$ model was investigated using Monte Carlo simulations on a lattice \cite{Bordag:2010ph}. It was found that for very small values of the coupling $\lambda\le 10^{-3}$ the phase transition is of first order, while for larger values the transition is of second order.

The above results indicate that it is important to take into account non-daisy like diagrams in order to capture the nature of the phase transition. Actually, as emphasized in \cite{Tetradis:1992xd}, the problems of the daisy and superdaisy resummations rely on the fact that, while efforts were made to include thermal effects in the effective quadratic coupling of the theory, nothing was done concerning the quartic coupling, which remained a constant. But in fact, as a result of the running, the coupling constant vanishes at $T_{\rm c}$ taming around this temperature the behavior of the resummed perturbation theory, whose expansion parameter $\lambda T/m_\textnormal{eff}$ would blow up for fixed $\lambda.$ An instructive comparison of these resummation methods with the evolution equation of the renormalization group method can be found in \cite{D'Attanasio:1996zt}. A successful description of the second order phase transition should be able to take into account the fact that the effective coupling constant exhibits a 4d behavior in the ultraviolet and a 3d one in the infrared. In the renormalization group approach the running effective coupling nicely interpolates between these two limits which allows for the correct description of the second order nature of the phase transition \cite{Elmfors:1992yn,Tetradis:1992xd,Liao:1995gt} and for the determination of the related critical exponents \cite{Tetradis:1993ts}. 

In this paper we study numerically the thermal phase transition of the one-component scalar field theory within the 2PI formalism, which is known to be a systematically improvable method to resum the perturbative series \cite{LW, Cornwall:vz}. We go beyond the lowest order (Hartree) approximation used in \cite{Reinosa:2011ut} by including in the 2PI effective action the field dependent setting-sun diagram. This is the simplest truncation of the 2PI functional that includes nonlocal contributions to the gap equation for the propagator, which we solve without further approximations. In particular we treat the momentum depedence of the propagator self-consistently. We properly address the issue of ultraviolet divergences which, after being regularized using a sharp cutoff, are removed using the renormalization method recently developed and applied in the context of the equilibrium 2PI formalism \cite{VanHees:2001pf,Blaizot:2003an,Berges:2005hc,Berges:2004hn}.

Our main result is that the transition turns into a second order type, as compared to the Hartree approximation, at least for the values of the parameters that we could access. Note that another attempt to discuss the order of the transition from higher contributions to the 2PI effective action was pursued in \cite{Bordag:2000tb} and in fact much more diagrams than the ones we shall consider here were included, namely all those which contribute at next-to-leading-order in the $1/N$ expansion of the $O(N)$ model. In this investigation however, the propagator equation which becomes momentum dependent at this level of approximation was solved with a momentum-independent ansatz, which lead to a slightly stronger first order phase than in the Hartree approximation, in disagreement with our present results which, although they concern the simplest nonlocal contribution to the 2PI effective effective action, involves a complete treatment of the corresponding momentum dependence.

We also compute several thermodynamical quantities, as well as the critical exponents. Our conclusion concerning the latter is that in the two-loop $\Phi$-derivable approximations, their values remain equal to those in a mean-field approximation. This is not surprising since, without the inclusion of the ``basketball'' diagram in the 2PI effective action, there is no wave-function renormalization in the gap equation, and thus no possibility for an anomalous dimension. It is nevertheless interesting to study what happens if one implements some ideas coming from the renormalization group approach and let the coupling run with the temperature. We find that the values of the critical exponents depart from their mean field values and that some of them can be even determined analytically, such as the critical exponent $\delta$ of the ``magnetization'' on the critical isotherm, which becomes equal to $5.$ As we shall see, this is related to the fact that, even though the approximation does not seem sufficient to generate non-analyticities in the field, the running coupling vanishes at the transition temperature. Some of the critical exponents have been studied using more elaborated truncations of the 2PI effective action by working directly in three dimensions and in the symmetric phase, see \cite{Alford:2004jj,Saito:2011xq}.

Owing to the length of the text, we provide below an itemized structure of the remainder of the paper in order to facilitate the orientation of the reader. 
\begin{itemize}
\item
Sec.~\ref{sec:definitions} introduces the model, the approximation and some basic objects of the 2PI formalism which are later used to renormalize the theory. In particular, we illustrate the known fact that in a given truncation of the 2PI effective action, there exist two inequivalent expressions for the two-point function and three inequivalent expressions for the four-point function. The cutoff regularization is also introduced and discussed in details.
\item
Sec.~\ref{sec:renorm_proc} motivates the need for an increased number of bare parameters when treating the renormalization of a given truncation of  the 2PI effective action. These bare parameters are fixed by means of both renormalization and ``consistency'' conditions, which we impose at a nonzero temperature $T_\star$ where the system is required to be in its symmetric phase. The consistency conditions ensure that certain features of the exact, untruncated 2PI formalism still hold at the renormalization point and also that, despite of their increased number, all bare parameters are fixed in terms of only two renormalized parameters: one renormalized mass $m_\star$ and one renormalized coupling $\lambda_\star$.  One nice feature of the two-loop approximation is that the bare parameters are given in terms of a finite number of perturbative sum-integrals and can thus be determined independently of the resolution of the gap equation. Finally some ideas borrowed from the renormalization group are implemented within our particular renormalization scheme. This allows us to define a RG-improved two-loop approximation which is later compared to the standard two-loop approximation. 
\item
Sec.~\ref{sec:transition} presents our numerical results on the phase transition both in the two-loop and in the RG-improved two-loop approximation. By selecting some points in parameter space, we find numerically that where in the Hartree approximation the phase transition is of the first order type, the inclusion of the setting-sun diagram at the level of the 2PI effective action turns the transition into a second order type, with mean-field exponents. The order of the phase transition is reflected also at the level of the bulk thermodynamic quantities, such as the heat capacity, speed of sound, and trace anomaly. We test the effects of the RG-improvement and find that the values of the critical exponents depart from their mean field value. 
\item
Sec.~\ref{sec:numerics} is devoted to details concerning the numerical method used to solve the model in the imaginary time formalism. Our method is based on the fact that, in the present approximation, the self-energy only receives logarithmic corrections at large external momentum. As a consequence the leading part of the self-consistent propagator is not modified as compared to the perturbative one. Owing to this fact, to each integral involving the self-consistent propagator we can subtract a similar integral involving the perturbative propagator. This leads to a sizeable acceleration of the convergence of the Matsubara sums and to an increase in accuracy in the determination of convolution-type integrals with the use of fast Fourier transformations. This subtraction method is implemented in the gap and field equations which are solved iteratively, as well as in other quantities studied numerically, like the curvature and the effective potential. We  investigate extensively the accuracy of our numerical method by testing the discretization and cutoff effects. Some related technical aspects are discussed in App.~\ref{app:pert}, where a collection of perturbative integrals is also given, and in App.~\ref{app:Matsubara}, which is devoted to the acceleration of the Matsubara sums. 
\item
Sec.~\ref{sec:renorm_proof} shows that the expressions for the bare parameters obtained from the renormalization and consistency conditions render finite the gap and field equations, as well as the effective potential. Some more technical parts are relegated to App.~\ref{app:th_exp}. This section has only theoretical relevance in that it shows what needs to be done in order to check explicitly the renormalizability of the model in the present approximation, but the finite equations derived here are not used to solve the model numerically. 
\item 
Sec.~\ref{sec:conclusion} is devoted to some conclusions and outlook.
\end{itemize}

\section{The two-loop $\Phi$-derivable approximation}\label{sec:definitions}
In this paper, we consider a real scalar $\varphi^4$ theory in four dimensions at finite temperature, defined by the Euclidean action
\beq\label{eq:action}
S[\varphi]\equiv\int_0^{1/T} d\tau\!\int d^3x\,\left(\frac{1}{2}(\partial_\tau\varphi)^2+\frac{1}{2}(\nabla\varphi)^2+\frac{m_0^2}{2}\,\varphi^2+\frac{\lambda_0}{4!}\,\varphi^4\right),
\eeq
where the inverse temperature sets the range of integration over the imaginary time. The parameters $m_0$ and $\lambda_0$ denote respectively the bare mass and the bare coupling. To ensure that the spectrum of the underlying Hamiltonian is bounded from below, one should restrict in principle to positive values of $\lambda_0$. We shall discuss this condition in more detail as we treat renormalization in Sec.~\ref{sec:renorm_proc}.

\subsection{Effective potential and gap equation}
\begin{figure}[!tbp] 
\begin{center}
\includegraphics[width=0.4\textwidth,angle=0]{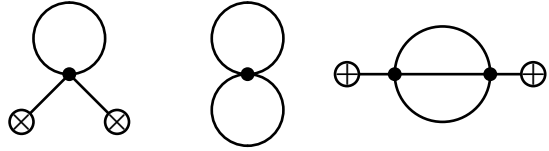}
\caption{Two-particle irreducible diagrams contributing to $\Phi[\phi,G]$ in the two-loop $\Phi$-derivable approximation. Plain lines represent the propagator $G$ and circled crosses represent the field $\phi$.\label{Fig:graphs}}
\end{center}
\end{figure}

The two-particle-irreducible (2PI) formalism provides a representation of the effective potential $\gamma(\phi)$ in terms of 2PI diagrams. It is obtained as the value taken by the functional
\beq\label{eq:gen}
\gamma[\phi,G]\equiv\frac{m_0^2}{2}\phi^2+\frac{\lambda_0}{4!}\phi^4+\frac{1}{2}\int_Q^T\big[\ln G^{-1}(Q)+(Q^2+m_0^2)G(Q)-1\big]+\Phi[\phi,G]
\eeq
at its stationary point $\smash{G=\bar G}$, that is $\gamma(\phi)$ is equal to $\gamma[\phi,\bar G]$ with $0=\delta\gamma[\phi,G]/\delta G|_{\bar G}$. In Eq.~(\ref{eq:gen}), the variable $\phi$ represents a homogeneous field configuration and $G(Q)\equiv G(i\omega_n,q)$ an even function of the Matsubara frequency $\omega_n\equiv2\pi nT$ and of the modulus of the three dimensional momentum $q\equiv|\q|$. We have also adopted the notation
\beq\label{eq:notation}
\int_Q^T f(Q)\equiv T\sum_n\int_q\,f(i\omega_n,q)\equiv T\!\!\sum_{n=-\infty}^{+\infty}\int \frac{d^3q}{(2\pi)^3}\,f(i\omega_n,q)\,.
\eeq
The functional derivative with respect to $G$, which appears in the definition of $\bar G$, is to be understood as the functional derivative in the space of functions $G(i\omega_n,\q)$ which are even with respect to $\omega_n$ and rotation invariant with respect to $\q$. However, in all expressions where such functional derivatives appear in this paper, they can be treated as normal, that is unconstrained, functional derivatives. Note also that our definition of the functional derivative in Fourier space implies a factor $(2\pi)^3/T$. With this convention, if ${\cal F}[G]$ is a functional of $G$ which we evaluate for $G=\bar G$, the following chain rule applies
\beq\label{eq:chain}
\frac{\partial{\cal F}[\bar G]}{\partial\phi}=\int_Q^T \left.\frac{\delta F[G]}{\delta G(Q)}\right|_{\bar G}\frac{\partial\bar G(Q)}{\partial\phi}\,.
\eeq
Finally, the functional $-\Phi[\phi,G]$ corresponds to the sum of all $0$-leg 2PI diagrams that one can draw in the ``shifted'' theory $S[\phi+\varphi]-S[\phi]-(\delta S/\delta\phi)\varphi$ at finite temperature, using the function $G$ in place of the free propagator. This functional cannot be computed exactly. So-called $\Phi$-derivable approximations consist in retaining in $\Phi[\phi,G]$ only certain classes of diagrams. In this paper, we consider the two-loop $\Phi$-derivable approximation:
\beq\label{eq:2loop}
\Phi[\phi,G]=\frac{\lambda_0}{4}\phi^2{\cal T}[G]+\frac{\lambda_0}{8}{\cal T}^2[G]-\frac{\lambda_0^2}{12}\phi^2{\cal S}[G]\,,
\eeq
which corresponds to the 2PI diagrams represented in Fig.~\ref{Fig:graphs}. We have introduced the notations
\beq
{\cal T}[G]\equiv \int_Q^T G(Q) \quad {\rm and} \quad {\cal S}[G]\equiv\int_Q^T\int_K^T G(Q)G(K)G(K+Q)
\eeq
for the ``tadpole'' and ``setting-sun'' sum-integrals, to be used throughout this work. Similarly, we introduce the notation
\beq
{\cal B}[G](K)\equiv\int_Q^T G(Q)G(Q+K)
\eeq
for the ``bubble'' sum-integral. The setting-sun sum-integral reads then ${\cal S}[G]=\int_Q^T G(Q){\cal B}[G](Q)$.\\

According to the above discussion, in order to evaluate the effective potential, one needs first to determine the propagator $\bar G$ from the stationarity condition $0=\delta\gamma[\phi,G]/\delta G|_{\bar G}$. This condition can be rewritten as $\bar G^{-1}_{\phi,\,T}(Q)\equiv Q^2+\bar M^2_{\phi,T}(Q)$ with $Q^2\equiv\omega_n^2+q^2$ and
\beq\label{eq:gap0}
\bar M^2_{\phi,T}(Q)=m_0^2+\left.\frac{2\delta\Phi}{\delta G(Q)}\right|_{\bar G_{\phi,T}}\,,
\eeq
which we refer to as the ``gap equation''. We have used momentarily the subscripts $\phi$ and $T$ to stress the fact that the propagator $\bar G_{\phi,T}(Q)$ and the momentum dependent mass $\bar M_{\phi,T}(Q)$ depend both on the field $\phi$ and on the temperature $T$. In what follows, we shall omit this notation unless specifically needed. A particular role will be played by the value of $\bar M^2(Q)$ at $Q=0$, which we denote more simply by $\bar M^2$. In the two-loop $\Phi$-derivable approximation, the gap equation reads
\beq\label{eq:gap}
\bar M^2(K)=m_0^2+\frac{\lambda_0}{2}\phi^2+\frac{\lambda_0}{2}\,{\cal T}[\bar G]-\frac{\lambda_0^2}{2}\phi^2\,{\cal B}[\bar G](K)\,,
\eeq
which we obtained from Eqs.~(\ref{eq:2loop}) and (\ref{eq:gap0}) by making use of the identities
\beq\label{eq:func_der}
\frac{\delta{\cal T}[G]}{\delta G(Q)}=1 \quad {\rm and} \quad \frac{\delta{\cal S}[G]}{\delta G(Q)}=3{\cal B}[G](Q)\,.
\eeq

\subsection{Field equation and geometry of the effective potential\label{sec:FE_geom}}
The phase transition will be studied by monitoring the position of the extrema of the effective potential as the temperature $T$ is lowered from an initial temperature $T_\star$ at which the system is required to be in its symmetric phase, see the discussion in Sec.~\ref{sec:transition}, down to zero temperature. The ``field equation'',  which codes the position of the extrema, is easily obtained if one makes use of the stationarity condition $0=\delta\gamma[\phi,G]/\delta G|_{\bar G}$ to express the first derivative of the effective potential as
\beq\label{eq:ustat}
\frac{\delta\gamma}{\delta\phi}=\left.\frac{\partial\gamma[\phi,G]}{\partial\phi}\right|_{\bar G} & = & m^2_0\phi+\frac{\lambda_0}{6}\phi^3+\left.\frac{\partial\Phi[\phi,G]}{\partial\phi}\right|_{\bar G}\nonumber\\
& = & \phi\left(m^2_0+\frac{\lambda_0}{6}\phi^2+\frac{\lambda_0}{2}\,{\cal T}[\bar G]-\frac{\lambda_0^2}{6}\,{\cal S}[\bar G]\right).
\eeq
Note that the field equation, which is obtained by equating this first derivative with zero, is coupled to the gap equation. Thus, for the purpose of determining the extrema of the effective potential, the gap and field equations need to be solved simultaneously.\\

Some valuable information can also be obtained from the field derivatives of the effective potential, in particular from the second and fourth derivatives at $\phi=0$, at least while the potential is defined around $\phi=0$. Taking a second derivative with respect to $\phi$ in Eq.~(\ref{eq:ustat}) and evaluating the result for $\phi=0$, we obtain
\beq\label{eq:Mhat}
\hat{M}^2_{\phi=0}\equiv\left.\frac{\delta^2\gamma}{\delta\phi^2}\right|_{\phi=0} & = & m^2_0+\frac{\lambda_0}{2}{\cal T}[\bar G_{\phi=0}]-\frac{\lambda_0^2}{6}\,{\cal S}[\bar G_{\phi=0}]\,.
\eeq
We shall later use this formula in order to define and determine the critical temperature in those cases where the system undergoes a second order phase transition. Note that we could obtain an expression for the second derivative $\smash{\hat{M}^2\equiv\delta^2\gamma/\delta\phi^2}$ for any value of the field but it is substantially more complicated than Eq.~(\ref{eq:Mhat}). Moreover, the definitions of $\bar M^2$ and $\hat M^2$ generalize in a straightforward way to any $\Phi$-derivable approximation and, if no approximation was considered, these two quantities would coincide. The fact that $\bar M^2$ and $\hat M^2$ are not equal in a given approximation needs to be regarded as a truncation artifact: here for instance, $\bar M^2_{\phi=0}-\hat M^2_{\phi=0}={\cal O}(\lambda_0^2)$ is a discrepancy that lies beyond the accuracy of the present approximation. This fact, and a similar one concerning the four-point function that we discuss now, is the main motivation for the renormalization procedure that we present in Sec.~\ref{sec:renorm_proc}.\\

Similarly, we can take three field derivatives on Eq.~(\ref{eq:ustat}). Evaluating the result for $\phi=0$, we obtain
\beq
\hat{V}_{\phi=0}\equiv\left.\frac{\delta^4\gamma}{\delta\phi^4}\right|_{\phi=0} & = & \lambda_0+3\int_Q^T \left[\frac{\lambda_0}{2}\left.\frac{\delta{\cal T}[G]}{\delta G(Q)}\right|_{\bar G_{\phi=0}}-\frac{\lambda_0^2}{6}\left.\frac{\delta{\cal S}[G]}{\delta G(Q)}\right|_{\bar G_{\phi=0}}\right]\!\left.\frac{\partial^2\bar G^2(Q)}{\partial\phi^2}\right|_{\phi=0}\,,
\eeq
where we have used Leibniz rule for computing multiple derivatives of the product of two functions, hence the factor of $3$, and the chain rule (\ref{eq:chain}) together with $\partial\bar G/\partial\phi|_{\phi=0}=0$. From Eq.~(\ref{eq:func_der}) and $\partial^2\bar G/\partial\phi^2|_{\phi=0}=-\bar G^2_{\phi=0}\partial^2\bar M^2/\partial\phi^2|_{\phi=0}$, we obtain finally
\beq\label{eq:Vhat0}
\hat{V}_{\phi=0}\equiv\left.\frac{\delta^4\gamma}{\delta\phi^4}\right|_{\phi=0} & = & \lambda_0-\frac{3}{2}\int_Q^T \Big[\lambda_0-\lambda_0^2\,{\cal B}[\bar G_{\phi=0}](Q)\Big]\bar G^2_{\phi=0}(Q)\!\left.\frac{\partial^2\bar M^2(Q)}{\partial\phi^2}\right|_{\phi=0}\,.
\eeq
The quantity $\partial^2\bar M^2/\partial\phi^2|_{\phi=0}$ obeys a linear integral equation which can be obtained from the gap equation (\ref{eq:gap}) using the same strategy as the one that leads to Eq.~(\ref{eq:Vhat0}). We obtain
\beq\label{eq:linear}
\left.\frac{\partial^2\bar M^2(K)}{\partial\phi^2}\right|_{\phi=0}=\lambda_0-\lambda_0^2\,{\cal B}[\bar G_{\phi=0}](K)-\frac{\lambda_0}{2}\int_Q^T \bar G^2_{\phi=0}(Q)\!\left.\frac{\partial^2\bar M^2(Q)}{\partial\phi^2}\right|_{\phi=0}.
\eeq
It will be convenient to introduce some additional notations. We shall write Eq.~(\ref{eq:Vhat0}) as
\beq\label{eq:Vhat}
\hat{V}_{\phi=0}=\hat\Lambda_{\phi=0}-\frac{3}{2}\int_Q^T \Lambda_{\phi=0}(Q)\bar G^2_{\phi=0}(Q)V_{\phi=0}(Q)\,,
\eeq
with $\hat\Lambda_{\phi=0}\equiv\lambda_0$, $\Lambda_{\phi=0}(K)\equiv\lambda_0-\lambda_0^2\,{\cal B}[\bar G_{\phi=0}](K)$ and
\beq
V_{\phi=0}(K)\equiv\left.\frac{\partial^2\bar M^2(K)}{\partial\phi^2}\right|_{\phi=0}\,.
\eeq
The quantity $V_{\phi=0}(K)$, whose value at $K=0$ we denote more simply by $V_{\phi=0}$, obeys the linear integral equation (\ref{eq:linear}) which we rewrite as
\beq\label{eq:V}
V_{\phi=0}(K)=\Lambda_{\phi=0}(K)-\frac{\bar\Lambda_{\phi=0}}{2}\int_Q^T \bar G^2_{\phi=0}(Q)V_{\phi=0}(Q)\,,
\eeq
with $\bar\Lambda_{\phi=0}\equiv\lambda_0$.\footnote{The reason for introducing two different notations $\hat\Lambda_{\phi=0}$ and $\bar\Lambda_{\phi=0}$ shall become clear in Sec.~\ref{sec:renorm_proc}.} This equation can be solved explicitly as\footnote{To see this, one writes first Eq.~(\ref{eq:V}) as
\beq
\int_Q^T\left[\delta^{(4)}(K-Q)+\frac{\bar\Lambda_{\phi=0}}{2}\bar G^2_{\phi=0}(Q)\right]V_{\phi=0}(Q)=\Lambda_{\phi=0}(K)\,,\nonumber
\eeq
where $\delta^{(4)}(K-Q)$ is the product of a Kronecker delta $\delta_{nm}$ for the frequencies, and a three dimensional Dirac distribution $\delta^{(3)}(\k-\q)$. This equation can then be easily inverted in terms of $\bar V_{\phi=0}$ because the definition (\ref{eq:Vbar}) is equivalent to
\beq\label{eq:V2}
\int_K^T\left[\delta^{(4)}(P-K)-\frac{\bar V_{\phi=0}}{2}\bar G^2_{\phi=0}(K)\right]\left[\delta^{(4)}(K-Q)+\frac{\bar\Lambda_{\phi=0}}{2}\bar G^2_{\phi=0}(Q)\right]=\delta^{(4)}(P-Q)\,.\nonumber
\eeq}
\beq\label{eq:V1}
V_{\phi=0}(K)=\Lambda_{\phi=0}(K)-\frac{\bar V_{\phi=0}}{2}\int_Q^T \bar G^2_{\phi=0}(Q)\Lambda_{\phi=0}(Q)\,,
\eeq
in terms of the quantity $\bar V_{\phi=0}$ such that
\beq\label{eq:Vbar}
\frac{1}{\bar V_{\phi=0}}=\frac{1}{\bar\Lambda_{\phi=0}}+\frac{1}{2}\,{\cal B}[\bar G_{\phi=0}](0)\,.
\eeq
As it was the case for $\bar M^2_{\phi=0}$ and $\hat{M}^2_{\phi=0}$, the definitions $\bar V_{\phi=0}$, $V_{\phi=0}$ and $\hat{V}_{\phi=0}$ can be generalized to non-vanishing values of the field and can be defined for any $\Phi$-derivable approximation. Moreover, when no approximation is considered, they coincide with each other and represent a unique definition for the four-point function at zero external momentum. This is one of the clues to understand the renormalization of $\Phi$-derivable approximations \cite{Berges:2005hc} which we illustrate in Secs.~\ref{sec:renorm_proc} in the case of the present two-loop approximation at finite temperature.

\subsection{A few words on regularization}\label{sec:reg}
The equations derived in the previous sections will be used throughout this work but strictly speaking they do not make sense in the absence of an ultraviolet regularization. Before any practical application, it is therefore mandatory to give them a precise meaning by choosing some regularization. To obtain a proper regularization of the 2PI functional, we start from the functional $W[J,K]$ defined by 
\beq\label{eq:W}
e^{W[J,K]}\equiv\int\mathcal{D}\varphi\,\exp\left\{-\frac{1}{2}\varphi\cdot (G_0R_\Lambda)^{-1}\cdot\varphi\right\}\times\frac{\int\mathcal{D}\varphi\,\exp\left\{-\frac{1}{2}\varphi\cdot (G_0R_\Lambda)^{-1}\cdot\varphi+S_{\rm int}[\varphi]+J\cdot\varphi+\frac{1}{2}\varphi\cdot K\cdot\varphi\right\}}{\int\mathcal{D}\varphi\,\exp\left\{-\frac{1}{2}\varphi\cdot (G_0R_\Lambda)^{-1}\cdot\varphi\right\}}\,\ \ \nonumber\\
\eeq
where we have introduced the notations $J\cdot\varphi=\int_x J(x)\varphi(x)$ and $\varphi\cdot K\cdot\varphi=\int_x\int_y \varphi(x)K(x,y)\varphi(y)$, with $\int_x\equiv \int_0^{1/T} d\tau\int d^3x$. We have written the quadratic part of the action $\varphi\cdot(G_0R_\Lambda)^{-1}\cdot\varphi$ in terms of a regulating function $R_\Lambda$ which in 3d-momentum space cuts off momenta $q\gtrsim\Lambda$ and obeys the property $R_\Lambda(q\ll\Lambda)\rightarrow 1$. In this work we shall restrict to a sharp regulating function $R_\Lambda(q)=\Theta(\Lambda-q)$. However, in the remainder of this section, we consider an arbitrary regulating function. We have also introduced a normalization such that the second factor of Eq.~(\ref{eq:W}) is completely regularized by the presence of $R_\Lambda$, at least perturbatively and for sources $K$ close to $K=0$ (this can be checked by expanding the second factor perturbatively since all the Feynman diagrams involve the fastly decreasing propagator $G_0R_\Lambda$). The first factor of Eq.~(\ref{eq:W}) requires its own regularization but this is straightforward since this factor is Gaussian. It can be written $\exp(-\gamma_0(m_0,\Lambda))$ where $\gamma_0(m_0,\Lambda)$ corresponds to the free energy per unit volume of the free theory (obtained for $S_{\rm int}[\varphi]=0$):
\beq
\gamma_0(m_0,\Lambda)=\int\frac{d^3q}{(2\pi)^3}\,R_\Lambda(q)\left[\varepsilon^{(0)}_q+2T\ln\big(1-e^{-\varepsilon^{(0)}_q/T}\big)\right]\,,
\eeq
with $\varepsilon^{(0)}_q\equiv\sqrt{q^2+m^2_0}$. After Legendre transformation of $W[J,K]$ and restriction to homogeneous field configurations, one obtains the ``regularized'' functional
\beq\label{eq:reg_gen}
\gamma[\phi,G]=\gamma_0(m_0,\Lambda)+\frac{m_0^2}{2}\phi^2+\frac{\lambda_0}{4!}\phi^4+\frac{1}{2}\int_Q^T\left[\ln G^{-1}(Q)-\ln \frac{Q^2+m^2_0}{R_\Lambda(q)}+\frac{Q^2+m_0^2}{R_\Lambda(q)}\,G(Q)-1\right]+\Phi[\phi,G]\,.
\eeq
What is meant by a regularized functional is not completely straightforward. In fact, a given functional of $G$ can be well defined for certain classes of propagators but not defined for others. The important fact about the functional (\ref{eq:reg_gen}) is that it is well defined for a class of propagators in the vicinity of its stationary point.\footnote{More rigorously stated, the functional has an extremum within the class of propagators for which it is defined. This is in one to one correspondence with the fact that the functional (\ref{eq:W}) is regularized in the vicinity of $K=0$.} In order to check this statement, note first that the gap equation, which defines the stationary point, reads
\beq\label{eq:reg_gap}
\bar G^{-1}(Q)=\frac{Q^2+m_0^2}{R_\Lambda(q)}+\bar\Pi(Q) \quad {\rm with} \quad \bar\Pi(Q)\equiv \left.\frac{2\delta\Phi}{\delta G(Q)}\right|_{\bar G}\,.
\eeq
One can convince oneself that, due to the presence of $R_\Lambda(q)$ and even though the Matsubara sums are not cut off by the regularization, any nonlocal contribution to the self-energy $\bar\Pi(Q)$ is suppressed\footnote{For instance, if one considers the perturbative bubble diagram ${\cal B}[G](K)$, after performing the Matsubara sum in Eq.~(\ref{eq:new}) of App.~\ref{app:pert}, it is clear that the regularized diagram behaves like $1/\omega^2$ at large $\omega$ and vanishes if $|{\bf k}|>2\Lambda$.} at large $Q$. Then, in this limit, the self-energy $\bar\Pi(Q)$ approaches a constant $\bar\Pi_\infty$ which is entirely determined by the first two diagrams contributing to $\Phi[\phi,G]$ (the only ones which give a local contribution to the self-energy), see Fig.~\ref{Fig:graphs}:
\beq
\bar\Pi(Q)\rightarrow \bar\Pi_\infty\equiv\frac{\lambda_0}{2}\phi^2+\frac{\lambda_0}{2}{\cal T}[\bar G] \quad {\rm as} \quad |Q|\rightarrow\infty\,.
\eeq
It follows that, at large $Q$:
\beq
\bar G(Q)=\frac{R_\Lambda(q)}{Q^2+m^2_0}-\frac{\bar\Pi_\infty R^2_\Lambda(q)}{(Q^2+m^2_0)^2}+\dots\,,
\eeq
where the second term is subleading with respect to the first one. From this we deduce first that all the diagrams appearing in the gap equation or in the effective potential through the functional $\Phi[\phi,G]$ are regularized, which was not obvious a priori. Moreover, the explicit sum-integral appearing in the 2PI functional (\ref{eq:reg_gen}) is well defined when evaluated for $G=\bar G$. This is because at large $Q$ we have:
\beq\label{eq:behavior}
\ln \bar G^{-1}(Q)-\ln \frac{Q^2+m^2_0}{R_\Lambda(q)}+\frac{Q^2+m_0^2}{R_\Lambda(q)}\,\bar G(Q)-1\sim\frac{1}{2}\frac{\bar\Pi^2_\infty R^2_\Lambda(q)}{(Q^2+m^2_0)^2}\,.
\eeq
For propagators $G$ in the vicinity of $\bar G$, we expect this behavior to be only slightly modified, which shows that the regularized 2PI functional (\ref{eq:reg_gen}) is well defined in the vicinity of $G=\bar G$, as announced above.\\

For practical purposes, it is convenient to consider the change of variables\footnote{Strictly speaking, if we want the exact identities $\hat M^2=\bar M^2$ and $\hat V=V=\bar V$ to hold true after this change of variables, we should also transform the field according to $\phi\rightarrow \phi \sqrt{R_\Lambda}$ in momentum space. Note however that since we restrict our study to homogeneous field configurations, the field in momentum space is concentrated around its zero mode component which is not affected by the change of variables since by assumption the regulator is such that $R_\Lambda(q\ll\Lambda)\rightarrow 1$.} $G\rightarrow GR_\Lambda$. In terms of this new variable, the regularized 2PI functional becomes
\beq\label{eq:reg_gen_2}
\gamma[\phi,G]=\gamma_0(m_0,\Lambda)+\frac{m_0^2}{2}\phi^2+\frac{\lambda_0}{4!}\phi^4+\frac{1}{2}\int_Q^T R_\Lambda(q)\big[\ln G^{-1}(Q)-\ln (Q^2+m^2_0)+(Q^2+m_0^2)G(Q)-1\big]+\Phi[\phi,GR]\,,\nonumber\\
\eeq
where we have introduced an additional $R_\Lambda(q)$ in the explicit sum-integral of Eq.~(\ref{eq:reg_gen_2}), although it is not needed a priori due to Eq.~(\ref{eq:behavior}). This is convenient however because the gap equation takes then the simple form (for $R_\Lambda(q)=\Theta(\Lambda-q)$, this equation needs to be understood for $q<\Lambda$)
\beq
\bar G^{-1}(Q)=Q^2+m^2_0+\left.\frac{2\delta\Phi}{\delta G(Q)}\right|_{\bar G R}\,.
\eeq
Moreover, the first derivative of the effective potential $\gamma(\phi)\equiv\gamma[\phi,\bar G]$ reads
\beq
\frac{\delta\gamma}{\delta\phi}=m^2_0\phi+\frac{\lambda_0}{6}\phi^3+\left.\frac{\partial\Phi}{\partial\phi}\right|_{\bar G R}\,.
\eeq
It follows that the formal expressions obtained in the previous sections for the gap and field equations and also the different $n$-point functions can be regularized by replacing each propagator $G$ by $GR$. From now on, we will thus consider that such a replacement has been done, but we shall leave the regulating function implicit. As far as the effective potential is concerned, it will be computed by evaluating the functional (\ref{eq:reg_gen_2}) for $G=\bar G$.  In fact, it is more convenient to use the following formula ($m_\star$ denotes for the moment an arbitrary parameter)
\beq
\int_Q^T R_\Lambda(q)\ln \frac{Q^2+m^2_0}{Q^2+m^2_\star}=\int_{m^2_\star}^{m^2_0}\,dM^2\,\int_Q^T \frac{R_\Lambda(q)}{Q^2+M^2} & = & \int\frac{d^3q}{(2\pi)^3}\,R_\Lambda(q)\int_{m^2_\star}^{m^2_0}\,dM^2\,\frac{1+2n_{\varepsilon_q}}{2\varepsilon_q}=\gamma(m_0,\Lambda)-\gamma(m_\star,\Lambda)\,,\nonumber\\
\eeq
to rewrite the regularized effective potential (\ref{eq:reg_gen_2}) as
\beq\label{eq:reg_gen_3}
\gamma[\phi,G]=\gamma_0(m_\star,\Lambda)+\frac{m_0^2}{2}\phi^2+\frac{\lambda_0}{4!}\phi^4+\frac{1}{2}\int_Q^T R_\Lambda(q)\big[\ln G^{-1}(Q)-\ln (Q^2+m^2_\star)+(Q^2+m_0^2)G(Q)-1\big]+\Phi[\phi,GR]\,.\nonumber\\
\eeq
As a final remark, note that the regularization that we have introduced, because it only cuts the modulus of the three-dimensional momentum, breaks explicitly the four dimensional rotation symmetry of the theory at zero temperature. However, since the operators $(\partial_\tau \varphi)^2$ and $(\nabla\varphi)^2$ do not require any renormalization in the present approximation,\footnote{At least as long as we restrict to homogeneous field configurations.} see below, the continuum results obtained as the cutoff is sent to infinity possess the four dimensional rotation symmetry at zero temperature.

\section{Renormalization procedure}\label{sec:renorm_proc}
As in the case of the Hartree approximation \cite{Reinosa:2011ut}, we will show that it is possible to adjust the dependence of the bare parameters in such a way that the results become insensitive\footnote{As explained in \cite{Reinosa:2011cs}, the discussion of divergences in a non-perturbative context such as the one considered here might be different from that in perturbation theory. In particular, certain divergences of the perturbative expansion do not appear as such after resummation. However they lead in general to the same difficulty, namely to the fact that one cannot define cutoff insensitive results with a high accuracy unless renormalization of the mass and the coupling is considered. Throughout this work the term ``divergence'' will be used in this somewhat extended meaning.} to the regulating scale $\Lambda$ as the latter is taken to infinity. In contrast to perturbation theory, and due to certain truncation artifacts which appear within $\Phi$-derivable approximations, we will need to introduce more bare parameters than usual. An important step will then be to fix all these parameters in terms of the usual number of renormalized parameters: one renormalized mass and one renormalized coupling.

\subsection{Multiply defined bare parameters}
A general remark is in order first. The two-loop approximation is not renormalizable a priori in the form we have presented it so far. To see why this is so, let us consider for instance the quantities $\bar M^2_{\phi=0}$ and $\hat M^2_{\phi=0}$. We have already seen that these two quantities differ by contributions of order $\lambda_0^2$. It follows that their quadratic divergences differ by terms of the same order. On the other hand, a closer look at Eq.~(\ref{eq:gap}) for $\phi=0$ and Eq.~(\ref{eq:Mhat}) shows that the bare mass squared $m^2_0$ enters both equations in exactly the same way, that is as a tree level term. Then, $\bar M^2_{\phi=0}$ and $\hat M^2_{\phi=0}$ cannot be renormalized simultaneously using the same bare mass: if one adjusts $m^2_0$ to absorb the quadratic divergence in one of the masses, there is an unbalanced quadratic divergence in the other one and vice-versa. Similar remarks apply to the different definitions of the four-point function at zero external momentum, $\bar V_{\phi=0}$, $V_{\phi=0}$ and $\hat{V}_{\phi=0}$ which, although they differ at order $\lambda_0^2$, involve the bare coupling $\lambda_0$ at tree level in an identical manner, leading once more to unbalanced divergences. Two attitudes are possible from this point on: one can either consider that $\Phi$-derivable approximations are ill-defined regarding the issue of renormalization, or one can try to cure these truncation artifacts and define a sensible renormalization scheme.\\

There is actually a simple solution to the problem of unbalanced divergences: in the case of the masses, one can slightly modify one of the two equations, Eq.~(\ref{eq:Mhat}) for instance, by introducing a second bare mass parameter $m_2$ in place of $m_0$.  Similarly, one can replace the tree level contribution $\lambda_0$ by $\lambda_2$ in $V_{\phi=0}$ and by $\lambda_4$ in $\hat{V}_{\phi=0}$. Despite the apparent simplicity of this way out, it is important to bear in mind that this procedure is only acceptable if it provides a way to determine $m_2$, $\lambda_2$ and $\lambda_4$ without introducing more physical or renormalized parameters than the ones which are usually present in a scalar $\varphi^4$ theory and if it ensures that the discrepancy between $m_0$ and $m_2$ and between $\lambda_0$, $\lambda_2$ and $\lambda_4$ disappears as the order of truncation is increased. We shall treat these matters in the next subsection. 
For the moment, we just note that all these changes can be formulated at the level of the regularized 2PI functional (\ref{eq:reg_gen_3}) which reads now
\beq\label{eq:reg_gen_4}
\gamma[\phi,G] = \gamma_0(m_\star,\Lambda)+\frac{m_2^2}{2}\,\phi^2+\frac{\lambda_4}{4!}\,\phi^4+\frac{1}{2}\int_Q^T\big[\ln G^{-1}(Q)-\ln (Q^2+m^2_\star)+(Q^2+m_0^2)G(Q)-1\big]+\Phi[\phi,G]\,,
\eeq
with
\beq
\Phi[\phi,G] & = & \frac{\lambda_2}{4}\,\phi^2{\cal T}[G]+\frac{\lambda_0}{8}{\cal T}^2[G]-\frac{\lambda_\star^2}{12}\,\phi^2{\cal S}[G]\,,
\eeq
and where, as explained in the previous section, each propagator $G(Q)$ in each diagram contributing to $\Phi[\phi,G]$ is multiplied by $R_\Lambda(q)=\Theta(\Lambda-q)$ and the explicit integral over $q$ in Eq.~(\ref{eq:reg_gen_4}) is cut off at the scale $\Lambda$. Note that there is an additional modification which we have not mentioned so far: the bare vertex which appears in the highest loop diagram, that is the setting-sun diagram, has been replaced by $\lambda_\star$, which will be identified below as the renormalized coupling at some renormalization scale. This replacement is again necessary if one wants to avoid unbalanced divergences: keeping the bare vertex in the setting-sun diagram would absorb divergences related to 2PI diagrams which are not present in the two-loop approximation. Actually, this is also what occurs in a perturbative calculation. However, in contrast with the perturbative case where replacing $\lambda_0$ by $\lambda_\star$ in the setting-sun diagram can be interpreted in terms of an expansion in powers of $\lambda_\star$, the situation is more subtle in the case of $\Phi$-derivable approximations and one needs to provide a consistent and systematic way to fix the bare coupling which appears in the setting-sun diagram and more generally in higher loop 2PI diagrams. Such a procedure exists and is pretty similar to the one we shall present below for fixing the additional bare couplings $\lambda_2$ and $\lambda_4$ but it lies slightly beyond the scope of the present paper. In what follows, we shall then admit that the replacement $\lambda_0\rightarrow\lambda_\star$ at the level of the setting-sun diagram is consistent at this order and refer to \cite{Berges:2005hc} for further details.\\

With all these modifications taken into account, the gap equation becomes
\beq\label{eq:gap2}
\bar M^2(K)=m_0^2+\frac{\lambda_2}{2}\,\phi^2+\frac{\lambda_0}{2}\,{\cal T}[\bar G]-\frac{\lambda^2_\star}{2}\,\phi^2\,{\cal B}[\bar G](K)\,,
\eeq
and the first derivative of the potential reads
\beq\label{eq:dg2}
\frac{\delta\gamma}{\delta\phi}=\phi\left(m^2_2+\frac{\lambda_4}{6}\phi^2+\frac{\lambda_2}{2}\,{\cal T}[\bar G]-\frac{\lambda_\star^2}{6}\,{\cal S}[\bar G]\right),
\eeq
from which one obtains the field equation
\beq\label{eq:field_update}
0=\bar\phi\left(m^2_2+\frac{\lambda_4}{6}\bar\phi^2+\frac{\lambda_2}{2}\,{\cal T}[\bar G_{\bar\phi}]-\frac{\lambda_\star^2}{6}\,{\cal S}[\bar G_{\bar\phi}]\right).
\eeq
The expression for the second derivative of the effective potential (the curvature) at $\phi=0$ is
\beq\label{eq:curvature_1}
\hat{M}^2_{\phi=0}=m^2_2+\frac{\lambda_2}{2}\,{\cal T}[\bar G_{\phi=0}]-\frac{\lambda_\star^2}{6}\,{\cal S}[\bar G_{\phi=0}]\,.
\eeq
Finally, the different definitions of the four-point function, that is Eqs.~(\ref{eq:Vbar}), (\ref{eq:V}) and (\ref{eq:Vhat}), remain unchanged if we replace the previous definitions of $\bar\Lambda_{\phi=0}$, $\Lambda_{\phi=0}(K)$ and $\hat\Lambda_{\phi=0}$ by\footnote{Note that $\bar\Lambda_{\phi=0}$ and $\hat\Lambda_{\phi=0}$ become two different quantities, which explains why we introduced two different notations in the first place.}
\beq\label{eq:updates}
\bar\Lambda_{\phi=0}\equiv\lambda_0\,, \quad \Lambda_{\phi=0}(K)\equiv\lambda_2-\lambda_\star^2\,{\cal B}[\bar G_{\phi=0}](K)\,, \quad {\rm and} \quad \hat{\Lambda}_{\phi=0}\equiv\lambda_4\,.
\eeq
Our results regarding renormalization concern this precise formulation of the two-loop $\Phi$-derivable approximation. We will show that it is possible to fix the dependence of the bare parameters $m_0,$ $m_2,$ $\lambda_0,$ $\lambda_2,$ and $\lambda_4$ with respect to the scale $\Lambda$ such that the solutions to the gap and field equations, and the effective potential converge as $\Lambda$ is sent to infinity. The proof is quite technical, in particular due to the fact that one needs to prove that the bare parameters can be taken independent of the field $\phi$ and of the temperature $T$. For this reason, we postpone the proof until Sec.~\ref{sec:renorm_proof}, where the interested reader can find all the details. For practical purposes, we only need to know that such a proof exists. The expressions for the bare parameters can then be obtained by imposing appropriate conditions, as we explain in the next two subsections. Note however that all the conditions cannot be independent because, despite the higher number of bare parameters required to absorb the ultraviolet divergences, the theory is to be parametrized in terms of the usual number of physical or renormalized parameters.

\subsection{Parametrization in terms of two renormalized parameters}\label{subsec:ren_cond}
Allowing for more bare parameters than usual is only acceptable if these are fixed in terms of the usual number of renormalized parameters and also if the renormalization procedure is such that the discrepancies between $m_2$ and $m_0$ as well as between $\lambda_4$, $\lambda_2$ and $\lambda_0$ are pushed to higher orders as one increases the number of loops of the $\Phi$-derivable approximation. In order to understand how this is possible within a given $\Phi$-derivable approximation, the crucial point is to remember that the need for multiple bare parameters arises from the existence of multiple definitions for the $n$-point functions, which coincide when no approximation is considered at all. We have seen in particular that it is possible to define the inverse two-point function at zero momentum in two different ways and the four-point function at zero momentum in three different ways. Now, if we imagine for a moment that these quantities were measurable at some temperature $T_\star$, there would be only one measured value for the mass at this temperature, and only one measured value for the coupling at this same temperature, in apparent contradiction with the multiplicity of definitions for the two- and four-point functions. It is then quite natural to adjust the different bare parameters in such a way that these truncation artifacts disappear at the temperature $T_\star$, where we would make contact with a measurement.\\

All the bare parameters that we introduced in the previous section can then be fixed through the conditions
\beq
\bar M^2_{\phi=0,\,T_\star}=\hat{M}^2_{\phi=0,\,T_\star}=m^2_\star \quad {\rm and} \quad \bar V_{\phi=0,\,T_\star}=V_{\phi=0,\,T_\star}=\hat{V}_{\phi=0,\,T_\star}=\lambda_\star\,.
\eeq
Note that we can arrange the previous conditions in two categories. Two renormalization conditions, for instance
\beq\label{eq:renorm}
\bar{M}^2_{\phi=0,\,T_\star}=m^2_\star \quad {\rm and} \quad \bar{V}_{\phi=0,\,T_\star}=\lambda_\star\,,
\eeq
and three ``consistency'' conditions
\beq\label{eq:consist2}
\hat{M}^2_{\phi=0,\,T_\star}=\bar{M}^2_{\phi=0,\,T_\star}\,, \quad V_{\phi=0,\,T_\star}=\bar V_{\phi=0,\,T_\star}\,, \quad {\rm and} \quad \hat{V}_{\phi=0,\,T_\star}=\bar V_{\phi=0,\,T_\star}\,.
\eeq
The consistency conditions do not involve any renormalized parameter. In this way, it is possible to fix the bare parameters in terms of the usual number of renormalized parameters. These conditions ensure also that the discrepancies between $m_0$ and $m_2$ or between $\lambda_0,$ $\lambda_2,$ and $\lambda_4$ are beyond the accuracy of the approximation at hand, as we will check on the explicit expressions for the bare parameters that we obtain below. For practical purposes, we shall use the consistency conditions in the form\footnote{Another interesting possibility is to impose the consistency conditions (\ref{eq:consist2}) at a temperature $\bar T_\star$ different from $T_\star$, while using the same renormalization conditions (\ref{eq:renorm}). This introduces $\bar T_\star$-dependencies which in some sense can be interpreted as scheme dependence. They could then be used to test the convergence of the $\Phi$-derivable expansion by comparing the sensitivity of the results to the scale $\bar T_\star$ between two orders of approximation, for instance between the present two-loop approximation and the three-loop approximation. Although interesting, this is beyond the scope of the present work.}
\beq\label{eq:consist}
\hat M^2_{\phi=0,\,T_\star}=m^2_\star\,, \quad V_{\phi=0,\,T_\star}=\lambda_\star\,, \quad {\rm and} \quad \hat{V}_{\phi=0,\,T_\star}=\lambda_\star\,,
\eeq
which are obtained trivially by combining Eqs.~(\ref{eq:renorm}) and (\ref{eq:consist2}). Note finally that $m^2_\star$ needs to be taken positive in order for the gap equation at $\phi=0$ and $T=T_\star$ to make sense. From the consistency conditions, it follows then that $\hat M^2_{\phi=0,T_\star}$ is positive: the effective potential is thus convex around $\phi=0$ at the renormalization temperature $T_\star$. We will see later that the effective potential is in fact globally convex at this temperature. Our system is then parametrized in the symmetric phase. We shall also see below that the sign of $\lambda_\star$ needs to be taken positive.

\subsection{Explicit expressions for the bare parameters}\label{subsec:exp_bare_param}
We now use the renormalization and consistency conditions in order to determine the expressions for the bare parameters $m_0,$ $m_2,$ $\lambda_0,$ $\lambda_2,$ and $\lambda_4$. As we show in Sec.~\ref{sec:renorm_proof}, these expressions are such that the solutions to the gap and field equations, as well as the effective potential (up to a field and temperature independent constant) become insensitive to $\Lambda$ at large $\Lambda$.\\

From Eq~(\ref{eq:gap2}) and the condition (\ref{eq:renorm}) for $\bar M^2$, we obtain 
\beq
m^2_\star=m^2_0+\frac{\lambda_0}{2}{\cal T}_\star[G_\star]\,,
\eeq
where ${\cal T}_\star[G_\star]\equiv \int_{Q_\star}^{T_\star} G_\star(Q_\star)$ and $G_\star(Q_\star)=1/(Q^2_\star+\bar M^2_{\phi=0,\,T_\star})=1/(Q^2_\star+m^2_\star)$ is the propagator at temperature $T_\star$ and $\phi=0$. The notation $Q_\star$ is used to emphasize the fact that the Matsubara frequencies are considered at the reference temperature $T_\star$, that is $Q_\star\equiv (i\omega^\star_n,q)$ with $\omega^\star_n\equiv 2\pi nT_\star$. Since $G_\star(Q_\star)$ is a free-type propagator, we obtain an almost explicit expression for $m_0$ in terms of $m_\star$:
\beq\label{eq:m0}
m^2_0=m^2_\star-\frac{\lambda_0}{2}{\cal T}_\star[G_\star]\,.
\eeq
From Eq.~(\ref{eq:curvature_1}) the condition (\ref{eq:consist}) for $\hat M^2$, we obtain similarly
\beq\label{eq:m2}
m^2_2=m^2_\star-\frac{\lambda_2}{2}{\cal T}_\star[G_\star]+\frac{\lambda^2_\star}{6}{\cal S}_\star[G_\star]\,,
\eeq
where ${\cal S}_\star[G_\star]\equiv\int_{Q_\star}^{T_\star}\int_{K_\star}^{T_\star}G_\star(Q_\star)G_\star(K_\star)G_\star(K_\star+Q_\star)$. We verify below that $\lambda_2-\lambda_0={\cal O}(\lambda^2_\star)$ from which it follows that $m^2_2-m^2_0={\cal O}(\lambda^2_\star)$, that is the discrepancy between the two bare masses is beyond the order of the present approximation. This feature generalizes to higher order truncations and is related to the use of the consistency conditions.\footnote{In fact there exists a systematically improvable class of $\Phi$-derivable approximations for which $\hat M^2_{\phi=0,T}=\bar M^2_{\phi=0,T}$ and thus $m^2_2=m^2_0$, see the discussion in \cite{Berges:2005hc}. Within this class of truncations, one has also $V_{\phi=0,T}=\bar V_{\phi=0,T}$ and thus $\lambda_2=\lambda_0$. However $\hat V_{\phi=0,T}\neq\bar V_{\phi=0,T}$ and thus $\lambda_4\neq\lambda_0$.}\\

We proceed similarly for the bare couplings. The condition (\ref{eq:renorm}) for $\bar V_{\phi=0}$ leads to
\beq\label{eq:l0}
\frac{1}{\lambda_0}=\frac{1}{\lambda_\star}-\frac{1}{2}\,{\cal B}_\star[G_\star](0)\,,
\eeq
where ${\cal B}_\star[G_\star](0)\equiv\int_{Q_\star}^{T_\star}G^2_\star(Q_\star)$. As the cutoff $\Lambda$ increases, there is a scale at which $\lambda_0$ diverges and above which it becomes negative. This is the Landau scale $\Lambda_{\rm p}$ defined by the condition
\beq
0=\frac{1}{\lambda_\star}-\int_{Q_\star}^{T_\star}\frac{\Theta(\Lambda_{\rm p}-q)}{(Q^2_\star+m^2_\star)^2}\,,
\eeq
where we have made the bubble sum-integral and the ultraviolet regulator explicit. If one wants to maintain $\lambda_0$ positive, one needs to choose $\lambda_\star$ positive and $\Lambda$ below the Landau scale $\Lambda_{\rm p}$. We shall work with values of the parameters such that $\Lambda_{\rm p}$ is much larger than all other scales in the problem, namely $m_\star,T_\star,T,\phi\ll\Lambda_{\rm p}$. Then our results will be pretty much insensitive to the cutoff $\Lambda$ in the regime $m_\star,T_\star,T,\phi\ll \Lambda<\Lambda_{\rm p}$. Note that, in the present two-loop approximation, it is mathematically possible to take $\Lambda>\Lambda_{\rm p}$ at the level of the renormalized quantities, and even consider their continuum limit as $\Lambda\rightarrow\infty$. The difference between this continuum values and the values obtained for $\Lambda<\Lambda_{\rm p}$ are pretty tiny.\footnote{In approximations where a Landau pole could appear as an actual pole of the propagator, the continuum limit does not exist. However, the effect of renormalization can still be understood as an insensitivity with respect to the cutoff, up to terms of the order of inverse powers of $\Lambda$, in the regime $m_\star,T_\star,T,\phi\ll\Lambda\ll \Lambda_{\rm p}$.}\\

The bare coupling $\lambda_2$ is determined from the condition (\ref{eq:consist}) for $V_{\phi=0}$ and from the condition (\ref{eq:renorm}) for $\bar V_{\phi=0}$. Using Eqs.~\eqref{eq:V} and \eqref{eq:updates} we obtain
\beq
\lambda_\star=\lambda_2-\lambda^2_\star\,{\cal B}_\star[G_\star](0)-\frac{\lambda_\star}{2}\int_{Q_\star}^{T_\star}G_\star^2(Q_\star)\big[\lambda_2-\lambda^2_\star\,{\cal B}_\star[G_\star](Q_\star)\big]\,.
\eeq
It is convenient to decompose $\lambda_2$ as $\lambda_2=\lambda_{2{\rm l}}+\delta\lambda_{2{\rm nl}}$ with
\beq\label{eq:l2nl}
\delta\lambda_{2{\rm nl}}=\lambda^2_\star{\cal B}_\star[G_\star](0)\,.
\eeq
It follows that
\beq
\lambda_\star & = & \lambda_{2{\rm l}}\left[1-\frac{\lambda_\star}{2}{\cal B}_\star[G_\star](0)\right]+\frac{\lambda^3_\star}{2}\int_{Q_\star}^{T_\star}G_\star^2(Q_\star)\big[{\cal B}_\star[G_\star](Q_\star)-{\cal B}_\star[G_\star](0)\big]\nonumber\\
& = & \frac{\lambda_{2{\rm l}}}{\lambda_0}\,\lambda_\star+\frac{\lambda^3_\star}{2}\int_{Q_\star}^{T_\star}G_\star^2(Q_\star)\big[{\cal B}_\star[G_\star](Q_\star)-{\cal B}_\star[G_\star](0)\big],
\eeq
where we have used Eq.~(\ref{eq:l0}). We arrive then at
\beq\label{eq:l2l}
\lambda_{2{\rm l}} & = & \lambda_0\left[1-\frac{\lambda^2_\star}{2}\int_{Q_\star}^{T_\star}G_\star^2(Q_\star)\big[{\cal B}_\star[G_\star](Q_\star)-{\cal B}_\star[G_\star](0)\big]\right].
\eeq
Finally, the condition (\ref{eq:consist}) for $\hat{V}_{\phi=0}$ given in Eq.~\eqref{eq:Vhat} with $\hat\Lambda_{\phi}=\lambda_4$ leads to
\beq\label{eq:l42}
\lambda_4=\lambda_\star+\frac{3}{2}\int_{Q_\star}^{T_\star}V_\star(Q_\star)G^2_\star(Q_\star)\Lambda_\star(Q_\star)\,,
\eeq
where\footnote{We use the fact that $V_{\phi=0}(K)-V_{\phi=0}(0)=\Lambda_{\phi=0}(K)-\Lambda_{\phi=0}(0)$.}
\beq
\Lambda_\star(K_\star) & \equiv & \Lambda_{\phi=0,\,T_\star}(K_\star)=\lambda_{2{\rm l}}-\lambda_\star^2\big[{\cal B}_\star[G_\star](K_\star)-{\cal B}_\star[G_\star](0)\big]\,,\label{eq:Vs}\\
V_\star(K_\star) & \equiv & V_{\phi=0,\,T_\star}(K_\star)=\lambda_\star-\lambda_\star^2\big[{\cal B}_\star[G_\star](K_\star)-{\cal B}_\star[G_\star](0)\big].\label{eq:Vren}
\eeq
Then
\beq
\lambda_4=\lambda_\star & + & \frac{3}{2}\,\lambda_\star\lambda_{\rm 2l}\,{\cal B}_\star[G_\star](0)-\frac{3}{2}(\lambda_{2{\rm l}}+\lambda_\star)\lambda^2_\star \int_{Q_\star}^{T_\star}G^2_\star(Q_\star)\big[{\cal B}_\star[G_\star](Q_\star)-{\cal B}_\star[G_\star](0)\big]\nonumber\\
& + & \frac{3}{2}\,\lambda^4_\star\int_{Q_\star}^{T_\star}G^2_\star(Q_\star)\big[{\cal B}_\star[G_\star](Q_\star)-{\cal B}_\star[G_\star](0)\big]^2\,.
\eeq
Using Eqs.~(\ref{eq:l0}) and (\ref{eq:l2l}), this becomes
\beq
\lambda_4=\lambda_\star & + & 3\frac{\lambda_{2{\rm l}}}{\lambda_0}(\lambda_0-\lambda_\star)+3(\lambda_{2{\rm l}}+\lambda_\star)\left(\frac{\lambda_{2{\rm l}}}{\lambda_0}-1\right)\nonumber\\
& + & \frac{3}{2}\,\lambda^4_\star\int_{Q_\star}^{T_\star}G^2_\star(Q_\star)\big[{\cal B}_\star[G_\star](Q_\star)-{\cal B}_\star[G_\star](0)\big]^2\,,
\eeq
and thus
\beq\label{eq:l4}
\lambda_4=-2\lambda_\star+3\frac{\lambda_{2{\rm l}}^2}{\lambda_0}+\frac{3}{2}\lambda^4_\star\int_{Q_\star}^{T_\star}G^2_\star(Q_\star)\big[{\cal B}_\star[G_\star](Q_\star)-{\cal B}_\star[G_\star](0)\big]^2\,.
\eeq
Note that $\lambda_2-\lambda_0={\cal O}(\lambda^2_\star)$, as announced above. Similarly $\lambda_4-\lambda_0={\cal O}(\lambda^2_\star)$. Moreover, since $\lambda_0$ increases strictly with $\Lambda$ in the interval $[0,\Lambda_{\rm p}]$ from the value $\lambda_\star$ at $\Lambda=0^+$ to $\infty$ at $\Lambda=\Lambda^+_{\rm p}$, we deduce using Eqs.~(\ref{eq:l2nl}), (\ref{eq:l2l}) and (\ref{eq:l4}) that the bare couplings $\lambda_2$ and $\lambda_4$ are positive for $\Lambda<\Lambda_{\rm p}$, that they are equal to $\lambda_\star$ at $\Lambda=0^+$ and that they diverge for $\Lambda=\Lambda^+_{\rm p}$. In the case of $\lambda_2$, this uses the fact that ${\cal B}_\star[G_\star](Q_\star)-{\cal B}_\star[G_\star](0)$ is negative, which we prove in App.~\ref{app:pert}. Thus, in the present approximation, although we had to introduce different bare couplings to cope with the problem of unbalanced divergences, the Landau scale at which all these bare couplings diverge is uniquely defined.\\

\begin{figure}[htbp] 
\begin{center}
\includegraphics[width=0.45\textwidth,angle=0]{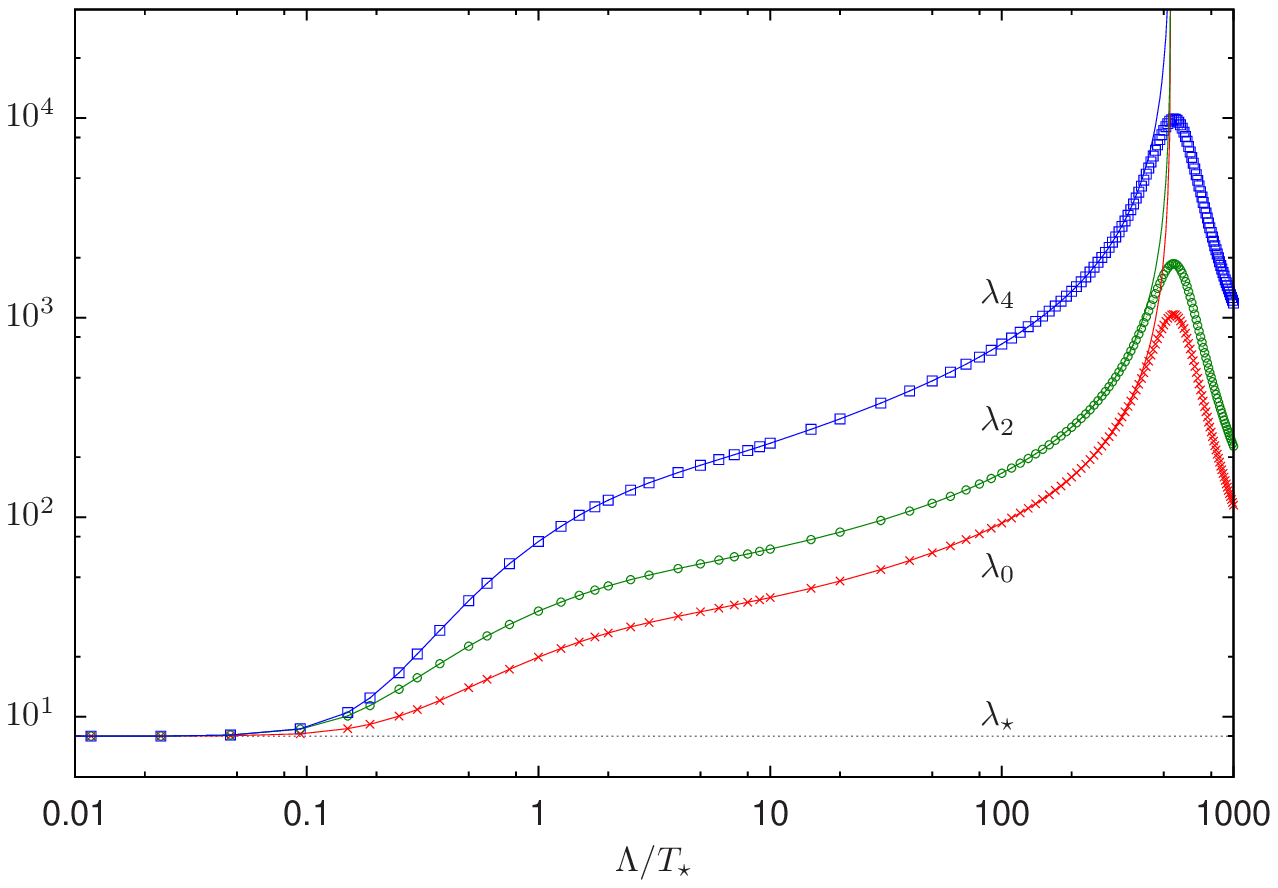}
\hspace*{0.7cm}
\includegraphics[width=0.46\textwidth,angle=0]{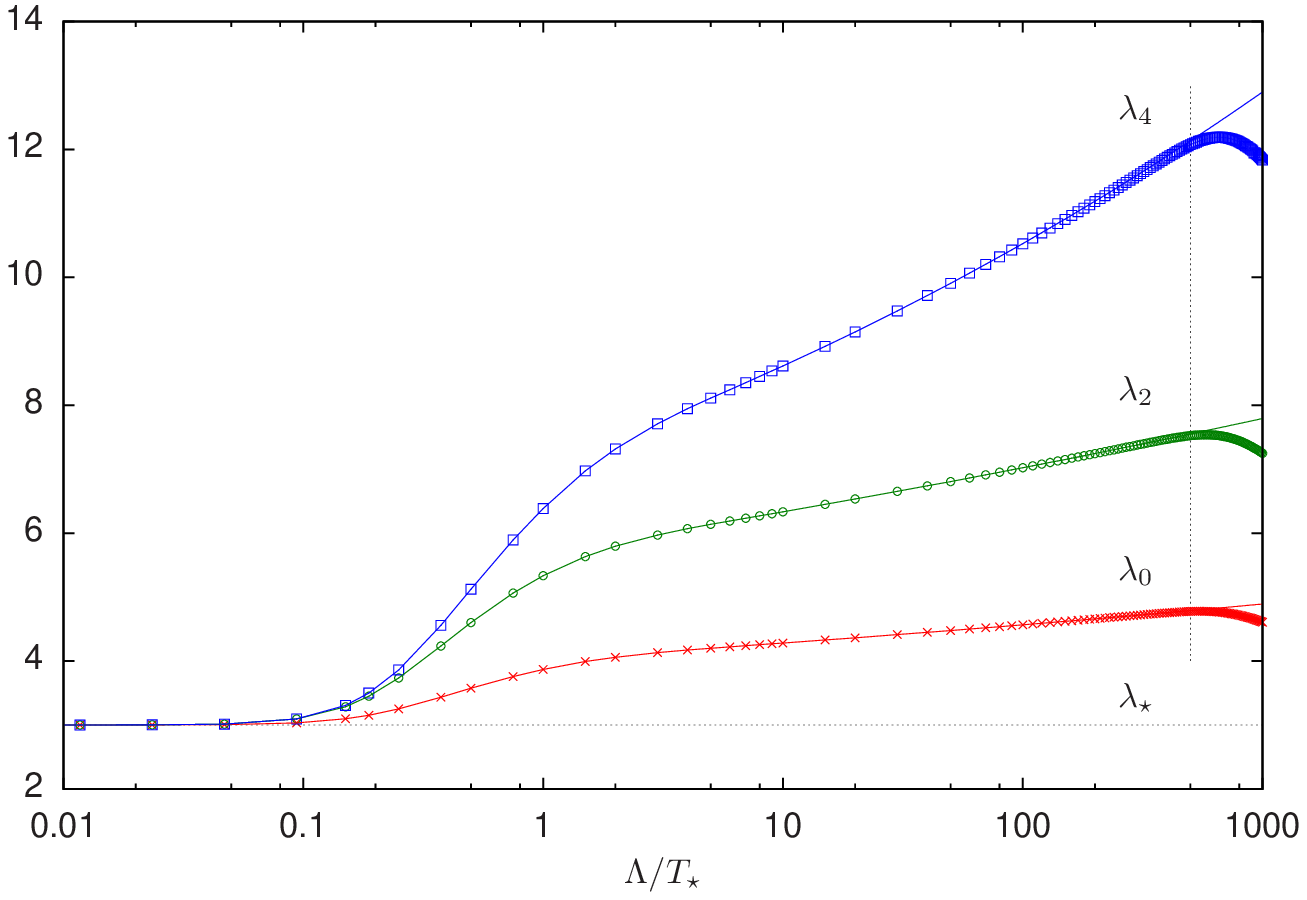}
\caption{Variation of the three bare couplings $\lambda_0,$ $\lambda_2,$ and $\lambda_4$ with the cutoff $\Lambda$, for $m_\star^2/T_\star^2=0.04$ and $T_\star=1$. The left panel corresponds to a renormalized coupling $\lambda_\star=8$ for which the Landau scale is $\Lambda_{\rm p}/T_\star\simeq 540$. The right panel corresponds to a renormalized coupling $\lambda_\star=3$ for which the Landau scale is $\Lambda_{\rm p}/T_\star\simeq e^{39}.$ The lines are obtained by performing exactly the Matsubara sum and evaluating the integrals over the modulus of the momentum using adaptive integration routines, except for the last term of Eq.~(\ref{eq:l4}) which is evaluated as a double sum. The points are obtained by evaluating the integrals in the expressions of the bare couplings as a double sum using $N_\tau=2^{10}$ non-negative Matsubara frequencies and $N_s=3\times 2^{10}$ values of the modulus of the the 3d-momentum. The discrepancy between the points and the corresponding line is related to the discretization of the momentum integrals. This will be used in Sec.~\ref{sec:numerics} in order to discuss discretization effects. Note also that the Matsubara sum in the expression of $\lambda_0$ was accelerated using Eq.~\eqref{eq:bub_improved}. \label{Fig:bare_couplings}}
\end{center}
\end{figure}

All the bare parameters involve perturbative sum-integrals in terms of the free-type propagator $G_\star$. In fact all the bare parameters but $\lambda_4$ can be reduced to the tadpole, bubble and setting-sun perturbative sum-integrals (and their mass derivatives), which makes their evaluation relatively easy: the Matsubara sums are computed exactly and the remaining integral over the modulus of the momentum is determined using adaptive integration routines. This is obvious for $m_0$, $m_2$ and $\lambda_0$ and $\delta\lambda_{2{\rm nl}}$. For $\lambda_{2{\rm l}}$, we can use the formula
\beq\label{eq:l2l2}
\lambda_{2{\rm l}}=\lambda_0\left[1+\frac{\lambda_\star^2}{2}\left({\cal B}_\star^2[G_\star](0)+\frac{1}{3}\frac{d {\cal S}_\star[G_\star]}{d m^2_\star}\right) \right]\,.
\eeq
In the case of $\lambda_4$, we can reduce it to
\beq\label{eq:l4_for_fig}
\lambda_4=\lambda_\star+3\frac{\lambda_2^2}{\lambda_0}-3\lambda_\star\left(3-2\frac{\lambda_\star}{\lambda_0}\right)^2+\frac{3}{2}\lambda_\star^4\int_{Q_\star}^{T_\star} G_\star^2(Q_\star){\cal B}_\star^2[G_\star](Q_\star).
\eeq
The last term cannot be reduced to simpler sum-integrals and is then evaluated directly as a double sum. The variation of the different bare couplings with the cutoff $\Lambda$ is shown in Fig.~\ref{Fig:bare_couplings}, for two different values of the renormalized coupling $\lambda_\star$. In the left panel, corresponding to a large value of the coupling $\lambda_\star=8$, the Landau scale is relatively close to the scales $T_\star$ or $m_\star$. We observe the divergence of the three bare couplings at the Landau scale. In the right panel, corresponding to a small value of the coupling $\lambda_\star=3$, the Landau scale is pretty far apart from the scales $T_\star$ and $m_\star$. For cutoff scales $\Lambda$ in the regime $m_\star,T_\star\ll\Lambda\ll\Lambda_{\rm p}$, we observe a logarithmic dependence of the bare couplings which mirrors the divergences of the gap and field equations absorbed by these bare parameters. In particular, the variation of $\lambda_4$ is rather important and shows that, if not properly renormalized, logarithmic divergences can lead to sizable cutoff dependencies, even if the coupling is small.\\

As a final remark note that starting from Eq.~(\ref{eq:reg_gen_4}) one can derive an expression for the difference $\Delta\gamma(\phi)=\gamma(\phi)-\gamma_\star(0),$ where $\gamma_\star(0)$ is the value of the potential at temperature $T_\star$ and at vanishing field.\footnote{This quantity is well defined because according to the renormalization condition for $\bar M^2_{\phi=0}$, $\bar M^2_{\phi=0,T}$ is defined for $T=T_\star$.} Hereinafter this difference will be referred to as the subtracted effective potential. To derive it, we use the expression (\ref{eq:dg2}) for $\delta\gamma/\delta\phi$ as well as the expression (\ref{eq:m0}) for $m^2_0$. Noting that $\gamma_\star(0)=\gamma_0(m_\star,\Lambda)-(\lambda_0/8) {\cal T}_\star^2[G_\star]$, it is then simple to arrive at the following expression: 
\beq\label{eq:pot_prat}
\Delta\gamma(\phi)=\gamma_0(m_\star,\Lambda)-\gamma_0^\star(m_\star,\Lambda) & + & \frac{1}{2}\int_Q^T\Big[\ln \bar G^{-1}(Q)-\ln G^{-1}_\star(Q)+(Q^2+m^2_\star)\bar G(Q)-1\Big]\nonumber\\
& - & \frac{\lambda_4}{4!}\phi^4+\frac{1}{2}\phi\,\frac{\delta\gamma}{\delta\phi}+\frac{\lambda_0}{8}\big[{\cal T}[\bar G]-{\cal T}_\star[G_\star]\big]^2.
\eeq
As far as the nature of the transition is concerned, it is enough to concentrate on $\Delta\gamma(\phi).$ The expression above is useful on a practical level because $\delta\gamma/\delta\phi$ needs to be computed anyway when solving the coupled system of gap and field equations and we can then use the same numerical routine. Moreover, Eq.~(\ref{eq:pot_prat}) involves a difference of tadpole sum-integrals which can be computed efficiently, as we explain in Sec.~\ref{sec:numerics}. We will also prove in Sec.~\ref{sec:renorm_proof} that $\Delta\gamma(\phi)$ is finite, whereas $\gamma(\phi)$ is only finite up to a temperature and field independent divergent constant.

\subsection{Renormalization group improvement}\label{sec:RG_theory}
In addition to solving the two-loop $\Phi$-derivable approximation, we shall also consider an ``improved'' two-loop approximation based on some ideas borrowed from the renormalization group and which we now explain.\\

In the renormalization procedure that we have presented in the previous sections, the temperature $T_\star$ played the role of a renormalization scale $\mu$. In the exact theory, the physical observables should not depend on $\mu$: any change in $\mu$ should be compensated by a ``running'' or ``flow'' of the renormalized parameters  $m_\star(\mu)$ and $\lambda_\star(\mu)$. In principle, if one is able to determine the running of the parameters, it is then possible to describe the same theory from different but equivalent points of view, each implying its own renormalization scale and the corresponding renormalized parameters. In particular, in calculations at finite temperature, one can choose a description in which the renormalization scale $\mu$ equals the temperature $T$.\\

The previous considerations become particularly interesting in the presence of some approximation because the different possible descriptions cease to be strictly equivalent. It can then happen that taking into account the running of the parameters leads to an ``improved'' approximation. Usually, the improvement is related to the fact that the running resums higher order contributions. In the present work, we shall see that the running will have somehow the opposite effect in the sense that it will remove certain fluctuations, namely fluctuations responsible for some of the artifacts of the $\Phi$-derivable approximation that we mentioned above.\\

In order to obtain the running of the renormalized parameters with the scale $T$ in the present approximation, we choose Eqs.~\eqref{eq:m0} and \eqref{eq:l0} and differentiate them with respect to $T_\star$ under the assumption that the bare parameters $m_0$ and $\lambda_0$ are fixed.\footnote{One can check that the corresponding differential equations are UV finite. This is not true if we would fix $m_0$ and $\lambda_4$ for instance. This is most certainly an artifact of the truncation.} Then, $m_\star(T)$ and $\lambda_\star(T)$ can be obtained by integrating the ordinary differential equations for $d\lambda_\star(T_\star)/d T_\star$ and $d m^2_\star(T_\star)/d T_\star$, starting from the initial temperature $T_\star$ at which we fix the value of the renormalized parameters: $m^2_\star(T_{\star})\equiv m^2_{\star}$ and $\lambda_\star(T_{\star})\equiv\lambda_{\star}$. In the present approximation, there is in fact an easier way to proceed. Indeed, by comparing Eqs.~\eqref{eq:m0} and \eqref{eq:l0} with 
\beq
\label{eq:gap_for_RG}
\bar M^2_{\phi=0}=m_0^2+\frac{\lambda_0}{2}{\cal T}[\bar G_{\phi=0}],\\
\frac{1}{\bar V_{\phi=0}}=\frac{1}{\lambda_0}+\frac{1}{2}{\cal B}[\bar G_{\phi=0}](0),
\label{eq:bV_for_RG}
\eeq
we see that, since $\bar M_{\phi=0,T_{\star}}=m_{\star}=m_\star(T_{\star})$ and $\bar V_{\phi=0,T_{\star}}=\lambda_{\star} =\lambda_\star(T_{\star}),$ the dependence of $m_\star(T)$ and $\lambda_\star(T)$ on $T$ is nothing but that of $\bar M_{\phi=0}$ and $\bar V_{\phi=0}$ on $T$. This simple fact provides us with the following recipe to implement the RG-improvement:
\begin{enumerate}
\item
solve the gap-equation \eqref{eq:gap_for_RG} for $\bar M_{\phi=0}$ in terms of the parameters $T_\star$, $m_\star$ and $\lambda_\star$;
\item
compute $\bar V_{\phi=0}$ from Eq.~\eqref{eq:bV_for_RG}, using the determined $\bar M_{\phi=0}$; 
\item
apply the replacements $T_\star\to T,$ $m_\star\to \bar M_{\phi=0},$ $\lambda_\star\to \bar V_{\phi=0}$ in every equation of interest.
\end{enumerate}
The replacements apply also to the bare parameters $m_2,$ $\lambda_2,$ and $\lambda_4$, which have to be redetermined and will be denoted $m^{\mbox{\tiny RG}}_2,$ $\lambda^{\mbox{\tiny RG}}_2,$ and $\lambda^{\mbox{\tiny RG}}_4$ when needed. The bare parameters $m_0$ and $\lambda_0$ do not need to be modified since they are invariant, by construction.\\

As an illustration of how the improvement works, let us consider the curvature of the effective potential. Before the improvement, it reads
\beq\label{eq:Mhat_unimprvd}
\hat{M}^2_{\phi=0}=m^2_\star+\frac{\lambda_2}{2}\,\big[{\cal T}[\bar G_{\phi=0}]-{\cal T}_\star[G_\star]\big]-\frac{\lambda_\star^2}{6}\,\big[{\cal S}[\bar G_{\phi=0}]-{\cal S}_\star[G_\star]\big]\,,
\eeq
where we have used the expression (\ref{eq:m2}) for $m^2_2$. After implementing the RG-improvement, it becomes
\beq\label{eq:RG_masses}
(\hat{M}^{\mbox{\tiny RG}}_{\phi=0})^2=\bar M^2_{\phi=0}+\frac{\lambda_2^{\mbox{\tiny RG}}}{2}\,\big[{\cal T}[\bar G_{\phi=0}]-{\cal T}[\bar G_{\phi=0}]\big]-\frac{\bar V_{\phi=0}}{6}\,\big[{\cal S}[\bar G_{\phi=0}]-{\cal S}[\bar G_{\phi=0}]\big]=\bar M^2_{\phi=0}\,.
\eeq
It follows that, in the RG-improved case, the two definitions of the mass coincide at $\phi=0$ for any value of the temperature (as long as the masses are defined) whereas this was only true for $T=T_\star$ in the non-improved case. The improvement has then restored a certain number of exact identities among the two possible definitions of the mass. Similar remarks apply to the three different definitions of the four-point function at $\phi=0$ and zero external momentum. In the RG-improved case they are identical for any temperature
\beq\label{eq:equality}
V^{\mbox{\tiny RG}}_{\phi=0}=\hat V^{\mbox{\tiny RG}}_{\phi=0}=\bar V_{\phi=0}\,.
\eeq
The equality $V^{\mbox{\tiny RG}}_{\phi=0}=\bar V_{\phi=0}$ is a particular case of the more general result 
\beq\label{eq:Vkimp}
V^{\mbox{\tiny RG}}_{\phi=0}(K)=\bar V_{\phi=0}-\bar V^2_{\phi=0}\big[{\cal B}[\bar G_{\phi=0}](K)-{\cal B}[\bar G_{\phi=0}](0)\big]\,.
\eeq
To obtain the latter, we start from Eq.~(\ref{eq:V2}) and apply the renormalization condition $V_{\phi=0,T_\star}=\lambda_\star$ to obtain
\beq\label{eq:eqeq}
V_{\phi=0}(K)=\lambda_\star+\Lambda_{\phi=0}(K)-\Lambda_\star(0)-\frac{\bar V_{\phi=0}}{2}\int_Q^T \bar G^2_{\phi=0}(Q)\Lambda_{\phi=0}(Q)+\frac{\lambda_\star}{2}\int_{Q_\star}^{T_\star} \bar G^2_\star(Q_\star)\Lambda_\star(Q_\star)\,.
\eeq
We note next that, under the improvement, $\Lambda_\star(K_\star)=\Lambda_{\phi=0,T_\star}(K_\star)$ as defined in Eq.~(\ref{eq:Vs}) becomes equal to $\Lambda^{\mbox{\tiny RG}}_{\phi=0}(K)$, from which it follows that the two integrals in (\ref{eq:eqeq}) cancel after the improvement. Equation (\ref{eq:equality}) is finally obtained by noticing that $\lambda_\star+\Lambda_{\phi=0}(K)-\Lambda_\star(0)=\lambda_\star-\lambda^2_\star[{\cal B}[\bar G_{\phi=0}](K)-{\cal B}_\star[G_\star](0)]$.  Similarly, using Eq.~\eqref{eq:Vhat} and the renormalization condition $\hat V_{\phi=0,T_\star}=\lambda_\star$, we obtain
\beq
\hat V_{\phi=0}=\lambda_\star-\frac{3}{2}\int_Q^T \Lambda_{\phi=0}(Q)\bar G^2_{\phi=0}(Q)V_{\phi=0}(Q)+\frac{3}{2}\int_{Q_\star}^{T_\star} \Lambda_\star(Q_\star)\bar G^2_\star(Q_\star)V_\star(Q_\star)\,.
\eeq
From the definition of $V_\star(K_\star)$ in Eq.~(\ref{eq:Vren}) and from Eq.~(\ref{eq:Vkimp}), it is easily checked that, $V_{\phi=0}(K)$ and $V_\star(K_\star)$ become equal under the improvement. Then, the two integrals in the previous equation cancel identically and $\hat V^{\mbox{\tiny RG}}_{\phi=0}=\bar V_{\phi=0}$. 

As we shall observe in Sec.~\ref{sec:transition}, the critical exponents, which are of the mean-field type in the two-loop approximation, are modified after the improvement is considered. In particular, the exponent $\delta$ gets closer (although it remains of the integer type) to its expected value in three dimensions. An unfortunate feature of the improvement is however that it is only defined in the symmetric phase: below a certain temperature $\bar T_{\rm c}<T_\star$, which will be identified later with the critical temperature in the RG-improved case, the solution of the gap equation at vanishing field $\bar M_{\phi=0}$ is not defined. Therefore, it will be only possible to determine the improved critical exponents from above the critical temperature. In particular, we will not be able to access the improved value for the exponent~$\beta$.

\subsection{Multiply defined four-point functions}\label{sec:Vs}
In the next section, we solve the gap and field equations, using the expressions Eqs.~(\ref{eq:m0}), (\ref{eq:m2}), (\ref{eq:l0}), (\ref{eq:l2nl}), (\ref{eq:l2l}) and (\ref{eq:l4}) for the bare parameters, both in the two-loop and in the RG-improved two-loop approximations, and use the corresponding effective potentials to discuss the characteristic features of the phase transition in the model. Before we do so, however, it is interesting to study the temperature dependence of the three four-point functions $\bar V_{\phi=0},$ $V_{\phi=0},$ and $\hat V_{\phi=0}$ defined at vanishing field and zero momentum. Since $\bar G_{\phi=0}$ is a free type propagator, the four-point functions at zero field are all given in terms of perturbative sum-integrals. In this perturbative setting, it is then easy to check that the four-point functions are renormalized by the bare couplings $\lambda_0$, $\lambda_2$ and $\lambda_4$ obtained in the previous section, without relying on the general proof given in Sec.~\ref{sec:renorm_proof}. Moreover, since we are only interested here in the continuum limit, we can determine the renormalized four-point functions using any regularization. We shall use dimensional regularization or cutoff regularization, depending on our convenience. More precisely, $\bar V_{\phi=0}$ and $V_{\phi=0}$, because they can be expressed solely in terms of tadpole, bubble and setting-sun sum-integrals (and their mass derivatives), see below, will be evaluated using dimensional regularization. In contrast, $\hat V_{\phi=0}$ cannot be completely reduced to these simple sum-integrals and we shall compute it using cutoff regularization.\\

The renormalized expression for $\bar V_{\phi=0}$ is trivially obtained by combining Eqs.~\eqref{eq:Vbar} and \eqref{eq:l0}
\beq\label{eq:Vbb}
\frac{1}{\bar V_{\phi=0}}=\frac{1}{\lambda_\star}+\frac{1}{2}\left[{\cal B}[\bar G_{\phi=0}](0)-{\cal B}_\star[G_\star](0)\right].
\eeq
Since ${\cal B}[\bar G_{\phi=0}]$ is a one-loop integral involving a free-type propagator, it is clear that there is no divergence in this formula, as it can also be explicitly checked by a direct calculation, for instance using dimensional regularization. To obtain an useful expression for $V_{\phi=0}$ we consider its difference with $\bar V_{\phi=0}$. Using Eqs.~\eqref{eq:V1}, \eqref{eq:Vbar}, and \eqref{eq:updates} and introducing the splitting $\lambda_2=\lambda_{2{\rm l}}+\delta\lambda_{2{\rm nl}}$, we arrive at
\beq
V_{\phi=0}-\bar V_{\phi=0} & = & \lambda_{\rm 2l}+\delta\lambda_{\rm 2nl}-\lambda_0-\lambda_\star^2{\cal B}[\bar G_{\phi=0}](0)-\frac{\bar V_{\phi=0}}{2}\int_Q^T\bar G^2_{\phi=0}(Q)\Big[\lambda_{\rm 2l}+\delta\lambda_{\rm 2nl}-\lambda_0-\lambda_\star^2{\cal B}[\bar G_{\phi=0}](Q)\Big]\nonumber\\
& = & \left(\frac{\lambda_{\rm 2l}}{\lambda_0}-1\right)\bar V_{\phi=0}+\delta\lambda_{\rm 2nl}-\lambda_\star^2{\cal B}[\bar G_{\phi=0}](0)-\frac{\bar V_{\phi=0}}{2}\int_Q^T\bar G^2_{\phi=0}(Q)\Big[\delta\lambda_{\rm 2nl}-\lambda_\star^2{\cal B}[\bar G_{\phi=0}](Q)\Big]\,.
\eeq
Using the expression for $\delta\lambda_{2{\rm nl}}$ in Eq.~(\ref{eq:l2nl}) as well as the expression for $\lambda_{2{\rm l}}$ in the form of Eq.~(\ref{eq:l2l2}), we obtain
\beq\label{eq:V_phi0_conv}
\frac{V_{\phi=0}}{\bar V_{\phi=0}} = 1-\frac{\lambda^2_\star}{2}\Big[{\cal B}[\bar G_{\phi=0}](0)-{\cal B}_\star[G_\star](0)\Big]{\cal B}_\star[G_\star](0)- \frac{\lambda^2_\star}{6}\left[\frac{d{\cal S}[\bar G_{\phi=0}]}{d\bar M^2_{\phi=0}}-\frac{d{\cal S}_\star[G_\star]}{dm^2_\star}\right]+\frac{2\lambda_\star}{\bar V_{\phi=0}} \left(1-\frac{\lambda_\star}{\bar V_{\phi=0}}\right).
\eeq
A direct calculation using dimensional regularization shows that the sum of the two square brackets in this formula is finite (see Eqs.~\eqref{eq:comb_D} and \eqref{eq:comb_D2}). As already mentioned, the four-point function $\hat V_{\phi=0}$ contains a three-loop integral which cannot be reduced to simpler sum-integrals. We shall use the following expression:
\beq\label{eq:V_hat_for_fig}
\hat V_{\phi=0}=\lambda_4+3\frac{\lambda_2^2}{\lambda_0^2}\left(\bar V_{\phi=0}-\lambda_0\right)+\lambda_\star^2\bar V_{\phi=0}\left[\frac{\lambda_\star^2}{12}\left(\frac{d{\cal S}[\bar G_{\phi=0}]}{d\bar M^2_{\phi=0}}\right)^2-\frac{\lambda_2}{\lambda_0}\frac{d{\cal S}[\bar G_{\phi=0}]}{d\bar M^2_{\phi=0}}\right]-\frac{3}{2}\lambda_\star^4\int_Q^T\bar G_{\phi=0}^2(Q){\cal B}^2[\bar G_{\phi=0}](Q),\ \ \ 
\eeq
obtained from Eq.~\eqref{eq:Vhat} by using Eqs.~\eqref{eq:V1} and \eqref{eq:updates}.\\

\begin{figure}[htbp] 
\begin{center}
\includegraphics[width=0.53\textwidth,angle=0]{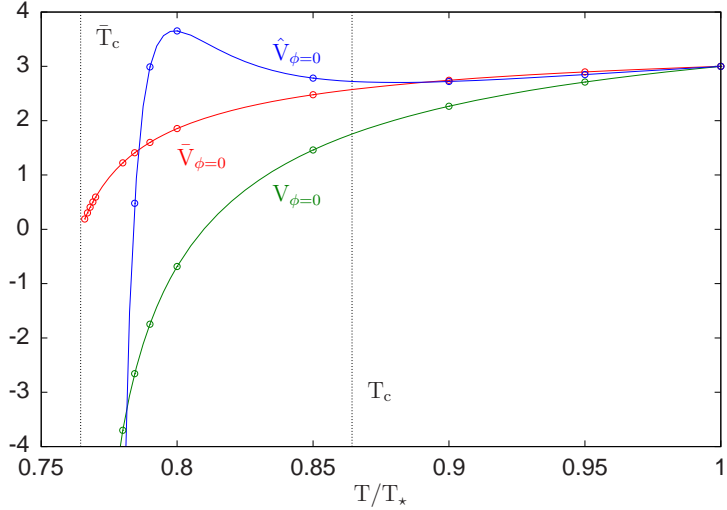}
\caption{Temperature dependence of the three four-point functions $\bar V_{\phi=0},$  $V_{\phi=0}$ and $\hat V_{\phi=0}$ at parameters $m^2_\star/T^2_\star=0.04$ and $\lambda_\star=3.$ In the case of $\bar V_{\phi=0}$ and $V_{\phi=0}$, the lines are obtained using adaptive integration routines to evaluate their expressions derived using dimensional regularization. In the case of $\hat V_{\phi=0}$ the line is obtained by evaluating the three-loop integral in the last term of Eqs.~\eqref{eq:V_hat_for_fig} and \eqref{eq:l4_for_fig} as a double sum ($N_\tau=2^{10}$ and $N_s=25\times 2^{10}$), while in all the other integrals, including those in the bare couplings, the Matsubara sum are done exactly and the momentum integral are evaluated with adaptive routines at cutoff $\Lambda/T_\star=100.$ The points are obtained by evaluating the integrals as a double sums using $N_\tau=2\times 2^{10}$ non-negative Matsubara frequencies, $N_s=13\times 2^{10}$ values of the modulus of the the 3-momentum, while decreasing the cutoff $\Lambda$ linearly from $\Lambda/T_\star=190$ at $T=T_\star$ to $\Lambda/T_\star=30$ at $T=\bar T_{\rm c}.$ \label{Fig:3V}}
\end{center}
\end{figure}

The variation with the temperature of the three four-point functions is presented in Fig.~\ref{Fig:3V}. Due to our choice of consistency conditions, the values of the four-point functions coincide at $T_\star,$ but in general they differ at other temperatures (with the exception of those values of $T$ where two of the curves cross each other). As we shall see later, for those values of the parameters chosen here, the system undergoes a second order phase transition at some temperature $T_{\rm c}$. Above $T_{\rm c}$, where the system is in the symmetric phase and the $n$-point functions are indeed defined at $\phi=0$, we observe that $\bar V_{\phi=0}$ and $\hat V_{\phi=0}$ stay pretty close to each other which shows that the violation of the exact identity $\bar V_{\phi=0}=\hat V_{\phi=0}$ is a mild one. The discrepancy is more important in the case of $V_{\phi=0}$ although the latter remains positive as long as $T>T_{\rm c}$. Note also that none of the four-point functions vanishes at~$T_{\rm c}$. 

In the non-improved two-loop approximation, the curves below $T_{\rm c}$ should not be taken too seriously because in the broken phase the $n$-point functions should be evaluated at the nontrivial minimum of the potential and not at $\phi=0$ (which is actually a maximum of the potential below $T_{\rm c}$). The reason why we are able to follow the four-point functions at $\phi=0$ below $T_{\rm c}$ is that the curvature of the potential $\hat M^2_{\phi=0}$ is different from the gap mass $\bar M^2_{\phi=0}$. There is then a range of temperatures $\bar T_{\rm c}<T<T_{\rm c}$ where, although the curvature turns negative, the gap mass remains positive making then possible the evaluation of $n$-point functions at $\phi=0$ in the broken phase. This range of temperatures becomes more interesting in the RG-improved two-loop approximation because in this case the critical temperature is $\bar T_{\rm c}$ and not $T_{\rm c}$ and it makes sense then to follow the four-point functions at $\phi=0$ down to $\bar T_{\rm c}$. In fact, since the three four-point functions become equal in this approximation, see Eq.~(\ref{eq:equality}), only the curve of $\bar V_{\phi=0}$ is relevant. We note then that, in the RG-improved two-loop approximation, the four-point function vanishes at the critical temperature $\bar T_{\rm c}$. As we shall see, this is directly connected to the modification of the exponent~$\delta$.

As a final remark, let us point out that studying the three four-point functions in the range $\bar T_{\rm c}<T<T_{\rm c}$ is interesting on numerical grounds for it gives valuable information concerning the discretization effects of the numerical method used to solve the model (see Sec.~\ref{sec:numerics} for details concerning numerics). Namely, some integrals are infrared divergent at $\bar T_{\rm c},$ and because of this $\bar V_{\phi=0}$ goes to $0$ while $V_{\phi=0}$ and $\hat V_{\phi=0}$ diverges negatively. There is a competition between the square of the derivative of the setting-sun integral times $\bar V_{\phi=0}$ and the last term of Eq.~\eqref{eq:V_hat_for_fig}, which both go as $\bar M^{-3}_{\phi=0}$ as the mass $\bar M_{\phi=0}$ goes to zero, and this competition determines whether $\hat V_{\phi=0}$ diverges negatively or positively at $\bar T_{\rm c}.$ In order to obtain the correct divergence of $\hat V_{\phi=0}$ numerically, a not too coarse discretization needs to be considered.

\section{Study of the phase transition}\label{sec:transition}

We now compute the effective potential and study how its shape changes as we lower the temperature $T$ from the renormalization temperature $T_\star$ down to $T=0.$ We shall first define the critical temperatures and evaluate them, then study the nature of the transition, followed by the thermodynamical observables and the critical exponents. All details regarding numerics are gathered in Sec.~\ref{sec:numerics}, where we explain in particular how to accelerate the convergence of Matsubara sums and how to achieve accurate convolution routines.

\subsection{Critical temperatures}
We shall see below that the effective potential is convex at the initial temperature $T_\star$, with a single minimum at $\phi=0$. In other words, the unique solution of the field equation is $\bar\phi=0$ and the system is in the symmetric phase. Note that this result is not obvious a priori: at the temperature $T_\star$, the renormalization and consistency conditions impose that the curvature of the potential is positive at $\phi=0$ and thus that the potential is convex in the vicinity of $\phi=0$, but there is no obvious reason why the potential should be globally convex.

\begin{figure}[!htbp] 
\begin{center}
\includegraphics[width=0.55\textwidth]{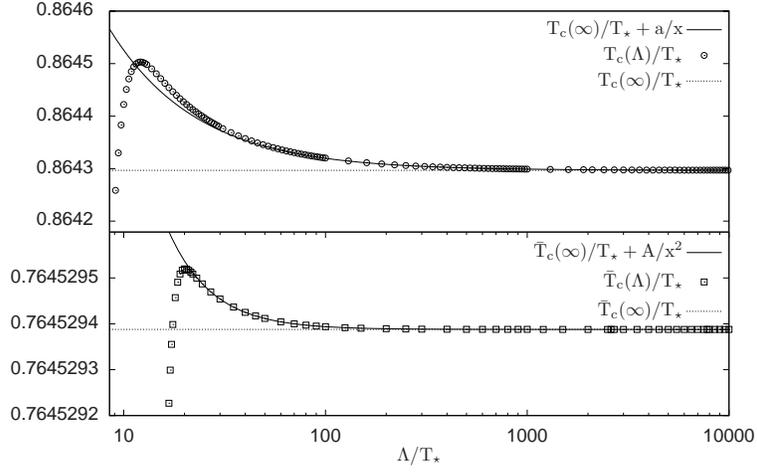}
\caption{Cutoff dependence of the critical temperatures $T_{\rm c}$ and $\bar T_{\rm c}$ (points) determined using numerical integration for parameters $m^2_\star/T_\star^2=0.04$ and $\lambda_\star=3,$ and their convergence towards the continuum values $T_{\rm c}(\infty)$ and $\bar T_{\rm c}(\infty).$ The different convergence rate shown by the fitted functions (solid lines), $1/\Lambda$ for $T_{\rm c}$ and $1/\Lambda^2$ for $\bar T_{\rm c},$ could be related to our choice of a sharp regulating function and the presence of a nonlocal sum-integral in the determination of $T_{\rm c}.$ \label{Fig:TcTcbar}}
\end{center}
\end{figure}

As we decrease the temperature away from $T_\star$, new extrema can appear, that is nontrivial solutions of the field equation. In particular, if a second order phase transition occurs at some critical temperature $T_{\rm c}$, nontrivial extrema are generated from $\phi=0$, because the curvature of the potential at $\phi=0$ vanishes and turns negative. The critical temperature $T_{\rm c}$ is then given by the equation
\beq\label{eq:def_Tc2}
\hat M^2_{\phi=0,T_{\rm c}}=0\,.
\eeq
This is an implicit equation for $T_{\rm c}$. Note however that since $\bar G_{\phi=0}$ is a free-type propagator, the determination of $T_{\rm c}$ only requires the calculation of perturbative sum-integrals. In this perturbative context, it is also relatively easy to prove that the curvature at $\phi=0$ and thus $T_{\rm c}$ possess a continuum limit, without relying on the general proof of renormalizability that we give in Sec.~\ref{sec:renorm_proof}. We start from the expression of the curvature at $\phi=0$ obtained from Eq.~\eqref{eq:curvature_1} by using the expression (\ref{eq:m2}) for $m_2$:
\beq\label{eq:curvature_2}
\hat M^2_{\phi=0}=m_\star^2+\frac{\lambda_2}{2}\big[{\cal T}[\bar G_{\phi=0}]-{\cal T}_\star[G_\star]\big]-\frac{\lambda^2_\star}{6} \big[{\cal S}[\bar G_{\phi=0}]-{\cal S}_\star[G_\star]\big].
\eeq
Introducing the splitting $\lambda_2=\lambda_{2{\rm l}}+\delta\lambda_{2{\rm nl}}$ and adding and subtracting an appropriate term\footnote{The reason for adding and subtracting this term is that the sum of the second and third lines of Eq.~(\ref{eq:toto}) is finite, as it can be checked by a direct calculation or by using the results of App.~\ref{app:th_exp}.} we obtain
\beq\label{eq:toto}
\hat M^2_{\phi=0}=m^2_\star & + & \frac{\lambda_{2{\rm l}}}{2}\big[{\cal T}[\bar G_{\phi=0}]-{\cal T}_\star[G_\star]\big]-\frac{\lambda_\star^2}{2}(\bar M^2_{\phi=0}-m^2_\star)\left[{\cal B}^2_\star[G_\star](0)+\frac{1}{3}\frac{d{\cal S}_\star[G_\star]}{dm^2_\star}\right]\nonumber\\
& + & \frac{\lambda^2_\star}{2}\,\Big[{\cal T}[\bar G_{\phi=0}]-{\cal T}_\star[G_\star]+(\bar M^2_{\phi=0}-m^2_\star){\cal B}_\star[G_\star](0)\Big]{\cal B}_\star[G_\star](0)\nonumber\\
& - & \frac{\lambda_\star^2}{6}\,\left[{\cal S}[\bar G_{\phi=0}]-{\cal S}_\star[G_\star]-(\bar M^2_{\phi=0}-m^2_\star)\frac{d{\cal S}_\star[G_\star]}{dm^2_\star}\right].
\eeq
Now, using Eq.~\eqref{eq:l2l2}, we arrive at
\beq
\hat M^2_{\phi=0}=m^2_\star & + & \frac{\lambda_0}{2}\big[{\cal T}[\bar G_{\phi=0}]-{\cal T}_\star[G_\star]\big]-\left[\bar M^2_{\phi=0}-m^2_\star-\frac{\lambda_0}{2}\big[{\cal T}[\bar G_{\phi=0}]-{\cal T}_\star[G_\star]\big]\right]\left(\frac{\lambda_{2{\rm l}}}{\lambda_0}-1\right)\nonumber\\
& + & \frac{\lambda^2_\star}{2}\,\Big[{\cal T}[\bar G_{\phi=0}]-{\cal T}_\star[G_\star]+(\bar M^2_{\phi=0}-m^2_\star){\cal B}_\star[G_\star](0)\Big]{\cal B}_\star[G_\star](0)\nonumber\\
& - & \frac{\lambda_\star^2}{6}\,\left[{\cal S}[\bar G_{\phi=0}]-{\cal S}_\star[G_\star]-(\bar M^2_{\phi=0}-m^2_\star)\frac{d{\cal S}_\star[G_\star]}{dm^2_\star}\right].
\eeq
Using the gap equation (\ref{eq:gap2}) at $\phi=0$ and the expression (\ref{eq:m0}) for $m^2_0$, one sees that the first line is simply equal to $\bar M^2_{\phi=0}$ which is convergent. Moreover, a direct calculation using the perturbative expressions for the tadpole, bubble and setting-sun sum integrals, shows that the sum of the second and the third line is also convergent, see App.~\ref{app:pert}. The convergent equation determining $T_{\rm c}$ reads then
\beq\label{eq:def_Tc}
0=\bar M^2_{\phi=0,T_{\rm c}} & + & \frac{\lambda^2_\star}{2}\,\Big[{\cal T}_{\rm c}[\bar G_{\phi=0,T_{\rm c}}]-{\cal T}_\star[G_\star]+(\bar M^2_{\phi=0,T_{\rm c}}-m^2_\star){\cal B}_\star[G_\star](0)\Big]{\cal B}_\star[G_\star](0)\nonumber\\
& - & \frac{\lambda_\star^2}{6}\,\left[{\cal S}_{\rm c}[\bar G_{\phi=0,T_{\rm c}}]-{\cal S}_\star[G_\star]-(\bar M^2_{\phi=0,T_{\rm c}}-m^2_\star)\frac{d{\cal S}_\star[G_\star]}{dm^2_\star}\right]
\eeq
with
\beq
\bar M^2_{\phi=0,T_{\rm c}}=m^2_\star+\frac{\lambda_\star}{2}\Big[{\cal T}_{\rm c}[\bar G_{\phi=0,T_{\rm c}}]-{\cal T}_\star[G_\star]+(\bar M^2_{\phi=0,T_{\rm c}}-m^2_\star){\cal B}_\star[G_\star](0)\Big].
\eeq
We solve this equation for $T_{\rm c}$ for different values of the cutoff $\Lambda$. The continuum limit $T_{\rm c}(\infty)$ can be computed using any regularization. We find it convenient to determine it using dimensional regularization.\footnote{In App.~\ref{app:pert}, using dimensional regularization, we provide an explicit continuum version the curvature $\hat M^2_{\phi=0}$, see Eqs.~\eqref{eq:comb_C} and \eqref{eq:comb_C2}.} Both for $T_{\rm c}(\Lambda)$ and $T_{\rm c}(\infty)$, the Matsubara sums are computed exactly and the remaining integrals over the modulus of the momentum are determined using adaptive numerical integration. These features allow for an accurate determination of $T_{\rm c}(\Lambda)$ and also for the study of the convergence of $T_{\rm c}(\Lambda)$ towards its continuum value $T_{\rm c}(\infty)$, as illustrated in Fig.~\ref{Fig:TcTcbar}. This represents a valuable information for the numerical resolution of the model since, due to memory limitations we cannot afford taking too large values of the cutoff while maintaining at the same time a good description of the infrared. We shall later use this accurate determination of $T_{\rm c}$ in order to test our numerical code.\\

Similarly, one can define a ``critical'' temperature $\bar T_{\rm c}$ for the gap mass, below which the gap equation at zero field has no solution \cite{Reinosa:2011ut}. It is defined by
\beq
\bar M^2_{\phi=0,\bar T_{\rm c}}=0\,.
\eeq
The way $\bar T_{\rm c}(\Lambda)$ approaches its continuum limit $\bar T_{\rm c}(\infty)$ is represented in Fig.~\ref{Fig:TcTcbar}. Note that when both $T_{\rm c}$ and $\bar T_{\rm c}$ exist, one has necessarily $T_{\rm c}\geq\bar T_{\rm c}$ since the very existence of $T_{\rm c}$ requires $\bar M^2_{\phi=0,T_{\rm c}}$ to be well defined, according to Eqs.~(\ref{eq:def_Tc}). However, these two temperatures do not coincide in general.\footnote{The cases for which $T_{\rm c}=\bar T_{\rm c}$ correspond to $m_\star=0$ and $\lambda_\star>0$ in which case $T_{\rm c}=\bar T_{\rm c}=T_\star$. If $m_\star=0$ and $\lambda_\star=0$, the temperatures $T_{\rm c}$ and $\bar T_{\rm c}$ are not determined since the gap mass is identically zero for any temperature and the potential is identically flat.} As already discussed in Sec.~\ref{sec:FE_geom}, one of the peculiarities of the 2PI formalism is that the different possible definitions of the two-point function do not necessarily coincide within a generic truncation. Because of this, in certain truncations, such as the two-loop approximation considered here, $\hat M^2_{\phi=0}$ is not equal to $\bar M^2_{\phi=0}$ and thus $T_{\rm c}\neq \bar T_{\rm c}$.\footnote{Note that there are other truncations such as the Hartree approximation, or the truncation that includes the ``basketball'' in $\Phi[\phi,G]$ which are such that $\hat M^2_{\phi}=\bar M^2_{\phi=0}$ and thus such that $T_{\rm c}=\bar T_{\rm c}$.} In general, the temperature $T_{\rm c}$, where the curvature of the effective potential at $\phi=0$ vanishes, either signals the vanishing of the field expectation value in a second order phase transition or it represents the lower spinodal temperature in a first order phase transition. The temperature $\bar T_{\rm c}$ resembles more the critical temperature in statistic physics, since the vanishing of the gap mass means enhanced fluctuations. Note also that below this temperature the potential at $\phi=0$ cannot be accessed because the gap equation does not admit a positive solution at $\phi=0.$

\subsection{Nature of the transition}
\begin{figure}[htbp] 
\begin{center}
\includegraphics[width=0.525\textwidth,angle=0]{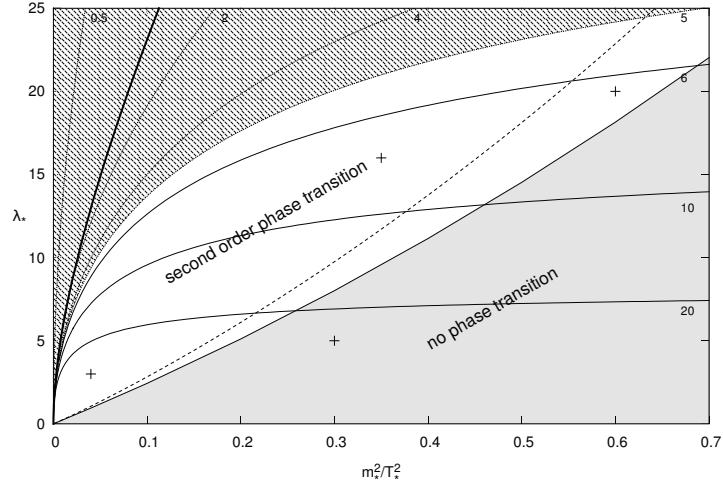}
\caption{The nature of the transition in a wide range of the parameter space inferred from the numerical study performed in the selected points indicated with a cross. In the two-loop approximation the boundary between the regions with no phase transition and second order phase transition is represented by the solid line. Along this line $T_{\rm c}=0.$ The dashed line is the boundary between the regions with no phase transition and first-order phase transition in the Hartree approximation. In the two-loop case this curve corresponds to $\bar T_{\rm c}=0.$ The meaning of the other curves is the same as in Fig.~4 of \cite{Reinosa:2011ut}: the labels indicate the value of $\ln(\Lambda_{\rm p}/m_\star),$ where $\Lambda_{\rm p}$ is the Landau pole. In the the region  $\ln(\Lambda_{\rm p}/m_\star)>5$ our results can be considered cutoff insensitive for a cutoff $\Lambda$ much larger than any physical scale but below the scale of the Landau pole. \label{Fig:params}}
\end{center}
\end{figure}

The existence of a solution to Eq.~(\ref{eq:def_Tc}) depends on the values of the parameters $m^2_\star/T^2_\star$ and $\lambda_\star$. By determining those values of the parameters for which $T_{\rm c}=0$, we can then separate the parameter space $(m^2_\star/T^2_\star,\lambda_\star)$ in two regions, corresponding to the white and grey areas depicted in Fig.~\ref{Fig:params}. A point in parameter space for which the system undergoes a second order phase transition belongs necessarily to the white region, for which a value of $T_{\rm c}$ can be defined. However, contrary to our discussion of the Hartree approximation \cite{Reinosa:2011ut}, we cannot draw analytical conclusions on the nature of the transition in one or the other region. We thus select a certain number of points in each region and investigate the nature of the transition numerically. In Fig.~\ref{Fig:params} the points that we have explored numerically are indicated with a cross.
\vglue2mm

\begin{figure}[!htbp] 
\begin{center}
\includegraphics[width=0.45\textwidth]{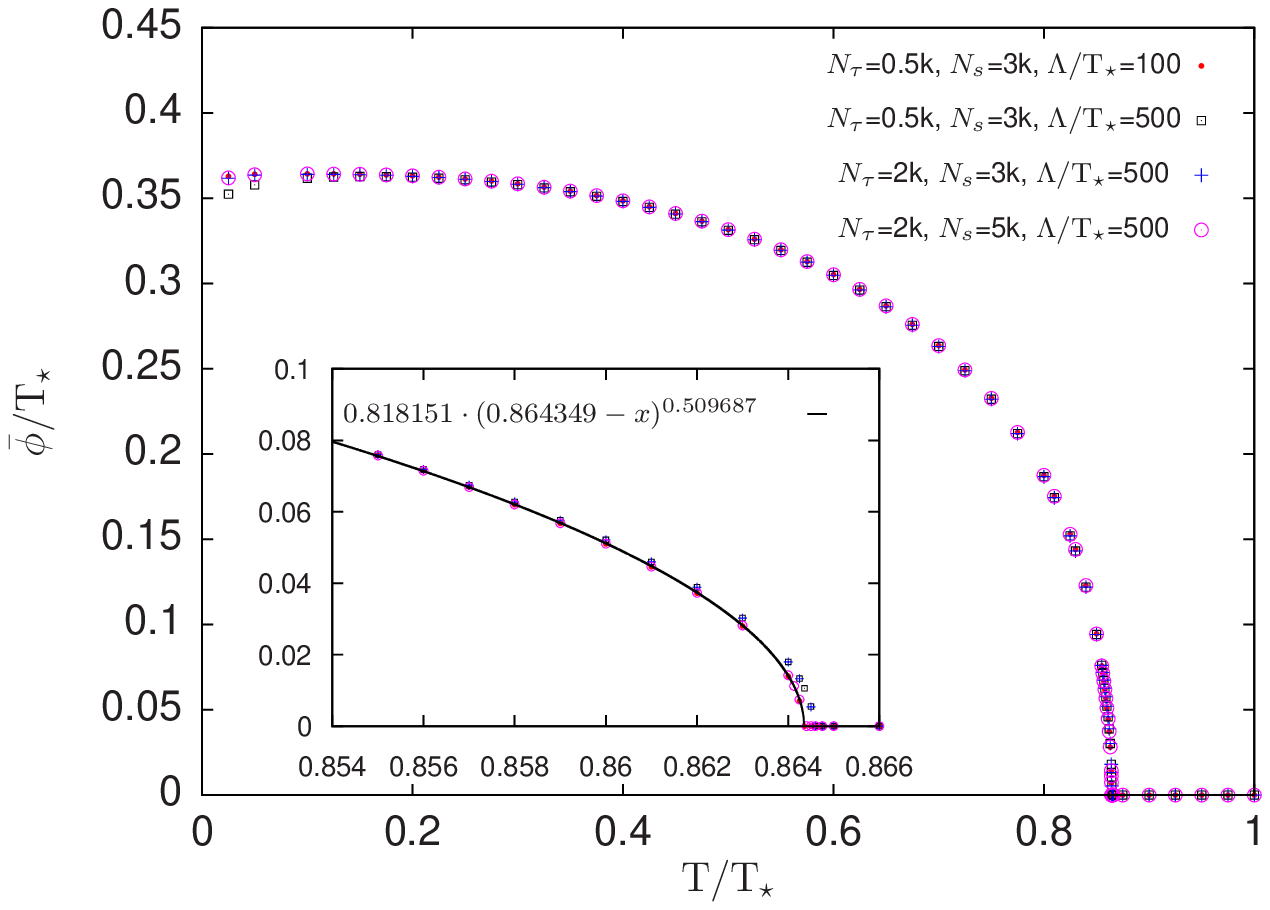}
\hspace*{0.5cm}
\includegraphics[width=0.45\textwidth]{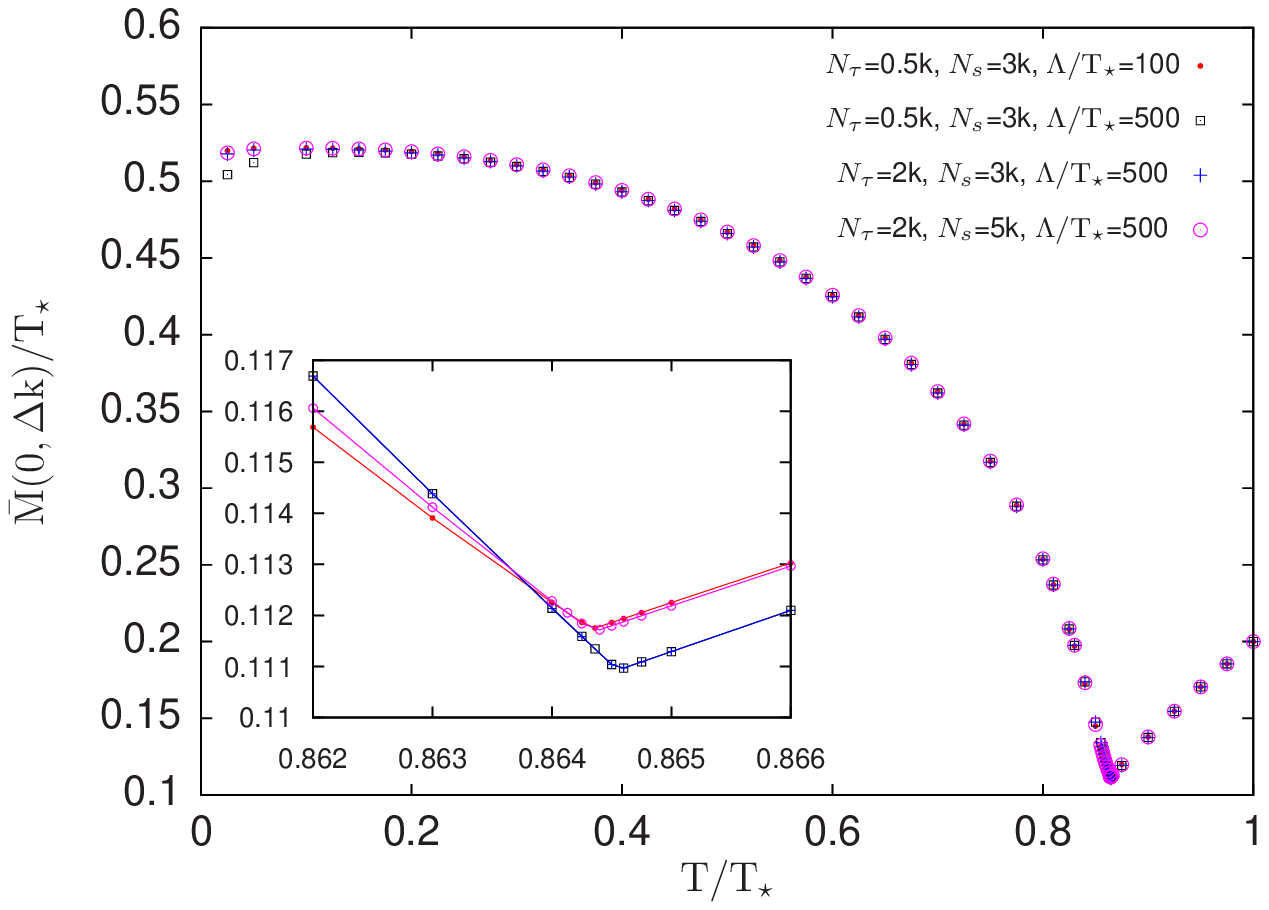}
\caption{Left panel: The variation of the order parameter $\bar\phi$ with the temperature. Right panel: The temperature dependence of the first bin of the self-energy. The minimum of this curve coincides with the temperature value $T_{\rm c},$ where $\bar\phi$ starts to develop a nonvanishing value. The insets show the discretization effects near $T_{\rm c}.$ The parameters are $m_\star^2/T_\star^2=0.04$ and $\lambda_\star=3.$ \label{Fig:phi-T}}
\end{center}
\end{figure}

\begin{figure}[!htbp] 
\begin{center}
\includegraphics[width=0.52\textwidth]{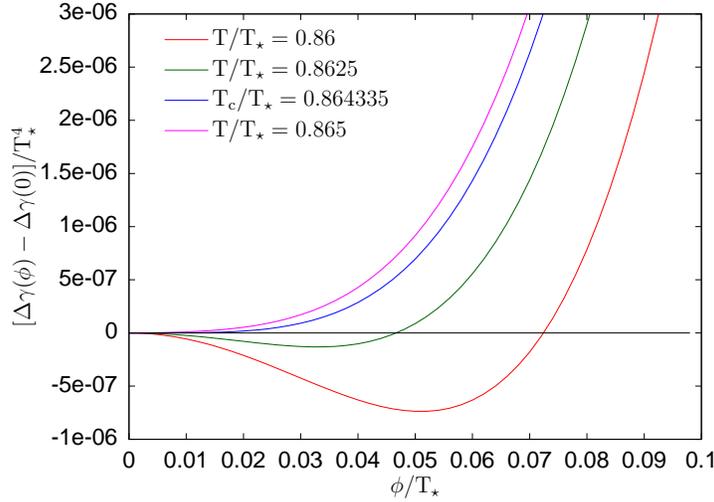}
\caption{The temperature evolution of the subtracted effective potential indicates a second order phase transition. The parameters are  $m_\star^2/T_\star^2=0.04$ and $\lambda_\star=3$ and the discretization is characterized by $\Lambda/T_\star=100,$ $N_\tau=512,$ and $N_s=3\times 2^{10}.$ \label{Fig:pot}} 
\end{center}
\end{figure}

We investigate first the nature of the phase transition for the parameters $m^2_\star/T_\star^2=0.04$ and $\lambda_\star=3$. For these parameters the phase transition was of the first order in the Hartree approximation, as one can see in Fig.~4 of \cite{Reinosa:2011ut}. These values of the parameters were also used in \cite{Arrizabalaga:2006hj}, where the model was solved in the same approximation, but in Minkowski space. Lowering the temperature from $T_\star,$ we follow the value of the order parameter $\bar\phi$ obtained from the solution of the coupled set of gap and field equations. We see in Fig.~\ref{Fig:phi-T} that $\bar\phi$ remains equal to zero down to some value of the temperature, which turns out to be $T_{\rm c}$, since at this temperature the curvature of the potential at $\phi=0$ vanishes, as it is clear from Fig.~\ref{Fig:pot}. Below $T_{\rm c}$, $\bar\phi$ starts to develop a non-vanishing value. No other extrema appear between $T_\star$ and $T_{\rm c}$, as it can be cross-checked by looking at the shape of the effective potential. This is a radically different behavior than the one obtained in the Hartree approximation, where some nontrivial extrema appeared and became the absolute ones before the curvature at $\phi=0$ could vanish.
\vglue2mm

In conclusion, the temperature variation of the field expectation value $\bar \phi$ shown in Fig.~\ref{Fig:phi-T} corroborated by the change of shape of the effective potential with the temperature shown in Fig.~\ref{Fig:pot} indicates a second order phase transition for the value of the parameters studied, in accordance with the result of \cite{Arrizabalaga:2006hj}. We observe a similar behavior for two other points chosen in the white region and we believe that this result generalizes to a large part of the white region. It would be interesting to study our approximation for very small couplings and see if this region is dominated by the Hartree approximation, which would imply a first order phase transition, in line with the Monte Carlo results of \cite{Bordag:2010ph}. This would be numerically challenging since for such small couplings, it would be difficult to differentiate a second order phase transition from a weakly first order one. For the point that we have tested in the grey region, the potential remains convex all the way down to $T=0.$
\vglue2mm

We can now go back to Fig.~\ref{Fig:params} and compare it to a similar figure obtained in the analytic investigation of the Hartree approximation in \cite{Reinosa:2011ut}. In this study the parameter space was also divided essentially in two regions. The separating boundary of the Hartree approximation is now represented in Fig.~\ref{Fig:params} by a dashed line and corresponds to those points for which $\bar T_{\rm c}$ vanishes. In other words, for points below the dashed line there is no $\bar T_{\rm c}$ and the potential can be evaluated at $\phi=0$ down to $T=0.$ For points above the dashed line, there is a $\bar T_{\rm c}$ and the potential cannot be evaluated in the vicinity of $\phi=0$ below $\bar T_{\rm c}$. As the analytic investigation of \cite{Reinosa:2011ut} revealed, in the Hartree approximation, points of the parameter space above the dashed boundary correspond to systems which undergo a first order phase transition and points below the dashed boundary correspond to systems that remain in the symmetric phase.\footnote{The dashed boundary is in fact a very narrow band along the boundary line of the Hartree approximation and completely indistinguishable from this line at the scale of the figure. For points inside this band, although nontrivial extrema develop between $T_\star$ and $T=0,$ the trivial minimum persists down to $T=0.$} The inclusion of the setting-sun diagram in the functional $\Phi[\phi,G]$ seems to change the nature of the transition to second order while enlarging the parameter space for which a transition occurs since the solid boundary line is pushed deeper in the no phase transition region of the Hartree case.\\

Similarly, we can study the effective potential using the RG-improved two-loop approximation. Note that since $\hat M^{\mbox{\tiny RG}}_{\phi=0}=\bar M_{\phi=0}$ (see Eq.~\eqref{eq:RG_masses}), the critical temperature which characterizes the sign change of the curvature at $\phi=0$ is also the critical temperature $\bar T_{\rm c}$ related to the vanishing of the gap mass. Since the gap mass at $\phi=0$ is not defined below $\bar T_{\rm c}$, neither is the running of the mass in our improved scheme which is thus only defined in the symmetric phase. For those points in parameter space that we tested, we observed either no transition or a second order phase transition at $\bar T_{\rm c}$ in the sense that the potential remained convex down to $\bar T_{\rm c}$ where its curvature at $\phi=0$ vanishes. In fact, the effective potential becomes very flat at $\bar T_{\rm c}$. This is because, the fourth derivative of the RG-improved effective potential at $\phi=0$ equals $\bar V_{\phi=0,T}$ and thus vanishes at $\bar T_{\rm c}$. Of course we cannot test what happens in the broken phase since our RG-improvement is defined only for $T\geq \bar T_{\rm c}$. Note also that the boundary between a second order phase transition and no phase transition coincides in this case with the boundary between a first order phase transition and no phase transition in the Hartree approximation. Note finally that the RG-improvement can also be applied to the Hartree approximation. In this case, one can show that the curvature of the potential for any value of the field changes from $\hat M^2_\phi=\bar M^2_\phi+(\bar V_\phi-\lambda_\star)\phi^2$, where $\bar V_\phi$ is the generalization of $\bar V_{\phi=0}$ to nonzero field, to $(\hat M^{\mbox{\tiny RG}}_\phi)^2=\bar M^2+(\bar V_\phi-\bar V_{\phi=0})\phi^2$. It can also be shown that $\bar V_\phi$ increases with $\phi^2$ from which it follows that $(\hat M^{\mbox{\tiny RG}}_\phi)^2\geq 0$ for any value of $\phi$ and down to the temperature $\bar T_{\rm c}$ at which $(\hat M^{\mbox{\tiny RG}}_{\phi=0})^2=\bar M^2_{\phi=0}=0$. Thus, in the RG-improved Hartree approximation, the potential remains convex down to the transition temperature.

\subsection{Thermodynamical properties}

In this subsection we study the bulk thermodynamic properties of the model based on the pressure and quantities derived from it, such as the interaction measure (trace anomaly), the heat capacity and the speed of sound.

The pressure is obtained from the subtracted effective potential given in Eq.~\eqref{eq:pot_prat} as 
\beq\label{eq:pressure}
p(T) = \lim_{T_0\to 0}\left[\Delta\gamma(\bar\phi)\big|_{T_0}-\Delta\gamma(\bar\phi)
\right].
\eeq
Actually, we cannot evaluate $\Delta\gamma(\bar\phi)$ exactly at $T_0=0$ because we can take into account only a limited number of Matsubara modes. Therefore, we are constrained to approximate the value of $\Delta\gamma(\bar\phi)$ at $T=0.$ In order to do so, we first determine $\Delta\gamma(\bar\phi)$ for smaller and smaller values of $T$ by progressively increasing $N_\tau,$ then we fit a high order polynomial to the available data points and obtain an estimation of $\Delta\gamma(\bar\phi)\big|_{T=0}$ by evaluating it at $T=0.$ The value is accepted if the scaled pressure $p/p_{\mbox{\tiny SB}}$ is a decreasing function of $T$ for $T\to 0,$ where $p_{\mbox{\tiny SB}}(T)=\pi^2 T^4/90$ is the pressure of the massless boson gas. In the opposite case, we increase $N_\tau,$ determine the subtracted potential for smaller $T$ and redo the fit.\footnote{For example at parameters $m_\star^2/T_\star^2=0.04$ and $\lambda_\star=3$ we used $\Lambda/T_\star=100$ and $N_s=3\times 2^{10}$ and increased $N_\tau$ from $512,$ used for $T/T_\star\ge 0.2,$ to $N_\tau=4\times 2^{10}$ for $0.1\le T/T_\star<0.2,$ and to $N_\tau=8\times 2^{10}$ for $0.06\le T/T_\star<0.1.$}

\begin{figure}[!tbp] 
\begin{center}
\includegraphics[width=0.8\textwidth]{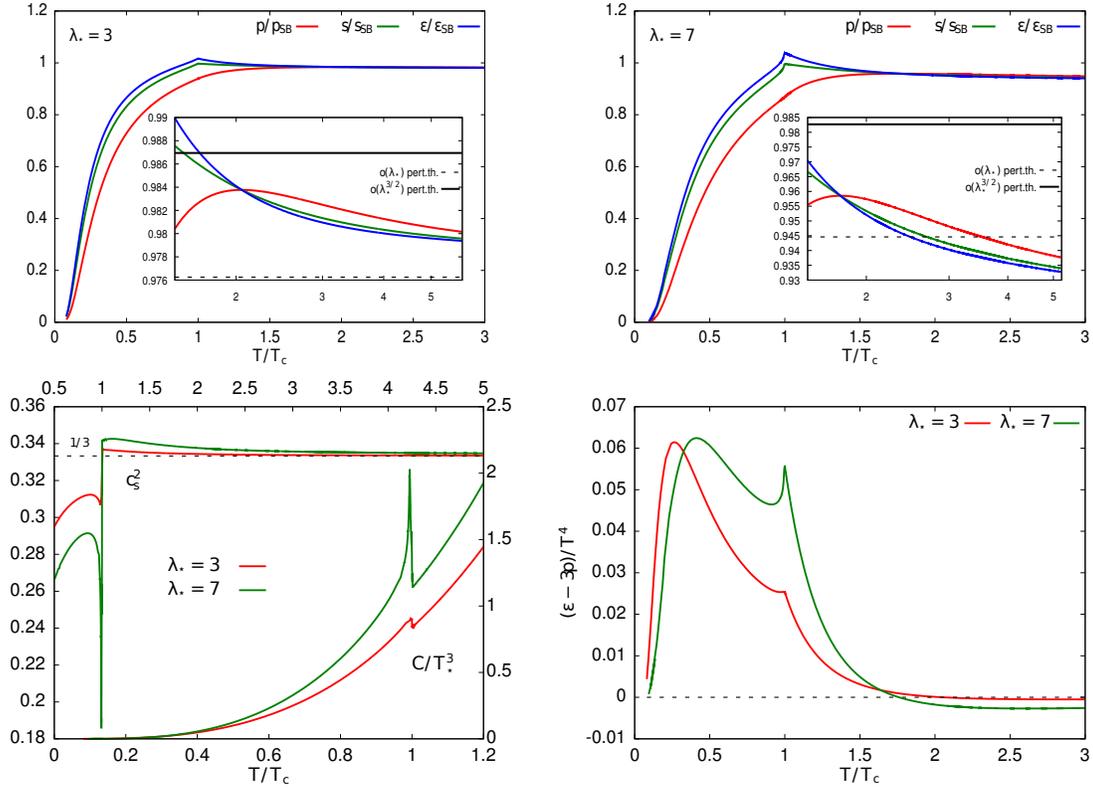}
\caption{Bulk thermodynamic quantities as a function of temperature for two different coupling values: the scaled pressure $p/p_{\mbox{\tiny SB}}$, entropy density $s/s_{\mbox{\tiny SB}}$ and energy density $\varepsilon/\varepsilon_{\mbox{\tiny SB}}$  (upper row), the square of the speed of sound and the heat capacity (lower left panel), the trace anomaly $(\varepsilon -3 p)/T^4$ (lower right panel). The insets in the upper panels show in a log-linear plot the perturbative results for the relative quantities obtained using a high temperature expansion up to and including ${\cal O}(\lambda_\star)$ and ${\cal O}(\lambda_\star^{3/2})$ (see Eq.~\eqref{eq:pert_HTE_p}). The upper and left axis of the plot in the lower left panel correspond to $c_{\rm s}^2,$ while the lower and right axis correspond to the heat capacity $C.$ The discretization parameters are those of Fig.~\ref{Fig:pot} and $m_\star^2/T_\star^2=0.04.$ \label{Fig:bulk}} 
\end{center}
\end{figure}

Having determined the pressure as a function of the temperature, the energy density is given by $\varepsilon=-p+Ts,$ where the entropy density is obtained using a numerical derivative as $s=d p/dT.$ The heat capacity $C=d\varepsilon /d T$ is obtained numerically as the second derivative of the pressure: $C=T d^2p/d T^2.$ The square of the speed of sound $c_{\rm s}^2=d p/d\varepsilon$ is determined from $c_{\rm s}^2=s/C$ and the trace anomaly of the energy momentum tensor $T^{\mu\nu}$ is obtained as $\Delta=T^\mu_{\,\,\mu}/T^4=(\varepsilon -3 p)/T^4$ or equivalently as $\Delta = Td(p/T^4)/d T.$ All these quantities displayed in Fig.~\ref{Fig:bulk} show nicely the second order nature of the transition: the scaled energy and entropy densities $\varepsilon/\varepsilon_{\mbox{\tiny SB}}$ and $s/s_{\mbox{\tiny SB}},$ and the trace anomaly display a cusp at $T_{\rm c},$ while the second derivative of the pressure with respect to the temperature is discontinuous, as displayed by the speed of sound and the heat capacity. The discontinuity is more pronounced at a larger value of the coupling. 

In the upper row of Fig.~\ref{Fig:bulk} we display the temperature dependence of the scaled pressure, entropy and energy densities calculated for two different couplings. These curves cross each other at the value of the temperature at which the scaled pressure has a maximum. This is because at this temperature, the interaction measure vanishes (since $p/p_{\mbox{\tiny SB}}\propto p/T^4$) and thus $\varepsilon=3p$ and $Ts=4p$. Moreover, since $\varepsilon_{\mbox{\tiny SB}}=3p_{\mbox{\tiny SB}}$ and $Ts_{\mbox{\tiny SB}}=4p_{\mbox{\tiny SB}}$, it follows that $p/p_{\mbox{\tiny SB}}=s/s_{\mbox{\tiny SB}}=\varepsilon/\varepsilon_{\mbox{\tiny SB}}$. In the insets of these plots we compare the temperature dependence of the pressure with the first terms in the perturbative expansion obtained at high temperature \cite{Arnold:1994ps} 
\beq\label{eq:pert_HTE_p}
p_{\rm pert}(T)=p_{\mbox{\tiny SB}}(T)\left[1-\frac{5\lambda_\star}{64\pi^2}+\frac{5\sqrt{6}\lambda^{3/2}_\star}{192\pi^3}+{\cal O}(\lambda^2_\star)\right],
\eeq
where the neglected higher order terms depend on the chosen renormalization scale. Note also that, although the formula was obtained in \cite{Arnold:1994ps} in the $\overline{\rm MS}$ scheme, one can use it with coupling $\lambda_\star$ because the differences between the two renormalization schemes appear only at higher order in the coupling. The pressure obtained in the current approximation at coupling $\lambda_\star=3$ is closer to the ${\cal O(\lambda_\star)}$ perturbative result for $T>3.5 T_{\rm c}.$ For the larger coupling constant, $\lambda_\star=7,$ the pressure goes below the ${\cal O(\lambda_\star)}$ perturbative result but for such high value of the coupling it makes less sense to compare to the perturbative expansion.

At high temperatures the trace anomaly vanishes and $\varepsilon/(3 p)$ goes to $1$, in such a way that, interestingly, $\varepsilon-3p$ is negative and its magnitude increases with the temperature. The fact that $\varepsilon/(3 p)\to 1$ is reflected in the square of the speed of sound, which approaches at high temperature the value $1/3,$ called the conformal limit because in a conformal invariant theory in three dimensions $c_{\rm s}^2=1/3.$  For low temperature the trace anomaly shows a bump, for both values of the coupling investigated. At the larger value of the coupling $\lambda_\star=7$ the cuspy structure becomes more prominent. These interesting features were already observed in \cite{Li:2009by}.

\subsection{Critical exponents}

There are six static critical exponents $\alpha,$ $\beta,$ $\gamma,$ $\delta,$ $\eta,$ and $\nu$, but, as a consequence of the static scaling hypothesis for the thermodynamic and correlation functions, which is verified in particular in the presence of a fixed point in the renormalization group flow \cite{stanley71}, there exist four scaling relations between them, so that only two of them are independent. Usually $\eta$ and $\nu$ are chosen and the other exponents can be determined from\footnote{The first and third scaling relations are the Josephson and Fisher identities and instead of the second and fourth one one could use equivalently the Widom and Rushbrooke relations: $\gamma=\beta(\delta-1)$ and $\alpha+2\beta+\gamma=2$ \cite{Goldenfeld:1992qy}.}
\beq\label{eq:scaling}
\alpha=2-d\nu\,,\quad \beta=(d-2+\eta)\,\frac{\nu}{2}\,,\quad\gamma=(2-\eta)\nu\,, \quad {\rm and} \quad \delta=\frac{d+2-\eta}{d-2+\eta}\,.
\eeq
Note however that there is a priori no reason why these relations should hold in a given approximation of the theory, such as for instance the two-loop $\Phi$-derivable approximation that we consider here. In what follows we determine the critical exponents in the two-loop and in the RG-improved two-loop $\Phi$-derivable approximations and discuss which of the scaling relations are fulfilled.

\subsubsection{Critical exponents in the two-loop approximation}
First of all note that there is a priori an ambiguity in the determination of certain critical exponents. For instance, in order to obtain the exponent $\eta$, we should study the behavior of the propagator at criticality. One possibility is to study $\bar G$. The corresponding critical temperature is $\bar T_{\rm c}$ and not $T_{\rm c}$ and the propagator should be evaluated at $\phi=0$ down to $\bar T_{\rm c}$.\footnote{If one evaluates $\bar G$ at $\phi=0$ only down to $T_{\rm c}>\bar T_{\rm c}$ and at $\bar G_{\bar\phi}$ for $T<T_{\rm c}$, $\bar G$ never reaches criticality, see Fig.~\ref{Fig:phi-T}.} But since $\bar G_{\phi=0}$ is local, we conclude that $\bar \eta^+=0$. We could instead consider the propagator obtained from the second derivative of the effective action, which generalizes the effective potential to non homogeneous configurations of the field. We would obtain a momentum dependent ``curvature''
\beq
\hat{M}^2_{\phi=0}(K)=K^2\delta Z+m^2_2+\frac{\lambda_2}{2}\,{\cal T}[\bar G_{\phi=0}]-\frac{\lambda_\star^2}{6}\,{\cal S}[\bar G_{\phi=0}](K)\,,
\eeq
where ${\cal S}[\bar G_{\phi=0}](K)$ is the momentum dependent setting-sun sum-integral with propagator $\bar G_{\phi=0}.$ At $T_{\rm c}$ this self-energy is critical, in the sense that its value for $K=0$ vanishes. However, since $\bar G_{\phi=0}$ is massive, the corresponding propagator shows no anomalous dimension. We conclude then that $\hat\eta^+=0$. Then, even though the definition of $\eta$ is ambiguous, in the present case, both approaches lead to the same result $\bar\eta^+=\hat\eta^+=0$, which coincides with that of the mean-field approximation.\\

Similar remarks apply to the critical exponent $\nu$. If we define the correlation length by $\bar\xi\propto\bar M_{\phi=0}^{-1}$, its scaling can be obtained by subtracting the renormalized gap equation at $\bar T_{\rm c}$ from the renormalized gap equation at temperature $T$, that is:
\beq
\bar M^2_{\phi=0}=\frac{\lambda_\star}{2}\big[{\cal T}[\bar G_{\phi=0}]-{\cal T}_{\bar T_{\rm c}}[G_0]+\bar M^2_{\phi=0}{\cal B}_\star[G_\star](0)\big]\, ,
\eeq
with $G_0(Q)\equiv 1/Q^2$. Using the high temperature expansion of the tadpole sum-integral given in Eq.~\eqref{eq:tadpole_HTE}, which is justified since $\bar M_{\phi=0}\rightarrow 0$ as $T\rightarrow \bar T_{\rm c}$, and neglecting the contributions of order $\bar M^2_{\phi=0}$, we obtain 
\beq\label{eq:HTE_mass}
\bar M_{\phi=0}\sim\frac{\pi}{3T}\left(T^2-\bar T_{\rm c}^2\right)\sim \frac{2\pi}{3}\big(T-\bar T_{\rm c}\big),
\eeq
from which it follows that $\bar\nu^+=1$. We can alternatively define the correlation length from $\hat\xi\propto\hat M_{\phi=0}^{-1}.$ The way the curvature vanishes at $T_{\rm c}$ is studied below when determining the exponent $\gamma$. We obtain that $\bar M^2_{\phi=0}$ vanishes linearly as $T-T_{\rm c}$ from which it follows that $\hat\nu^+=1/2$. In order to solve this ambiguity, note that the nature of the transition is determined from the change of shape of the potential at $T_{\rm c}$. The relevant value for the critical exponent is thus $\hat\nu^+=1/2$, which is again equal to the value obtained in the mean field approximation.\\

\begin{figure}[!htbp] 
\begin{center}
\includegraphics[width=0.42\textwidth]{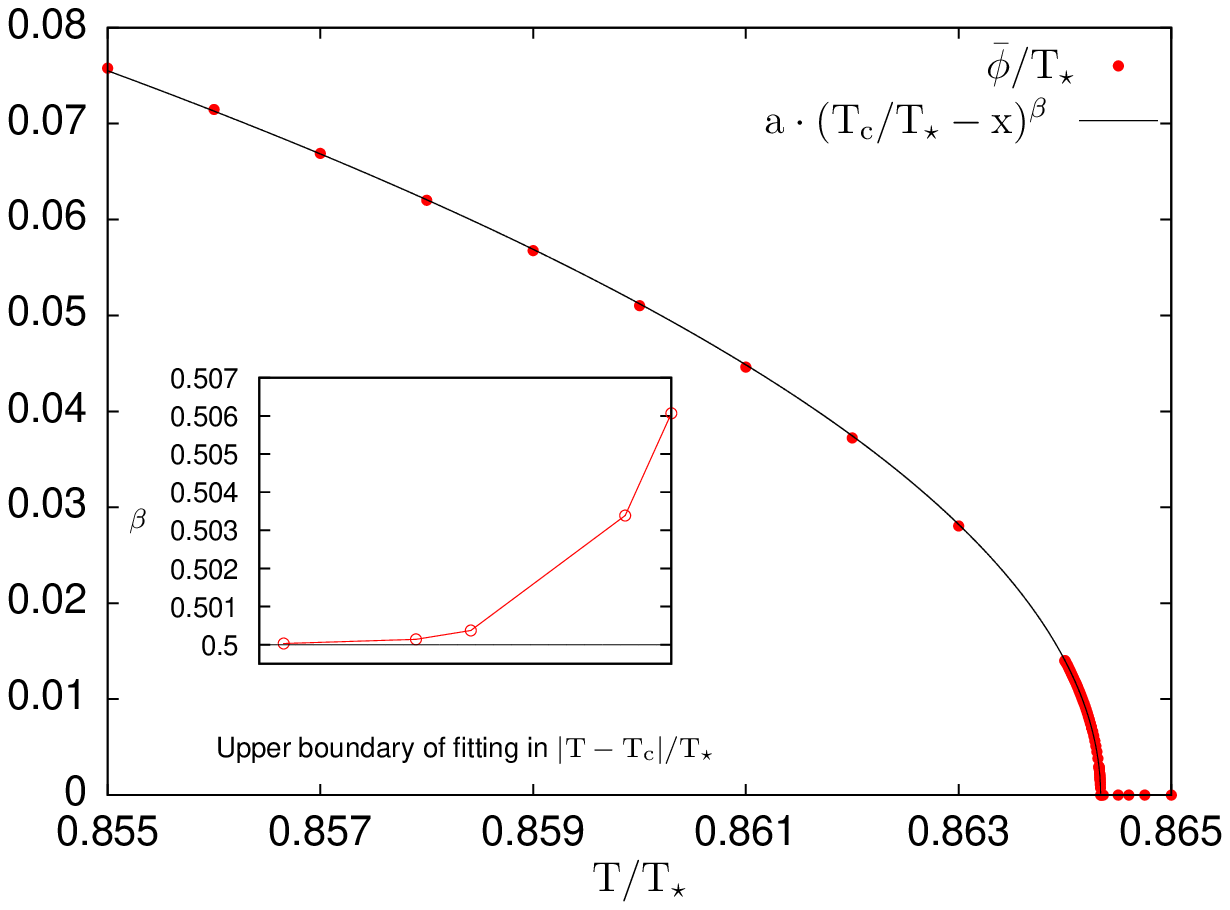}
\hspace*{0.5cm}
\includegraphics[width=0.42\textwidth]{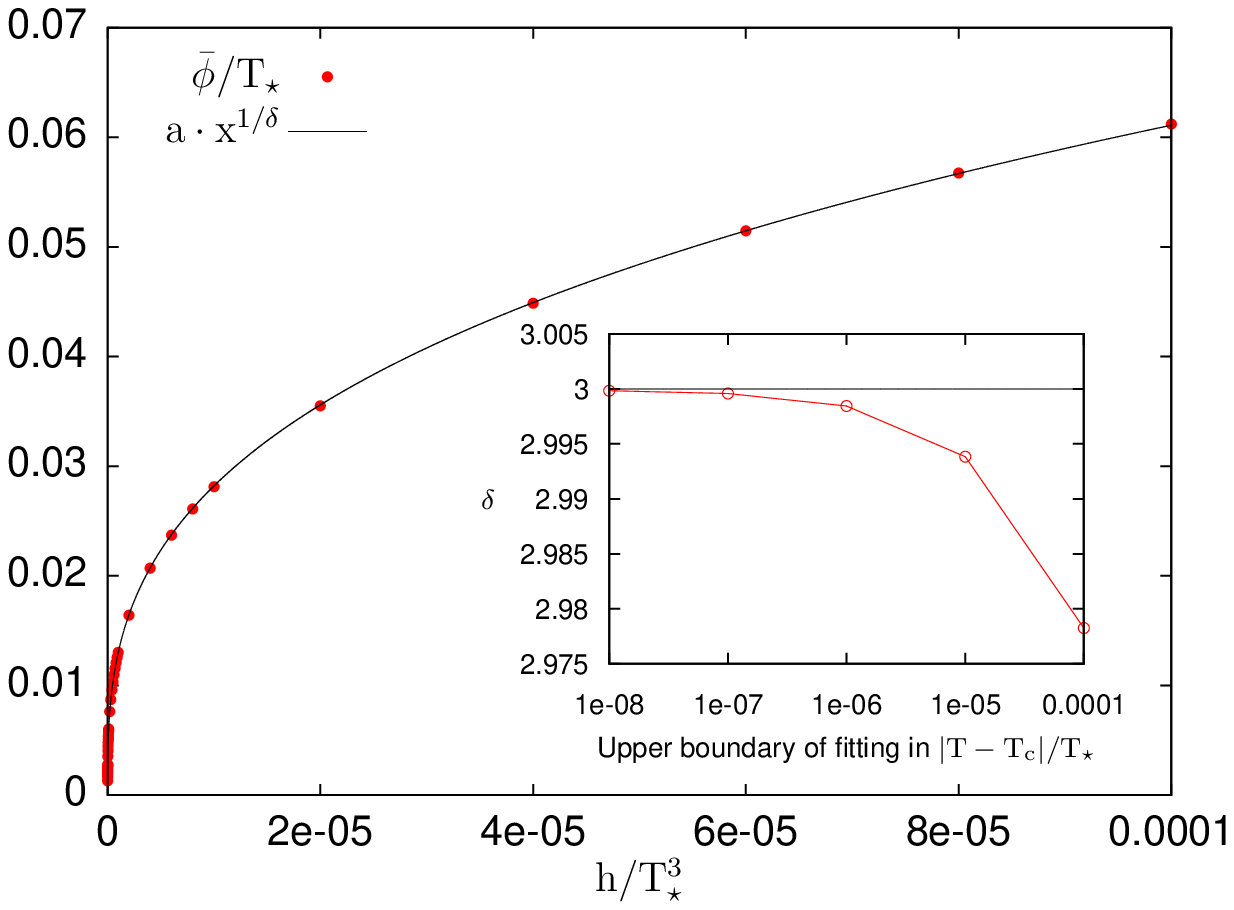}
\caption{Numerical determination of the critical exponents $\beta$ (left panel) and $\delta$ (right panel) through fits to the field expectation value $\bar\phi.$ The parameters and discretization used are the same as in Fig.~\ref{Fig:pot}. The inserts show the dependence of the the fitted values of $\beta$ and $\delta$ on the upper boundary of the fitting window. As the fitting window shrinks, the fitted exponents converge to the mean field values $1/2$ and $3,$ respectively.\label{Fig:crexp_bd}}
\end{center}
\end{figure}

The critical exponent $\beta$ by fitting $\bar\phi$ to $|T_{\rm c}-T|^\beta$. This requires first an accurate determination of $T_{\rm c}$ from our numerical results.\footnote{This value of $T_{\rm c}$ is not the same numerically than the one obtained from Eq.~(\ref{eq:def_Tc}) using accurate numerical integration of perturbative integrals. We will later use these two different ways to obtain $T_{\rm c}$ in order to test our numerical procedure.} We could proceed by locating precisely the temperature at which $\bar\phi$ starts developing a non-zero value. However, since the temperature derivative of $\bar\phi$ is infinite at $T^-_{\rm c},$ it is easier to determine the value of $T_{\rm c}$ by locating the minimum of the self-energy at the lowest available momentum and frequency: indeed the self-energy reaches a minimum value when $\bar\phi$ starts to develop a nonvanishing value. This is shown in the inset of the right panel of Fig.~\ref{Fig:phi-T}. Once $T_{\rm c}$ has been determined the exponent $\beta$ can be fitted. As shown in the Fig.~\ref{Fig:crexp_bd}, the fit is compatible with the mean-field value $\beta=1/2$. A similar method is used to determine the exponent $\delta$. We introduce an external field $h$ (this amounts to shifting the effective potential by $-h\phi$), we set the temperature $T$ to the numerically determined value of $T_{\rm c}$ and fit $\bar\phi$ to $h^{1/\delta}$. The results are compatible with the mean-field value $\delta=3$, see Fig.~\ref{Fig:crexp_bd}.\\

\begin{figure}[!htbp] 
\begin{center}
\includegraphics[width=0.46\textwidth]{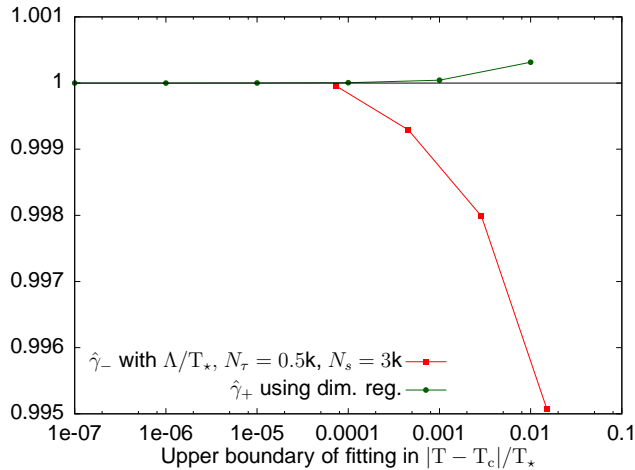}
\caption{Numerical determination of the critical exponent $\hat\gamma_-$ (lower curve) in the broken symmetry phase and $\hat\gamma_+$ (upper curve) in the symmetric phase by fitting $a(T-T_{\rm c})^{\hat\gamma_\pm}$ to the curvature. In the symmetric phase the momentum independent curvature is determined using dimensional regularization. The critical exponent converges to the value $1$ when shrinking the size of the fitting window around $T_{\rm c}$. The parameters are $m^2_\star/T_\star^2=0.04$ and $\lambda_\star=3.$ \label{Fig:gamma}} 
\end{center}
\end{figure}

In order to obtain $\gamma$, we fit the susceptibility $\chi\equiv \partial\bar\phi/\partial h$ at $h=0$ to a power law $|T_{\rm c}-T|^{-\bar\gamma}$. Because, in the exact theory
\beq
\left.\frac{\partial\bar\phi}{\partial h}\right|_{h=0}=\left.\left(\frac{\delta^2\gamma}{\delta \phi^2}\right)^{-1}\right|_{\bar\phi}=(\hat M^2_{\bar\phi})^{-1}\,,
\eeq
we can also fit the inverse curvature of the potential to $|T_{\rm c}-T|^{-\hat\gamma}$. Note that in a given truncation, such as the approximation considered here, there is an ambiguity in the determination of $\gamma$ because there is no reason a priori why $\bar\gamma$ should equal $\hat\gamma$. Our numerical results for $\hat\gamma$ are again compatible with the mean-field value $\hat\gamma_-=\hat\gamma_+=1$, see Fig.~\ref{Fig:gamma}. Note that $\hat\gamma_+$ was obtained using dimensional regularization. Indeed, as we already discussed, in the symmetric phase, the formula for the curvature at $\phi=0$ involves only perturbative integrals which can be evaluated using dimensional regularization. Of course, since the curvature is finite, its continuum result does not depend on the regularization chosen to obtain it. The determination of $\bar\gamma^+$ and $\bar\gamma^-$ would be numerically more involved.\\

Finally, the heat capacity has already been determined in the previous section together with other thermodynamical quantities, see Fig.~\ref{Fig:bulk}. It presents a discontinuity at $T_{\rm c},$ as it is the case in the mean-field approximation. To this behavior, one attributes conventionally the value $\alpha=0$ for the critical exponent $\alpha$. To summarize, in the two-loop $\Phi$-derivable approximation, the critical exponents coincide with those in the mean field approximation. In a sense, although it predicts the correct order of the transition, the two-loop approximation is not enough to produce non-analyticities in the effective potential which would modify the Ginzburg-Landau picture.

\subsubsection{RG-improved critical exponents}\label{sec:RG_num}

As already explained, the RG improvement that we have introduced is only applicable for temperatures above the transition temperature which in this case is equal to $\bar T_{\rm c}$. We can only determine the critical exponents by approaching the transition from the symmetric phase. In particular, we cannot access the exponent $\beta$.\\

An interesting feature of the RG-improved approximation is that, in the symmetric phase, there is no difference between $\hat M^{\mbox{\tiny RG}}_{\phi=0}$ and $\bar M_{\phi=0}$. The determination of $\nu_{\mbox{\tiny RG}}^+$ is then not ambiguous and coincides with that of $\bar\nu^+$ in the previous section. Then $\nu_{\mbox{\tiny RG}}^+=1$ which differs from the mean field value $1/2$. The value of $\eta$ remains equal to $0$.\\

In order to determine the exponents $\delta_{\mbox{\tiny RG}}$ and $\gamma_{\mbox{\tiny RG}}^+$, we can take advantage of some simplifications which occur in the RG-improved field equation at $\bar T_{\rm c}$. Remember first that the RG-improved equation is obtained by applying the replacements $m_\star\rightarrow \bar M_{\phi=0}$ and $\lambda_\star\rightarrow\bar V_{\phi=0}$. Because $\bar V_{\phi=0}$ goes to zero as $T$ approaches the transition temperature $\bar T_{\rm c}$, we will be able to neglect a certain number of contributions. Moreover, since $\bar M_{\phi=0}$ goes also to zero, we will be able to use high temperature expansions for some integrals calculated in dimensional regularization. We use, in particular, the expansion of the tadpole
\beq\label{eq:tadpole_HTE}
{\cal T}[\bar G_{\phi=0}] = \frac{T^2}{12}-\frac{\bar M_{\bar\phi=0} T}{4\pi}-\frac{\bar M^2_{\bar\phi=0}}{16\pi^2}\left[\frac{1}{\epsilon}+\ln\frac{\mu^2}{4\pi T^2}+\gamma_{\mbox{\tiny E}}\right]+{\cal O}\left(\frac{\bar M^4_{\bar\phi=0}}{T^2}\right)\,,
\eeq
from which we obtain
\beq\label{eq:B_HTE}
B[\bar G_{\phi=0}](0)\sim \frac{T}{8\pi \bar M_{\phi=0}} \quad \mbox{and} \quad \bar V_{\phi=0}\sim16\pi\frac{\bar M_{\phi=0}}{T}\,,
\eeq
as well as \cite{Parwani:1991gq}
\beq
\label{eq:S&dS_HTE}
{\cal S}[\bar G_{\phi=0}]\sim-\frac{T^2}{32\pi^2}\log\frac{\bar M^2_{\phi=0}}{T^2}
\quad\mbox{and}\quad
\frac{d{\cal S}[\bar G_{\phi=0}]}{d \bar M^2_{\phi=0}}\sim-\frac{T^2}{32\pi^2 \bar M^2_{\phi=0}}\,.
\eeq

\begin{figure}[!htbp] 
\begin{center}
\includegraphics[width=0.52\textwidth]{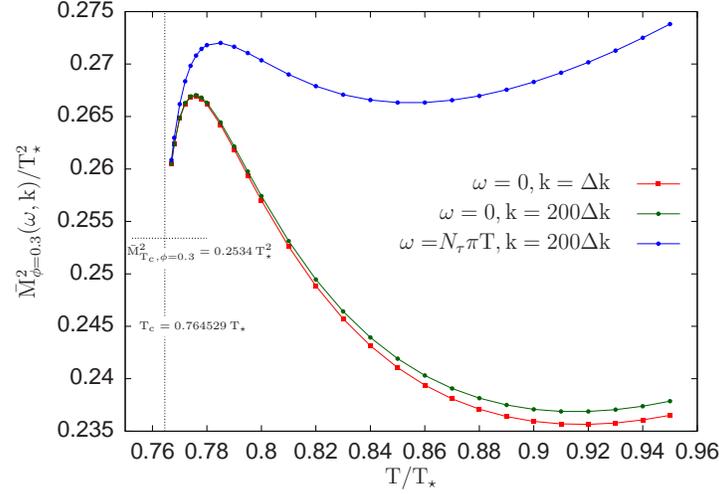}
\caption{Flattening of the momentum dependent self-energy at constant $\phi,$ as one approaches $\bar T_{\rm c}$ in the RG-improved case. The model parameters are $m_\star^2/T_\star^2 = 0.04,$ $\lambda_\star = 3$ and the discretization is characterized by $\Lambda = 50,$ $N_\tau = 2\times 2^{10}$, $N_s = 26\times 2^{10}.$ \label{Fig:RG-flat}} 
\end{center}
\end{figure}

In order to obtain the RG-improved gap and field equations, we can apply the above-mentioned replacements in Eqs.~\eqref{eq:gap2} and \eqref{eq:field_update}. The expressions for the bare couplings $m_2^2$, $\lambda_{2{\rm l}}$ and $\delta\lambda_{2{\rm nl}}$ become
\beq
(m_2^{\mbox{\tiny RG}})^2 & = & m^2_0-\frac{\lambda_2-\lambda_0}{2} {\cal T}[\bar G_{\phi=0}]+\frac{\bar V^2_{\phi=0}}{6}{\cal S}[\bar G_{\phi=0}]\,,\\
\lambda_{2{\rm l}}^{\mbox{\tiny RG}} & = & \lambda_0\left[1+\frac{\bar V^2_{\phi=0}}{2}\left( {\cal B}^2[\bar G_{\phi=0}](0)+\frac{1}{3}\frac{d {\cal S}[\bar G_{\phi=0}]}{d \bar M^2_{\phi=0}}\right)\right],\\
\delta\lambda_{2{\rm nl}}^{\mbox{\tiny RG}} & = & \bar V_{\phi=0}^2{\cal B}[\bar G_{\phi=0}]\,,
\eeq 
where we have used the expressions (\ref{eq:m0}), (\ref{eq:m2}), (\ref{eq:l2nl}) and (\ref{eq:l2l2}) for $m^2_0,$ $m^2_2,$ $\delta\lambda_{\rm 2nl},$ and $\lambda_{2{\rm l}}$. Using Eqs.~(\ref{eq:B_HTE}) and (\ref{eq:S&dS_HTE}), we find the following behaviors for these parameters as we approach $\bar T_{\rm c}:$
\beq
(m^{\mbox{\tiny RG}}_2)^2\rightarrow m^2_0-\frac{1}{3}\lambda_0{\cal T}_{\bar T_{\rm c}}[G_0]\,, \quad \lambda_{2{\rm l}}^{\mbox{\tiny RG}}\rightarrow\frac{5}{3}\lambda_0\,, \quad \delta\lambda_{2{\rm nl}}^{\mbox{\tiny RG}}\sim 32\pi\frac{\bar M_{\phi=0}}{\bar T_{\rm c}}\to 0\,.
\eeq
A similar analysis can be done for  $\lambda_4$ which is expressed in terms of $\lambda_0$ and $\lambda_{2{\rm l}}$ in Eq.~\eqref{eq:l4}. The last integral of \eqref{eq:l4} involves a three-loop sum-integral which we do not compute and whose high temperature expansion is not known to us. Therefore, we evaluated this integral numerically at constant temperature and found that its value goes as $\bar M^{-3}_{\phi=0}$ as the mass $\bar M_{\phi=0}$ goes to zero. Since this integral is multiplied by $\bar V^4_{\phi=0}(T),$ it gives no contribution as $T\to T_{\rm c}.$ For the other integrals the HTE is known. Using Eqs.~\eqref{eq:B_HTE} and (\ref{eq:S&dS_HTE}) we arrive finally at
\beq
\lambda^{\mbox{\tiny RG}}_4\to \frac{25}{3}\lambda_0\,. 
\eeq
Using these replacements, one obtains the following field equation (coupled to the gap equation) in the presence of the external field $h$
\beq
h&=&\bar\phi\left[\frac{2}{3}\left(m_0^2+\frac{\lambda}{2} {\cal T}_{\bar T_{\rm c}}[G_0]\right)+\frac{5}{3}\bar M^2_{\bar \phi,\bar T_{\rm c}}\right],\label{eq:field_imp}\\
\bar M^2_{\bar \phi,\bar T_{\rm c}}&=&m_0^2+\frac{\lambda_0}{2}\left[\frac{5}{3}\bar\phi^2+{\cal T}_{\bar T_{\rm c}}[\bar G_{\bar\phi,\bar T_{\rm c}}] \right],\label{eq:gap_imp}
\eeq
where $G_0(Q)\equiv 1/Q^2$. In obtaining these equations we have also used the fact that since the setting-sun sum-integral in the field equation and the bubble sum-integral in the gap equation are multiplied by $\bar V_{\phi=0}$ their contribution vanishes at $\bar T_{\rm c}$. This can also be checked numerically. For instance, in Fig.~\ref{Fig:RG-flat}, we show the flattening of the momentum dependent gap mass $\bar M^2_{\phi=0}(K)$ as we approach $\bar T_{\rm c}$ due to the fact that the nonlocal contribution to the gap equation vanishes in this limit.\\

The round bracket in the field equation (\ref{eq:field_imp}) is just $\bar M^2_{\phi=0,\bar T_{\rm c}}=0$. Using Eq.~\eqref{eq:m0} with $T_\star$ replaced by $\bar T_{\rm c}$ and $m_\star$ replaced by $0$ to express $m_0^2$, and Eq.~\eqref{eq:l0} at the reference temperature $T_\star$ to express $\lambda_0,$ one obtains the following renormalized equations:
\beq
h=\frac{5}{3}\bar\phi\,\bar M^2_{\bar \phi,\bar T_{\rm c}} \quad \mbox{and} \quad \bar M^2_{\bar\phi,\bar T_{\rm c}}=\frac{\lambda_{\star}}{2}\left(\frac{5}{3}\bar\phi^2+{\cal T}_{\bar T_{\rm c}}[\bar G_{\bar \phi,\bar T_{\rm c}}] - {\cal T}_{\bar T_{\rm c}}[G_0] + \bar M^2_{\bar \phi,\bar T_{\rm c}} {\cal B}_\star[G_\star](0)\right)\,.
\label{eq:marko-field}
\eeq
As $h\to 0$, $\bar\phi\to 0$ and thus $\bar M^2_{\bar\phi,\bar T_{\rm c}}\to 0,$ which justifies a high temperature expansion. Using the first terms in the expansion of the tadpole sum-integral given in Eq.~\eqref{eq:tadpole_HTE}, the gap equation becomes quadratic:
\beq\label{eq:gap_marko}
\bar M_{\bar \phi,\bar T_{\rm c}}^2\left[1-\frac{\lambda_\star}{2}
\left({\cal B}_\star^{(1)}[G_\star](0)-\frac{1}{8\pi^2}\left(\gamma_{\mbox{\tiny E}}+\ln\frac{m_\star}{4\pi \bar T_{\rm c}}\right)\right)
\right]
+\frac{\lambda_\star \bar T_{\rm c}}{8\pi} \bar M_{\bar \phi,\bar T_{\rm c}} 
-\frac{5}{6}\lambda_\star\bar\phi^2=0\,,
\eeq
where terms of order $\bar M_{\bar \phi,\bar T_{\rm c}}^4/\bar T_{\rm c}^2$ and higher were neglected. At lowest order, one can neglect the terms of order $\bar M_{\bar \phi,\bar T_{\rm c}}^2$ and obtain $\bar M_{\bar \phi,\bar T_{\rm c}}\sim 20\pi\bar\phi^2/(3 T_{\rm c}).$ Plugging this result into the field equation in Eq.~\eqref{eq:marko-field}, one obtains the analytic value $\delta_{\mbox{\tiny RG}}=5.$ We shall not present the numerical determination of $\delta_{\mbox{\tiny RG}},$ for the simple reason that the HTE is very accurate in the region of $h$ used for the numerical determination of $\delta$ in the case without RG-improvement (see Fig.~\ref{Fig:crexp_bd}), and thus the solution of the gap equation in \eqref{eq:marko-field} is very accurately approximated by the solution of the quadratic equation \eqref{eq:gap_marko}. The value $\delta_{\mbox{\tiny RG}}=5$ can be understood as follows: assuming that the potential admits a Taylor expansion around $\phi=0$, the field equation at $T=\bar T_{\rm c}$ and for small $h$, reads
\beq
h  & = & \bar M^2_{\phi=0,\bar T_{\rm c}}\bar\phi+\frac{1}{3!}\bar V_{\phi=0,\bar T_{\rm c}}\bar\phi^3+\frac{1}{5!}\left.\frac{\delta^6\gamma}{\delta\phi^6}\right|_{\phi=0,\bar T_{\rm c}}\bar\phi^5+{\cal O}(\bar\phi^7)\,,
\eeq
where we have used that the second and fourth derivatives of the potential at $\phi=0$ coincide with $\bar M^2_{\phi=0}$ and $\bar V_{\phi=0}$. Using that $\bar M^2_{\phi=0,\bar T_{\rm c}}=0$ and $\bar V_{\phi=0,\bar T_{\rm c}}=0$, and assuming that the sixth derivative does not vanish, we obtain $h\propto \bar\phi^5$ and then $\delta^{\mbox{\tiny RG}}=5$. Thus, although the RG-improvement ensures that $\bar V_{\phi=0,\bar T_{\rm c}}=0$, which is a necessary condition for $\delta$ to be larger than $3$, the two-loop approximation is not sufficient to generate nonanalyticities in the field dependence of the potential which would yield a noninteger value for $\delta$.\\

The value of $\hat\gamma^+_{\mbox{\tiny RG}}$ can be determined analytically with a similar calculation, since it is given by the way the curvature at zero $\bar M^2_{\phi=0,T}$ behaves as we approach $\bar T_{\rm c}$. We have already seen that $\bar M_{\phi=0}\propto T-\bar T_{\rm c}$. It follows that $\hat\gamma^+_{\mbox{\tiny RG}}=2.$ The numerical determination of $\hat\gamma^+_{\mbox{\tiny RG}}$ is again simple and does not warrant a presentation. Similarly, one can determine analytically $\bar\gamma^+_{\mbox{\tiny RG}}$ and one finds $\bar\gamma^+_{\mbox{\tiny RG}}=\hat\gamma^+_{\mbox{\tiny RG}}$.\\

Concerning the heat capacity, this can be determined numerically down to $\bar T_{\rm c}$ through the formula $\displaystyle C=-T\partial^2\gamma(\bar\phi)/\partial T^2,$ by applying to the effective potential the method described in Sec.~\ref{sec:RG_theory}. Around $\bar T_{\rm c}$ an analysis based on the high temperature expansion reveals that the heat capacity behaves as $C_+\simeq a_+ + b_+(T-\bar T_{\rm c})$ with the constant $a_+=2\pi^2 \bar T_{\rm c}/15$ and $b_+=34\pi^2 \bar T^3_{\rm c}/135,$ independently of the value of the coupling. Unfortunately, we cannot conclude on the value of $\alpha_{\mbox{\tiny RG}}$ because we do not know whether there is a jump in the value of heat capacity at $\bar T_{\rm c}$. The only thing we can state is that the heat capacity does not diverges as we approach $\bar T_{\rm c}^+$.\\

Note finally that in the RG-improved case the last two of the four scaling relations (\ref{eq:scaling}) are fulfilled with $d=3$. The two other scaling relations cannot be checked for we cannot access $\alpha$ or $\beta$.

\section{Generalities concerning the numerical implementation}\label{sec:numerics}
The resolution of the gap and field equations or the evaluation of the effective potential, together with the determination of the bare parameters, involve various sum-integrals. Local sum-integrals of the form
\beq
{\cal V}[f]\equiv \int_Q^T f(Q)
\eeq
and nonlocal sum-integrals
\beq\label{eq:convol}
{\cal C}[f,g](K)\equiv \int_Q^T f(Q)\,g(K-Q)
\eeq
in the form of convolutions. We now explain how these various sum-integrals are discretized in view of their practical evaluation. We also present a method which allows to increase the rate of convergence of the discretized sum-integrals towards their exact result, leading to a sizable improvement in accuracy or computational speed.

\subsection{Discretization of the sum-integrals\label{ss:dis-sum-int}}

Let us start with the nonlocal sum-integrals because, as we will see, this puts some restrictions on the choice of the discretization. Sum-integrals of this type will be evaluated using Fast Fourier Transform algorithms. A convolution such as (\ref{eq:convol}) can be written as
\beq\label{eq:nl}
{\cal C}[f,g]={\cal F}\big[{\cal F}^{-1}[f]{\cal F}^{-1}[g]\big]\,,
\label{Eq:conv}
\eeq
where we have introduced the Fourier transform operator ${\cal F}$ and its inverse ${\cal F}^{-1}$ defined by
\beq
{\cal F}[f](i\omega_n,\q) & \equiv & \int_0^{\beta}d\tau\int d^3x\,e^{i\omega_n\tau-i\q\cdot{\bm x}}\,f(\tau,{\bm x})\,,\\
{\cal F}^{-1}[f](\tau,{\bm x}) & \equiv & T\!\!\sum_{n=-\infty}^{\infty}\int\frac{d^3q}{(2\pi)^3} e^{-i\omega_n\tau+i\q\cdot {\bm x}}\,f(i\omega_n,\q)\,,
\eeq
with $\beta=1/T$. The functions $f(i\omega_n,\q)$ that we will have to deal with are invariant both under $i\omega_n\rightarrow-i\omega_n$ and under rotations (they only depend on the modulus $q=|\q|$ of the momentum). It follows that their Fourier transforms ${\cal F}[f](\tau, {\bm x})$ are invariant both under $\tau\rightarrow \beta-\tau$ and under rotations (they only depend on the modulus $x=|{\bm x}|$). Similar remarks apply to the inverse Fourier transforms. Using these properties, we arrive at
\beq
q{\cal F}[f](i\omega_n,q) & = & 2\pi \left[2\int_0^{\beta/2} d\tau \cos(\omega_n\tau) \left(2\int_0^\infty d x\,\sin(q x)\,x f(\tau,x)\right)\right],
\label{Eq:F_forw}\\
x{\cal F}^{-1}[f](\tau,x) & = & \frac{1}{4\pi^2}\left[2\int_0^\infty d q\,\sin(q x) \left(T q f(0,q)+2 T\sum_{n=1}^{\infty}\cos(\omega_n\tau)\,q f(i\omega_n,q)\right)\right],
\label{Eq:F_back}
\eeq
where we need only the Matsubara frequencies $\omega_n=2\pi nT$ with $n\ge0.$ We can rewrite this as
\beq\label{eq:for}
{\cal F}[f]_\bullet=2\pi\,({\cal F}_c\otimes {\cal F}_s)[f_\bullet] \quad {\rm and} \quad {\cal F}^{-1}[f]_\bullet=\frac{1}{2\pi}({\cal F}_c^{-1}\otimes {\cal F}_s^{-1})[f_\bullet]\,,
\eeq
where the notation $f_\bullet$ means that the function $f$ is multiplied by the modulus of its 3d argument, for instance $f_\bullet(i\omega_n,q)=qf(i\omega_n,q)$, and we have introduced
\beq\label{eq:Fc}
{\cal F}_c[f](i\omega_n)=2\int_0^{\beta/2}d\tau\,\cos(\omega_n\tau)\,f(\tau) \quad {\rm and} \quad {\cal F}_c^{-1}[f](\tau)=Tf(0)+2T\sum_{n=1}^\infty \cos(\omega_n\tau)f(i\omega_n)\,,
\eeq
as well as
\beq\label{eq:Fs}
{\cal F}_s[f](q)=2\int_0^\infty dx\,\sin(qx)\,f(x) \quad {\rm and} \quad {\cal F}_s^{-1}[f](x)=\frac{2}{2\pi}\int_0^\infty dq\,\sin(qx)\,f(q)\,.
\eeq
Note that if $\tilde f(q)$ is the 3d Fourier transform of a rotational invariant function $f(x),$ that is $\tilde f(q)=\int d^3x f(|{\bm x}|) e^{-i\q\cdot{\bm x}},$ then $\tilde f_\bullet=2\pi\mathcal{F}_{s}[f_\bullet],$ and in turn $f_\bullet=\frac{1}{2\pi}\mathcal{F}_{s}^{-1}[\tilde f_\bullet].$

We have thus reduced the evaluation of the convolution to the evaluation of sine and cosine transforms whose discretized versions (DST and DCT) can be performed efficiently using one of the variants implemented in numerical libraries.\footnote{We use the routines of the Fastest Fourier Transform in the West (FFTW) library \cite{FFTW3}, which contain a factor of 2 in the formulas of the DST and DCT transformations. This is the reason for separating factors of 2 in Eqs.~\eqref{Eq:F_forw} and \eqref{Eq:F_back}.} These variants differ in the type of boundary condition used when the original data is extended in view of performing on it the discrete fast Fourier transformation. As explained in Appendix~E of \cite{Borsanyi:2008ar}, for the rotation invariant part, we use a discretization which avoids potential singularities as $x\to 0$ and $q \to 0,$ in that it does not store on the grid zero momentum and direct space values, and which matches the boundary conditions of the DST-II and DST-III formulas for the sine and inverse sine transforms, respectively (note however that the method that we put forward in the next section allows to reduce considerably the appearance of singularities in the UV). In momentum space, the highest stored momentum is the cutoff $\Lambda$ and the grid is defined as $k_{\tilde k}=(\tilde k+1) \Delta k,$ with $\tilde k=0\dots N_s-1$ and $\Delta k=\Lambda/N_s$ the lattice spacing in momentum space, while in direct space, the grid is defined as $x_s=(s+\frac{1}{2})\Delta x,$ with $s=0\dots N_s-1$ and $\Delta x$ the direct space lattice spacing satisfying $\Delta x\Delta k=\pi/N_s.$ We retain $N_\tau-1$ positive Matsubara frequencies and the static mode $\omega_n=0,$ so that the available Matsubara frequencies are $\omega_n=(2\pi/\beta)n=n\Delta\omega,$ with $n=0\dots N_\tau-1.$ The corresponding temporal grid is defined as $\tau_t=(t+\frac{1}{2})\Delta\tau$ with $t=0\dots N_\tau-1$ and $\Delta\tau$ the temporal lattice spacing such that $\Delta\tau\Delta\omega=\pi/N_\tau$. One can see that with this discretization, the discrete version of the cosine and inverse cosine transforms appearing in Eq.~(\ref{eq:Fc}) are the DCT-II the DCT-III, respectively.\\

In order to write the discretized version of the nonlocal sum-integral (\ref{eq:nl}) in a compact way, we first introduce a shorthand notation for the sequence of discrete sine and cosine transforms which acts on an $N_\tau \times N_s$ array in which we store the values of the Matsubara frequencies and the modulus of the momenta.  We define the following forward and backward discrete transforms
\beq
{\cal F}_{N_\tau,N_s}[f(t,s)](n,\tilde k) &\equiv& \textrm{DCT-II}_{t}\big[\textnormal{DST-II}_{s}[f(t,s)](t,\tilde k) \big](n,\tilde k),\\
{\cal F}^{-1}_{N_\tau,N_s}[f(n,\tilde k)] (t,s) &\equiv& \textrm{DCT-III}_{n}\big[\textnormal{DST-III}_{\tilde k}[f(n,\tilde k)](n,s) \big](t,s),
\eeq
where $n,t=0\dots N_\tau-1$ and $\tilde k, s=0\dots N_s-1,$ and the array $f[t][s]$ is denoted for simplicity as $f(t,s).$ The index indicates the part of the array on which the transformation acts. Note also that ${\cal F}_{N_\tau,N_s}$ and ${\cal F}^{-1}_{N_\tau,N_s}$ are inverse to each other up to a numerial constant: ${\cal F}^{-1}_{N_\tau,N_s}[{\cal F}_{N_\tau,N_s}[f]]=4N_\tau N_s f$. This comes from the fact that DST-III and DCT-III are the inverses of DST-II and DST-II up to factors $2N_\tau$ and $2N_s$ respectively. With the notation above it is easy to see, by using Eqs.~\eqref{Eq:F_forw} and \eqref{Eq:F_back}, that the discretized version of the convolution reads 
\beq
{\cal C}_{N_\tau,N_s}[f,g](n,\tilde k)=\frac{c}{\tilde k+1}{\cal F}_{N_\tau,N_s}\left[\frac{1}{s+\frac{1}{2}}{\cal F}^{-1}_{N_\tau,N_s}\big[(\tilde p+1)f(m,\tilde p)\big](t,s) \cdot {\cal F}^{-1}_{N_\tau,N_s}\big[(\tilde p+1)g(m,\tilde p)\big](t,s)\right](n,\tilde k),
\label{Eq:conv_d}
\eeq
where $n,m,t=0\dots N_\tau-1,$ $\tilde k,\tilde p, s=0\dots N_s-1,$ and the prefactor $\displaystyle c=T^2\Delta\tau (\Delta k)^3/(8\pi^3)$ contains the dimensionfull quantities arising from the discretization of the integrals.\\

Next, we turn to the sum-integrals of the local type. Having stored on the grid $N_\tau-1$ positive frequencies and the static mode $\omega_n=0,$ this sum-integral will be approximated numerically as
\beq\label{eq:vol_d}
{\cal V}_{N_\tau,N_s}[f]&\equiv T&\sum_{n=-N_\tau+1}^{N_\tau-1}\left[\int\frac{d^3q}{(2\pi)^3}\right]_{N_s}f(i\omega_n,q)\nonumber\\
&=&T\left[\int\frac{d^3q}{(2\pi)^3}\right]_{N_s}f(0,q)+2T\sum_{n=1}^{N_\tau-1}\left[\int\frac{d^3q}{(2\pi)^3}\right]_{N_s}f(i\omega_n,q)
\,,
\eeq
where the notation $[...]_{N_s}$ refers to some quadrature rule, in practice we use the trapezoidal rule. After the exact evaluation of the angular integrals, one applies the extended trapezoidal rule \cite{NR} for the integral over $q$ in the interval $[0,\Lambda].$ To obtain the formula, we include first the zero momentum (not contained by our momentum grid) in the sequence of points on the abscissa. Then, having $N_s+1$ equally spaced points, we apply the trapezoid rule on the $N_s$ intervals $(0,\Delta k), \dots, ((N_s-1)\Delta k,N_s \Delta k)$ and obtain explicitly
\beq\label{eq:trapezoid}
\left[\int\frac{d^3q}{(2\pi)^3}\right]_{N_s} f(q) = \frac{(\Delta k)^3}{2\pi^2} \left[\frac{N_s^2}{2}f_{N-1}+\sum_{q=0}^{N_s-2}(q+1)^2 f_q\right].
\eeq

\subsection{Increasing the rate of convergence of sum-integrals\label{ss:increase-conv}}
In this section we take advantage of the fact that the asymptotic behavior of the propagator $\bar G(Q)$ is exactly given by $1/Q^2$ in order to accelerate the convergence of the discretized sum-integrals towards their exact result. In order to illustrate the method we consider the tadpole sum-integral ${\cal T}[\bar G]$ first. The most straightforward way to compute the latter would be as
\beq\label{eq:tad_naive}
{\cal T}[\bar G]\simeq {\cal V}_{N_\tau,N_s}[\bar G]\,.
\eeq
The error that one should expect from this type of approximation is studied in detail in App.~\ref{app:Matsubara}. It is shown in particular that the error related to the finite number of Matsubara frequencies is directly connected with the rate at which the summand $\bar G(Q)=\bar G(i\omega_n,q)$ approaches zero at large $n$. Then, if in one way or another we are able to reorganize the evaluation of ${\cal T}[\bar G]$ in terms of sum-integrals involving summands which decrease faster than $\bar G(Q)$ at large $n$, we will certainly reduce the error. Consider then the identity
\beq
{\cal T}[\bar G]=\big[{\cal T}[\bar G]-{\cal T}[G_\star]\big]+{\cal T}[G_\star]=\int_Q^T \delta\bar G(Q)+{\cal T}[G_\star]\,.
\eeq
The second term ${\cal T}[G_\star]$ involves the free-type propagator $G_\star$ and the corresponding sum can be computed almost exactly,\footnote{The Matsubara sum can be performed analytically and the momentum integral can be computed numerically using accurate adaptive integration routines.} see App.~\ref{app:pert}. The first term involves a Matsubara sum whose summand $\delta \bar G(Q)=\bar G(Q)-G_\star(Q)$ decreases faster than $\bar G(Q)$ at large $n$. Then, if we approximate the tadpole sum-integral ${\cal T}[\bar G]$ by
\beq\label{eq:tad_improved}
{\cal T}[\bar G]\simeq {\cal V}_{N_\tau,N_s}[\delta\bar G]+{\cal T}[G_\star]\,,
\eeq
we obtain an evaluation of ${\cal T}[\bar G]$ which is more accurate than (\ref{eq:tad_naive}) for the same number of Matsubara frequencies.\\

The same strategy can be applied to the bubble sum-integral. We can of course use the straightforward approximation
\beq\label{eq:naive_summand}
{\cal B}[\bar G](K)\simeq {\cal C}_{N_\tau,N_s}[\bar G,\bar G](K)\,.
\eeq
But we can instead reorganize the calculation of ${\cal B}[\bar G](K)$ first, according to
\beq\label{eq:id2}
{\cal B}[\bar G](K) & = &\big[{\cal B}[\bar G](K)-{\cal B}[G_\star](K)\big]+{\cal B}[G_\star](K)\nonumber\\
& = & \int_Q^T \big[\bar G(Q)\bar G(K-Q)-G_\star(Q)G_\star(K-Q)\big]+{\cal B}[G_\star](K)\nonumber\\
& = & \int_Q^T G_\star(Q)\delta\bar G(K-Q)+\int_Q^T \delta\bar G(Q)\bar G(K-Q)+{\cal B}[G_\star](K)\nonumber\\
& = & \int_Q^T \big[G_\star(Q)+\bar G(Q)\big]\delta\bar G(K-Q)+{\cal B}[G_\star](K)\,,
\eeq
where we have used $\bar G(Q)=G_\star(Q)+\delta\bar G(Q)$ as well as the change of variables $Q\rightarrow K-Q$. The benefit of the last expression is that it involves a contribution ${\cal B}[G_\star](K)$ which can be determine almost exactly, see App.~\ref{app:pert}, and a contribution in the form of a convolution with an integrand which decreases faster in the UV than the original integrand. Our final approximation for the bubble sum-integral is then
\beq\label{eq:final-approx}
{\cal B}[\bar G](K)\simeq {\cal C}_{N_\tau,N_s}[G_\star+\bar G,\delta\bar G]+{\cal B}[G_\star](K)\,.
\eeq
This is a better approximation than (\ref{eq:naive_summand}) for the same number of Matsubara frequencies.\\

Finally, consider the setting-sun sum-integral ${\cal S}[\bar G]=\int_Q^T \bar G(Q){\cal B}[\bar G](Q)$. The straightforward approximation would be
\beq\label{eq:SS_naive}
{\cal S}[\bar G]\simeq {\cal V}_{N_\tau,N_s}\big[\bar G\,{\cal C}_{N_\tau,N_s}[\bar G,\bar G]\big]\,.
\eeq
Instead, we write
\beq
{\cal S}[\bar G] & = & \int_Q^T \bar G(Q)\big[{\cal B}[\bar G](Q)-{\cal B}[G_\star](Q)\big]+\int_Q^T \bar G(Q){\cal B}[G_\star](Q)\nonumber\\
& = & \int_Q^T \bar G(Q)\big[{\cal B}[\bar G](Q)-{\cal B}[G_\star](Q)\big]+\int_Q^T \delta\bar G(Q){\cal B}[G_\star](Q)+{\cal S}[G_\star]\,.
\eeq
The first term involves a summand which decreases faster than the original one $\bar G(Q){\cal B}[\bar G](Q)$. Moreover, the inner sum (and the corresponding momentum integral) appears as a convolution and can thus be treated efficiently using Fast Fourier Transform algorithms. In the second term, the summand decreases again faster than the original summand and it contains a factor ${\cal B}[G_\star](Q)$ which can be determined almost exactly. Finally the last term ${\cal S}[G_\star]$ can be determined almost exactly, see App.~\ref{app:pert}. Our approximation for the setting-sun sum-integral will then be
\beq
{\cal S}[\bar G] \simeq {\cal V}_{N_\tau,N_s}\big[\bar G\,{\cal C}_{N_\tau,N_s}[G_\star+\bar G,\delta\bar G]\big]+{\cal V}_{N_\tau,N_s}\big[\delta \bar G\,{\cal B}[G_\star]\big]+{\cal S}[G_\star]\,.
\eeq

\subsection{Optimized equations and bare parameters}
In this section we gather the different equations that need to be solved and the expression of the subtracted effective potential and its curvature at $\phi=0$ which needs to be evaluated and, using the operators ${\cal V}_{N_\tau,N_s}$ and ${\cal C}_{N_\tau,N_s}$, we put them in a form which is ready for numerical implementation. The first equation to be solved is the gap equation which reads
\beq\label{eq:gap_num}
\bar M^2(K)=m^2_\star & + & \frac{\lambda_{2{\rm l}}}{2}\phi^2+\frac{\lambda_0}{2}{\cal V}_{N_\tau,N_s}[\delta\bar G]+\frac{\lambda_0}{2}\big[{\cal T}[G_\star]-{\cal T}_\star[G_\star]\big]\nonumber\\
& - & \frac{\lambda^2_\star}{2}\phi^2{\cal C}_{N_\tau,N_s}\big[G_\star+\bar G,\delta\bar G\big](K)-\frac{\lambda^2_\star}{2}\phi^2\big[{\cal B}[G_\star](K)-{\cal B}_\star[G_\star](0)\big],
\eeq
where we have used the expressions for $\delta\lambda_{2{\rm nl}}$ and $m^2_0$ and it is clear from Eq.~(\ref{eq:gap_num}) that these are computed almost exactly. The other bare parameters relevant for the gap equation are computed as
\beq\label{eq:l0_num}
\frac{1}{\lambda_0}=\frac{1}{\lambda_\star}-\frac{1}{2}{\cal V}_{N_\tau,N_s}[G_\star^2]
\eeq
and
\beq\label{eq:l2l_num}
\lambda_{2{\rm l}}=\lambda_0\left[1-\frac{\lambda^2_\star}{2}\,{\cal V}_{N_\tau,N_s}\Big[G_\star^2(Q_\star)\big[{\cal B}_\star[G_\star](Q_\star)-{\cal B}_\star[G_\star](0)\big]\Big]\right].
\eeq
Note that we could compute these bare parameters almost exactly as well. However, the proof of renormalization of Sec.~\ref{sec:renorm_proof} reveals that these parameters absorb divergences in ${\cal V}_{N_\tau,N_s}[\delta\bar G]$. It is thus natural to compute $\lambda_0$ and the outer sum-integral of $\lambda_{2{\rm l}}$ with the same $N_\tau$ and $N_s$. In contrast, the inner sum-integral of $\lambda_{2{\rm l}}$ is computed almost exactly for it has to do with the last term of Eq.~(\ref{eq:gap_num}) which is determined almost exactly. The gap equation can be coupled to the field equation in order to determine the extrema of the effective potential. In the presence of an external source $h$, the discretized form of the field equation reads
\beq\label{eq:field_num}
& & h=\bar \phi\left\{m^2_\star+\frac{\lambda_4}{6}\bar \phi^2+\frac{\lambda_2}{2}{\cal V}_{N_\tau,N_s}[\delta\bar G]+\frac{\lambda_2}{2}\,\big[{\cal T}[G_\star]-{\cal T}_\star[G_\star]\big]\right.\nonumber\\
& & \hspace{1.5cm} \left.-\frac{\lambda^2_\star}{6}\Big[{\cal V}_{N_\tau,N_s}\big[\bar G\,{\cal C}_{N_\tau,N_s}[G_\star+\bar G,\delta\bar G]\big]+{\cal V}_{N_\tau,N_s}\big[\delta \bar G\,{\cal B}[G_\star]\big]\Big]-\frac{\lambda^2_\star}{6}\big[{\cal S}[G_\star]-{\cal S}_\star[G_\star]\big]\right\},
\eeq
where $\lambda_2$ is evaluated as $\lambda_2=\lambda_{2{\rm l}}+2\lambda_\star\big(1-\lambda_\star/\lambda_0\big)$ and the bare coupling $\lambda_4$ is computed as
\beq\label{eq:l4_num}
\lambda_4=-2\lambda_\star+3\frac{\lambda_{2{\rm l}}^2}{\lambda_0}+\frac{3}{2}\lambda^4_\star{\cal V}_{N_\tau,N_s}\Big[G^2_\star(Q_\star)\big[{\cal B}_\star[G_\star](Q_\star)-{\cal B}_\star[G_\star](0)\big]^2\Big]\,.
\eeq
Here, again, the appearance of the difference of bubbles is due to the almost exactly determined last term of Eq.~(\ref{eq:gap_num}) (see the steps leading from Eqs.~\eqref{eq:abc} to \eqref{eq:jsp}), therefore it is natural to perform these integrals almost exactly in the expression of $\lambda_4,$ too. The discretized effective potential reads
\beq
\Delta\gamma(\phi)& =&-h\phi -\frac{\lambda_4}{4!}\phi^4 + \frac{1}{2}\phi\frac{\delta\gamma}{\delta\phi}+\frac{1}{2\pi^2}\int_0^{\Lambda} d q\, q ^2\left[T\ln \Big(1-e^{-\epsilon^\star_q/T}\Big) - T_\star\ln \Big(1-e^{-\epsilon^\star_q/T_\star}\Big)\right]
\nonumber\\
&+&\frac{1}{2} {\cal V}_{N_\tau,N_s}\Big[\ln \bar G^{-1} - \ln G_\star^{-1} -\big(\bar M^2 - m^2_\star\big)\bar G \Big]
+\frac{\lambda_0}{8} \Big[{\cal V}_{N_\tau,N_s}[\delta \bar G] + \big[{\cal T}[G_\star]-{\cal T}_\star[G_\star]\big]\Big]^2,
\label{eq:pot_num}
\eeq
with $\epsilon_q^\star=\sqrt{q^2+m_\star^2}$. The derivative $\delta\gamma/\delta\phi$ of the effective potential in the formula above stands for an expression similar to the right hand side of Eq.~\eqref{eq:field_num}, but with $\bar\phi$ replaced by $\phi,$ while the coupling $\lambda_0$ and $\lambda_4$ are computed as in Eqs.~\eqref{eq:l0_num} and \eqref{eq:l4_num}, respectively. Finally, the discretized form of the curvature at vanishing field is given by
\beq\label{eq:Mhat_dis}
\hat M^2_{\phi=0} & = & m^2_\star+\frac{\lambda_2}{2}{\cal V}_{N_\tau,N_s}[\delta\bar G_{\phi=0}]+\frac{\lambda_2}{2}\,\big[{\cal T}[G_\star]-{\cal T}_\star[G_\star]\big]\nonumber\\
& - & \frac{\lambda^2_\star}{6}\Big[{\cal V}_{N_\tau,N_s}\big[\bar G_{\phi=0}\,{\cal C}_{N_\tau,N_s}[G_\star+\bar G_{\phi=0},\delta\bar G_{\phi=0}]\big]+{\cal V}_{N_\tau,N_s}\big[\delta \bar G_{\phi=0}\,{\cal B}[G_\star]\big]\Big]-\frac{\lambda^2_\star}{6}\big[{\cal S}[G_\star]-{\cal S}_\star[G_\star]\big],
\eeq
where $\bar G_{\phi=0}(Q)=1/(Q^2+\bar M^2_{\phi=0}),$ with $\bar M^2_{\phi=0}$ obtained from
\beq
\bar M^2_{\phi=0} = m^2_\star & + & \frac{\lambda_0}{2}{\cal V}_{N_\tau,N_s}[\delta\bar G_{\phi=0}]+\frac{\lambda_0}{2}\big[{\cal T}[G_\star]-{\cal T}_\star[G_\star]\big].
\eeq

\subsection{On the iterative solution of the equations}
The solution of the gap equation \eqref{eq:gap_num} at fixed value of the field, as well as the solution of the coupled set of gap and field equations \eqref{eq:gap_num} and \eqref{eq:field_num} are obtained using iterations. We illustrate now this iterative procedure in the case of the solution of the field equation at vanishing external field $h$ and comment on the convergence of the algorithm at different values of the coupling constant. 

First, we determine the bare couplings $\lambda_0,$ $\lambda_{2{\rm l}},$ and $\lambda_4$ given in Eqs.~\eqref{eq:l0_num}, \eqref{eq:l2l_num}, and \eqref{eq:l4_num}, and evaluate those perturbative integrals in the gap and field equations which are defined at temperature $T_\star.$ In case the of $\lambda_0,$ the convergence of the Matsubara sum is improved as described in Appendix~\ref{app:Matsubara} (see Eq.~\eqref{eq:bub_improved}). The explicit expressions of the integrals evaluated using the adaptive numerical integration routines of the GNU Scientific Library (GSL)~\cite{gsl} are given in Appendix~\ref{app:pert}. The quantities determined in this way will not change during the iterative process which, after the initialization of the propagator $\bar G$ with $G_\star$, consists of the following two steps:
\begin{enumerate}
\item 
In Eq.~\eqref{eq:field_num} the double sums and the perturbative integrals defined at temperature $T$ are computed with the actual propagator $\bar G(i\omega_n,k).$ If the sum of $\bar\phi$-independent terms in the curly brackets is negative, then $\bar\phi\ne0$ is expressed by equating the expression between curly brackets with zero, if the sum is positive, then the trivial solution $\bar \phi=0$ is considered. 
\item 
Using the value of $\bar \phi$ obtained in step $1.$, $\bar M^2(i\omega_n,k)$ is determined from Eq.~\eqref{eq:gap_num}, by evaluating the double sums and the integrals with the actual propagator $\bar G,$ then the propagator is updated with 
\beq
\bar G_{\rm update}(i\omega_n,k)=\left(\omega_n^2+k^2+\bar M^2(i\omega_n, k)\right)^{-1}\,.
\eeq
\end{enumerate}
These two steps are iterated until the procedure converges to the desired accuracy. We monitored the value of the propagator at the lowest available frequency and momentum and stopped the iteration when both $\left|\bar G^{(i)}(0,\Delta k)-\bar G^{(i+1)}(0,\Delta k)\right|/\bar G^{(i+1)}(0,\Delta k)<10^{-8}$ and $\left|\bar \phi^{(i)}-\bar \phi^{(i+1)}\right|/\bar\phi^{(i+1)}<10^{-8}$ were satisfied. When this algorithm is used to determine $\bar\phi(T)$ by changing the temperature, then the iteration starts with the converged propagator obtained at the previous value of the temperature. In some cases, like the determination of the critical exponent $\delta,$ one has to work with nonvanishing external source $h.$ In this case $\bar\phi$ is obtained in step 1. by solving a cubic equation in the field.

This simple iterative procedure fails to converge for large enough $\lambda_\star$ ($\sim 9$), because the corrections are too large in each iteration. In this case somehow, the first iterative step bring us out of the attraction domain of the solution, if such a domain exists at all. This mostly happens when solving the gap-equation at fixed value of the field. When both the gap and field equations are solved iteratively, the procedure eventually converges for even larger couplings, but the number of iterations increases with the value of the coupling. In order to increase the speed of convergence or to achieve convergence at all, one follows the procedure used in \cite{Berges:2004hn} and modifies the value of the updated propagator with a weighted average between the old value of the propagator and the iterated value. In this case, in the $(i+1)th$ iteration the following updating method is used 
\beq
\bar G^{(i+1)}_{\rm update}(i\omega_n,k)=\alpha\left[\omega_n^2+k^2+\bar M^2(i\omega_n,k)^{(i+1)}\right]^{-1}+(1-\alpha)G^{(i)}(i\omega_n,k)\,,
\eeq
where $\alpha\in(0,1]$ and $\bar M^2(i\omega,k)^{(i+1)}$ is calculated with $G^{(i)}.$ For large values of the coupling, $\lambda_\star\ge9,$ one needs $\alpha<1$ to achieve convergence at $m^2_\star/T_\star^2=0.04$ and $T_\star=1.$ For small couplings, $\alpha=1$ ensures the fastest convergence, and in an intermediate range, $7.5<\lambda_\star<9,$ the use of $\alpha\neq 1$ increases the speed of convergence.

\subsection{Cutoff convergence, discretization effects and the role of improvements \label{ss:discretization}}

From the proof of renormalizability given in Sec.~\ref{sec:renorm_proof} we know that our results should becomes insensitive to the cutoff when the latter is taken large. However, since the proof is based on certain arguments that we can only verify using some perturbative estimates, it is interesting to check the cutoff insensitivity numerically, within a given accuracy. To do so, we have to pay particular attention to the discretization because we have to ensure that the latter does not distort the physics neither in the ultraviolet nor in the infrared regime. This means that as we increase the cutoff we need a good resolution in momentum space, that is small lattice spacing $\Delta k$, and also enough Matsubara modes taken into account. This represents a challenge for the judicious use of the available memory.
Note that the implementation of the numerical improvements described above helps in this respect for the same accuracy can be achieved with less discretization points. For the same number of discretization points, the improved code is more computer time demanding because it involves the accurate numerical evaluation of perturbative integrals. However, in order to reach the same level of accuracy, the non-improved code needs to be run with a higher number of finer discretization, which requires an increased computer time as well.

In what follows we shall illustrate on some physical quantities the effect of the discretization related to the use of the fast Fourier transforms to compute the convolution integrals, the extended trapezoidal rule for the momentum integrals and the finite number of Matsubara frequencies. After showing that these discretization effects are under control we will show also that the quantities of interest calculated with our best algorithm converge with increasing cutoff. 

\subsubsection{Discretization errors due to the use of FFT}

The discretization errors related to the use of the fast Fourier transformation  for the evaluation of the convolution integral can be easily illustrated with the help of the exact three dimensional convolution
\beq\label{eq:bub3d_exact}
\int\frac{d^3 q}{(2\pi)^3} G(q) G(p-q) = \frac{1}{4\pi p} \arctan\left(\frac{p}{2 M}\right),
\eeq 
where $G(p) = 1/(p^2+ M^2)$. This simple example is also relevant for our four dimensional study because it corresponds to the contribution of the static mode at finite temperature. Even though the  integral in Eq.~(\ref{eq:bub3d_exact}) is convergent, it will be interesting to calculate it using the same regulator as the one used in the four dimensional case. Using similar techniques as in Appendix~\ref{app:pert}, we arrive at
\beq\label{eq:bub3d_cutoff}
{\cal C}_\Lambda[G](p)=\mathop{\int_{|q|<\Lambda}}_{|p-q|<\Lambda}G(q) G(p-q)=\frac{1}{8\pi^2 p}\int_0^\Lambda d k k\, G(k) \ln\left(\frac{{\rm min}^2(k+p,\Lambda)+M^2}{(p-k)^2+M^2}\right).
\eeq
which can be evaluated accurately using adaptive integration routines.\\

\begin{figure}[!htbp] 
\begin{center}
\includegraphics[width=0.52\textwidth]{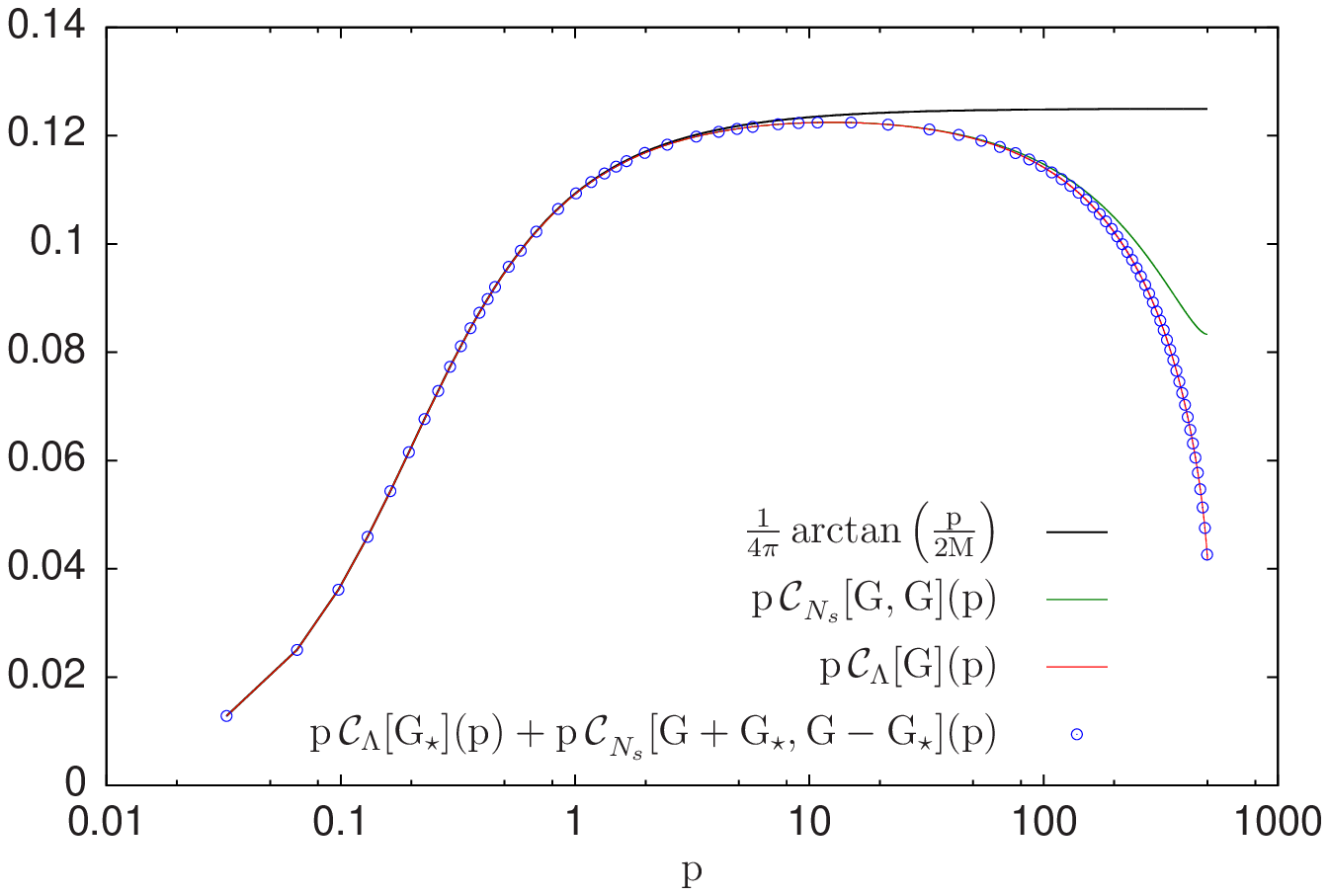}
\caption{Comparison of different methods of evaluating the perturbative bubble integral in 3d to the exact result. For more explanation see the text. The mass parameters are $M^2=0.01$ and $m_\star^2=1$ and the cutoff used was $\Lambda=500$ in some arbitrary units. For the FFT we used $N_s=15\times 2^{10}$  modulus of the momenta. \label{Fig:3dconv}}
\end{center}
\end{figure}

We can use the two results (\ref{eq:bub3d_exact}) and (\ref{eq:bub3d_cutoff}) to benchmark our method for evaluating convolution integrals and also to test how the continuum limit is approached. The different ways of computing the three dimensional bubble integral are plotted in Fig.~\ref{Fig:3dconv}. Note first that the bubble integral ${\cal C}_\Lambda[G]$ in the presence of a cut-off deviates from the continuum result already for values of the momentum much below the cutoff: at $p=\Lambda/10$ the deviation is already of $5\%.$ The result of a naive convolution on the momentum intervall $[0,\Lambda]$ using discrete sine transforms stays close to ${\cal C}_\Lambda[G]$ up to $p\simeq\Lambda/2$ (interestingly, it is closer to the continuum result for larger values of the momenta but this is a numerical artifact whose sign cannot be controlled in general). Instead, if we use the improved formula Eq.~\eqref{eq:final-approx} (in three dimensions), we can reproduce ${\cal C}_\Lambda[G](p)$ on the whole range of available momenta, up to $p=\Lambda$. This is related to the fact that, in the improved formula \eqref{eq:final-approx}, one of the functions to be convolved decreases faster in the UV than in the original convolution: $\delta G(q)\sim q^4$ instead of $G(q)\sim 1/q^2$. The overall picture remains the same when the cutoff is increased. 

\subsubsection{Discretization errors due to the use of a finite number of Matsubara modes}
In Fig.~\ref{Fig:TcTcbar}, the temperature $T_{\rm c}$ for which the curvature at $\phi=0$ vanishes, was determined for different values of the cutoff, by evaluating some perturbative integrals accurately (after the Matsubara sums were performed exactly, the remaining integrals were performed using adaptive integration routines). We can use these values as a benchmark to test the accuracy of $T_{\rm c}$ obtained using the discretized version of its defining equation and to illustrate the effect of the improvements on the numerical procedure. Here we focus on the discretization effects related to the use of a finite number of Matsubara sums using thee different levels of improvements.  
\vglue3mm

\begin{figure}[!htbp]
\begin{center}
\includegraphics[width=0.46\textwidth]{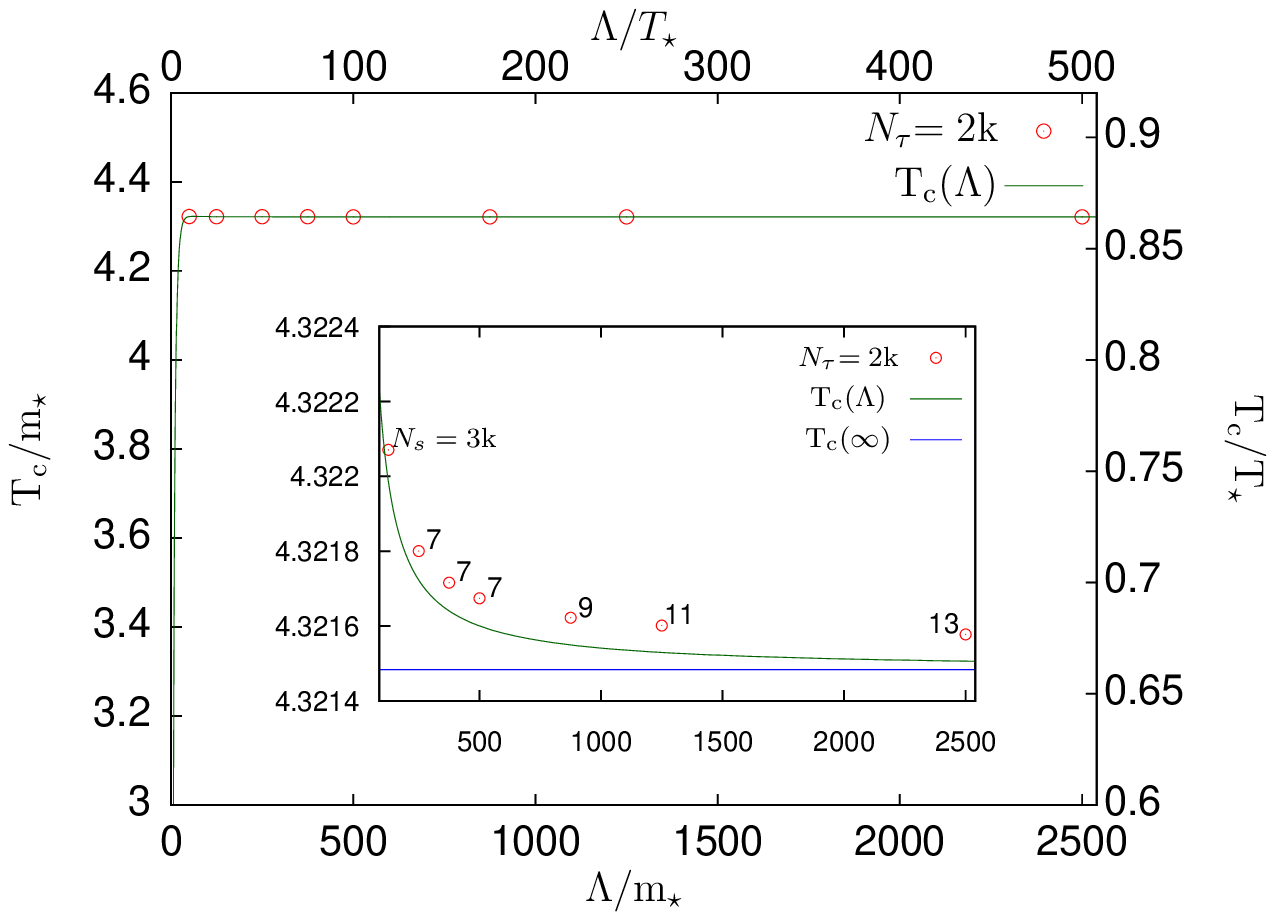}
\hspace*{0.5cm}
\includegraphics[width=0.46\textwidth,angle=0]{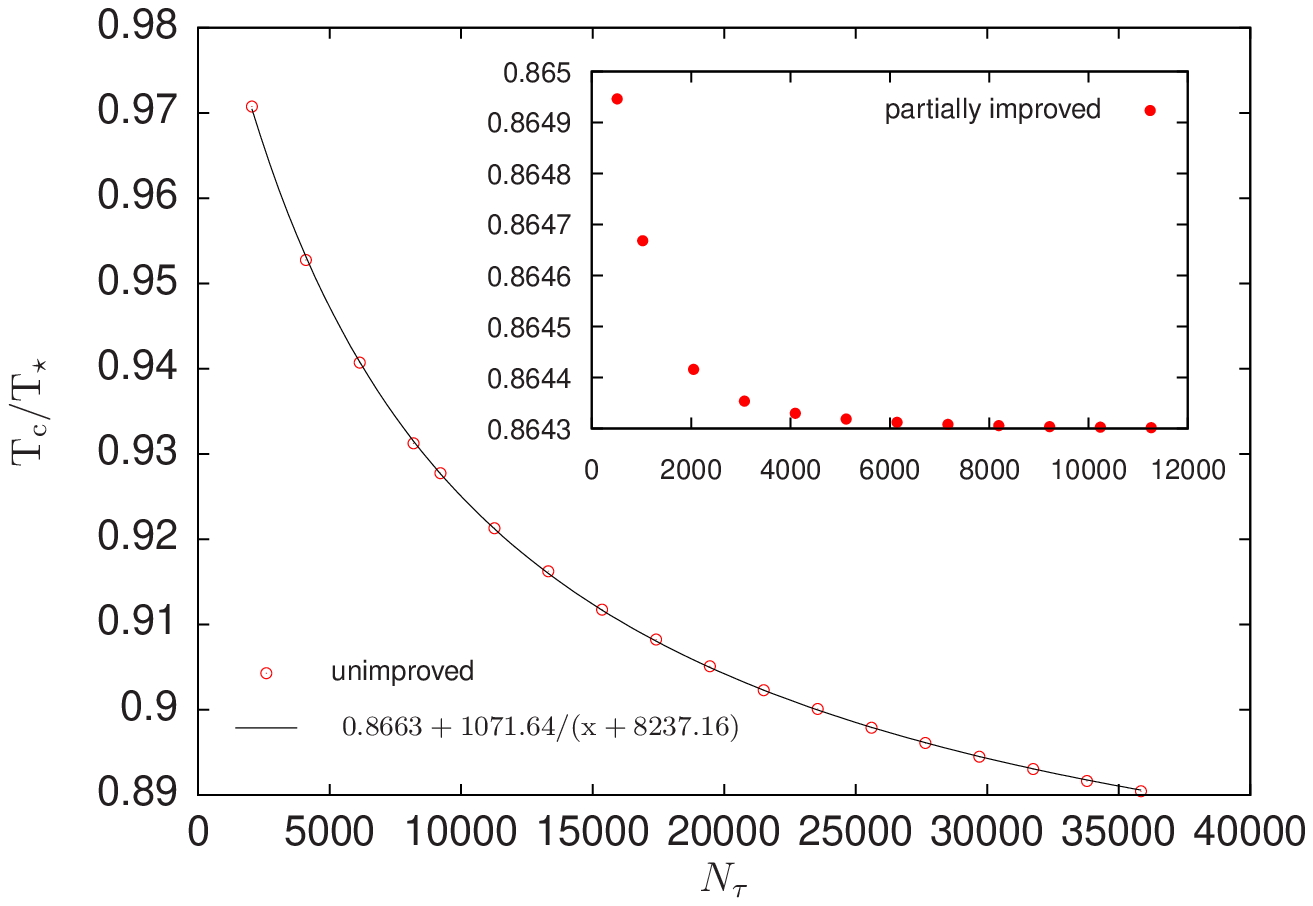}
\caption{Illustration of the effects of the improvements by comparing  $T_{\rm c}$ determined numerically at parameters $m_\star^2/T_\star^2=0.04$ and $\lambda_\star=3$ using fully improved, partially improved, and unimproved codes (see the text for explanations). Left panel: Solution of the defining equation \eqref{eq:def_Tc2} for $T_{\rm c},$ obtained with the fully improved code and with curvature $\hat M^2_{\phi=0}$ determined from the discretized expression \eqref{eq:Mhat_dis} (points), in comparison with the solution of \eqref{eq:def_Tc}, obtained by evaluating the perturbative integrals with adaptive routines (solid line). Right panel: Dependence of $T_{\rm c}$ on the number $N_\tau$ of Matsubara frequencies used to evaluate the sum-integrals in the expression \eqref{eq:Mhat_unimprvd} for $\hat M^2_{\phi=0}$ in case of the completely unimproved code and of the partially improved code with $\Lambda/T_\star=100$  and $N_s=3\times 2^{10}$ (inset). In case of the unimproved code, an asymptotic value of $T_{\rm c}$ can be extracted with a fit, as shown by the line. \label{Fig:TcTcbar2}}
\end{center}
\end{figure}

The {\it unimproved} code avoids the use of any adaptive numerical integration and uses instead the most straightforward discretization for the quantities appearing in the expression \eqref{eq:Mhat_unimprvd} of the curvature. The convolution is evaluated using fast Fourier transforms with the formula \eqref{Eq:conv_d} and all the momentum independent sum-integrals are approximated with a double sum: a sum over a finite number of Matsubara frequencies and a summation over a finite number of the modulus of the momentum, using the extended trapezoidal formula according to Eqs.~(\ref{eq:vol_d}) and (\ref{eq:trapezoid}). The momentum dependent bubble integral appearing in the setting-sun integral \eqref{eq:SS_naive} and the expressions \eqref{eq:l2l_num} and \eqref{eq:l4_num} for the bare couplings $\lambda_{2\rm{l}}$ and $\lambda_4$ are evaluated as a convolution, cf. Eq.~\eqref{eq:naive_summand}. The difference in the {\it partially improved} code is that it uses an accelerated Matsubara sum in the tadpole integral and in the bubble integral with zero external momentum appearing both in the expression of the bare quantities and in that of the curvature itself (see Eqs.~\eqref{eq:Tad_improved} and \eqref{eq:bub_improved}). The {\it fully improved} code uses the type of improvement done in the partially improved case cf. \eqref{eq:bub_improved} only in the sum-integral appearing in the expression \eqref{eq:l0_num} of the bare coupling $\lambda_0,$ but, as a major improvement, it uses the subtraction procedure described in Sec.~\ref{sec:numerics} which involves perturbative integrals evaluated using adaptive routines and leads to modified formulas, with improved convergence for the tadpole, as well as bubble and setting sun integrals, in which the convolution is applied to functions which decrease much faster in the UV than the original propagators.\\

The results for $T_{\rm c}$ obtained within these three levels of discretization are shown in Fig.~\ref{Fig:TcTcbar2}. In the plot on the left the result of the fully improved code (points) shows very good agreement with the accurate result of Fig.~\ref{Fig:TcTcbar} (line). As shown in the inset the test of the convergence of $T_{\rm c}$ to the  continuum result as the cut-off is increased required the increase of $N_s$. The discrepancy between the points and the cutoff result $T_{\rm c}(\Lambda)$ is mainly due to the evaluation of the convolution integral with Fourier techniques. Although barely visible in the inset, this discrepancy decreases with increasing values of the cutoff and $N_s.$ The scaling used in the left axis makes possible a direct comparison of this figure to Fig.~2 of \cite{Arrizabalaga:2006hj}, where the same quantity was obtained by solving the model within the same 2PI approximation, but in Minkowski space. Note that in that reference $T_{\rm c}$ (denoted there as $\hat T_{\rm c}$) slightly increases with the cutoff. This is not a shortcoming of the renormalization procedure, because here we have applied exactly the same method which leads to the same relations between the bare and renormalized quantities, rather it is probably a discretization artifact of the numerical method used in \cite{Arrizabalaga:2006hj}.  

The plot on the right of Fig.~\ref{Fig:TcTcbar2} shows $T_{\rm c}$ obtained with the unimproved and partially improved code (inset). The result of the partially improved code are acceptable if $N_\tau$ is increased by a factor of 5 compared to that obtained with the fully improved code. The values given by the unimproved code are far away from the true ones, even for huge values of $N_\tau.$ However, due to the decrease of the results with $N_\tau,$ an acceptable asymptotic value for $T_{\rm c}$ can be extracted through a fit. 

The comparison presented in Fig.~\ref{Fig:TcTcbar2} shows that the acceleration of Matsubara sums discussed in Appendix~\ref{app:Matsubara} is an important ingredient of the numerical method used to obtain accurate results. 

%
\subsubsection{Discretization errors due to the use of the trapezoidal rule}

The effect of the discretization related to the use of the trapezoidal rule to perform local type integrals can be easily seen by comparing the values of $\lambda_0$ and $\lambda_2$ evaluated accurately using adaptive numerical integration, with those obtained for a given discretization, that is for fixed values of $N_s$ and $N_\tau.$ The comparison is shown in Fig.~\ref{Fig:bare_couplings}. Since the acceleration of the Matsubara sum given in \eqref{eq:bub_improved} is implemented in the expression of $\lambda_0,$ $N_\tau$ does not play practically any role, and thus the comparison tell us up to which value of $\Lambda$ the discretization in momentum space is acceptable. Based on this figure we concluded that $N_s=3\times 2^{10}$ is enough for $\Lambda/T_\star=100,$ but for $\Lambda/T_\star=500$ it is not sufficient to obtain accurate results.

A second example where one can see clearly the effect of the discretization of the momentum integrals is the variation of $V_{\phi=0}$ and $\bar V_{\phi=0}$ with the temperature from $T_\star$ down to $\bar T_{\rm c},$ as shown in Fig.~\ref{Fig:3V}. There we compared the values obtained using the numerical method used in the fully improved code with those obtained by evaluating the perturbative integrals adaptively. We saw that in order to be able to obtain for a given discretization the temperature dependence of $V_{\phi=0}$ and $\bar V_{\phi=0}$ with Fourier techniques, one has to decrease the value of $\Lambda$ as one approaches $\bar T_{\rm c}$, because as a rule of thumb a good description requires to have the lattice spacing in momentum space smaller than the propagator mass, that is $\Delta k=\Lambda/N_s<\bar M_{\phi=0}$.\\

\begin{figure}[!htbp] 
\begin{center}
\includegraphics[width=0.50\textwidth]{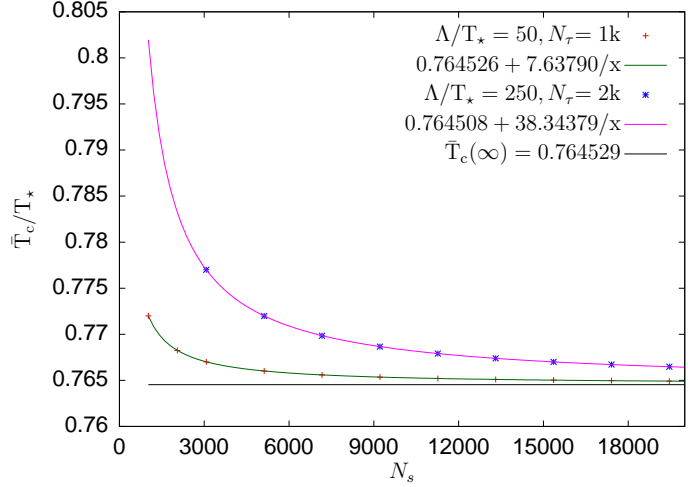}
\caption{$\bar T_{\rm c}$ obtained from solving \eqref{eq:Tcbar_l0} as a function of $N_s$ at two different values of $\Lambda$. The curve with the smaller $\Lambda/N_s$ converges faster.\label{Fig:barTc_num}}
\end{center}
\end{figure}

As a third example, if one tries to determine $\bar T_{\rm c}$ from the discretized version of its defining equation in the fully improved case
\beq
\label{eq:Tcbar_l0}
m_\star^2+\frac{\lambda_0}{2}{\cal V}_{N_\tau,N_s}[\delta \bar G_0]+\frac{\lambda_0}{2}\big[{\cal T}[G_\star]-{\cal T}_\star[G_\star]\big]=0,
\eeq
where $\delta G_0=\bar G_0-G_\star,$ with the massless propagator $G_0(Q)=1/Q^2$, one runs into difficulties related to the fact that one cannot resolve the infrared behavior of the double-sum, which would require a momentum lattice spacing smaller than the mass scale. The best one can do here is to fix the value of the cutoff and increase $N_s,$ that is determine $\bar T_{\rm c}$ for smaller and smaller values of the lattice spacing in momentum space $\Delta k=\Lambda/N_s.$ The value of $N_\tau$ does not play a big role here, as we have tested by using $N_\tau=2^{10}$ and $N_\tau=2\times 2^{10}.$ As shown in Fig.~\ref{Fig:barTc_num}, $\bar T_{\rm c}$ decreases as $1/N_s$. This allows to determine quite accurately through a fit the critical temperature, even from the discretized version of the defining equation.\\

\begin{figure}[!htbp] 
\begin{center}
\includegraphics[width=0.46\textwidth]{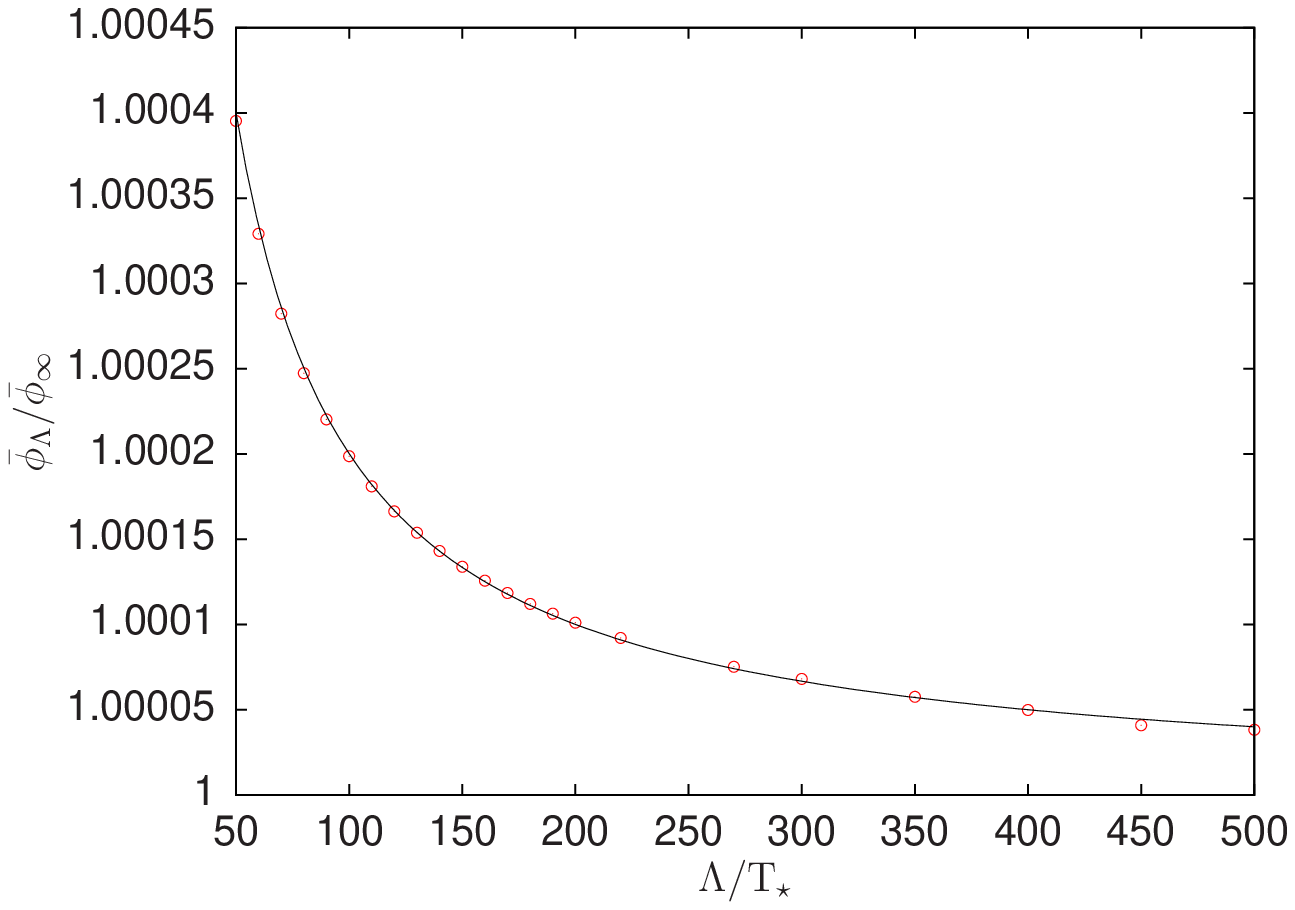}
\hspace*{0.5cm}
\includegraphics[width=0.46\textwidth]{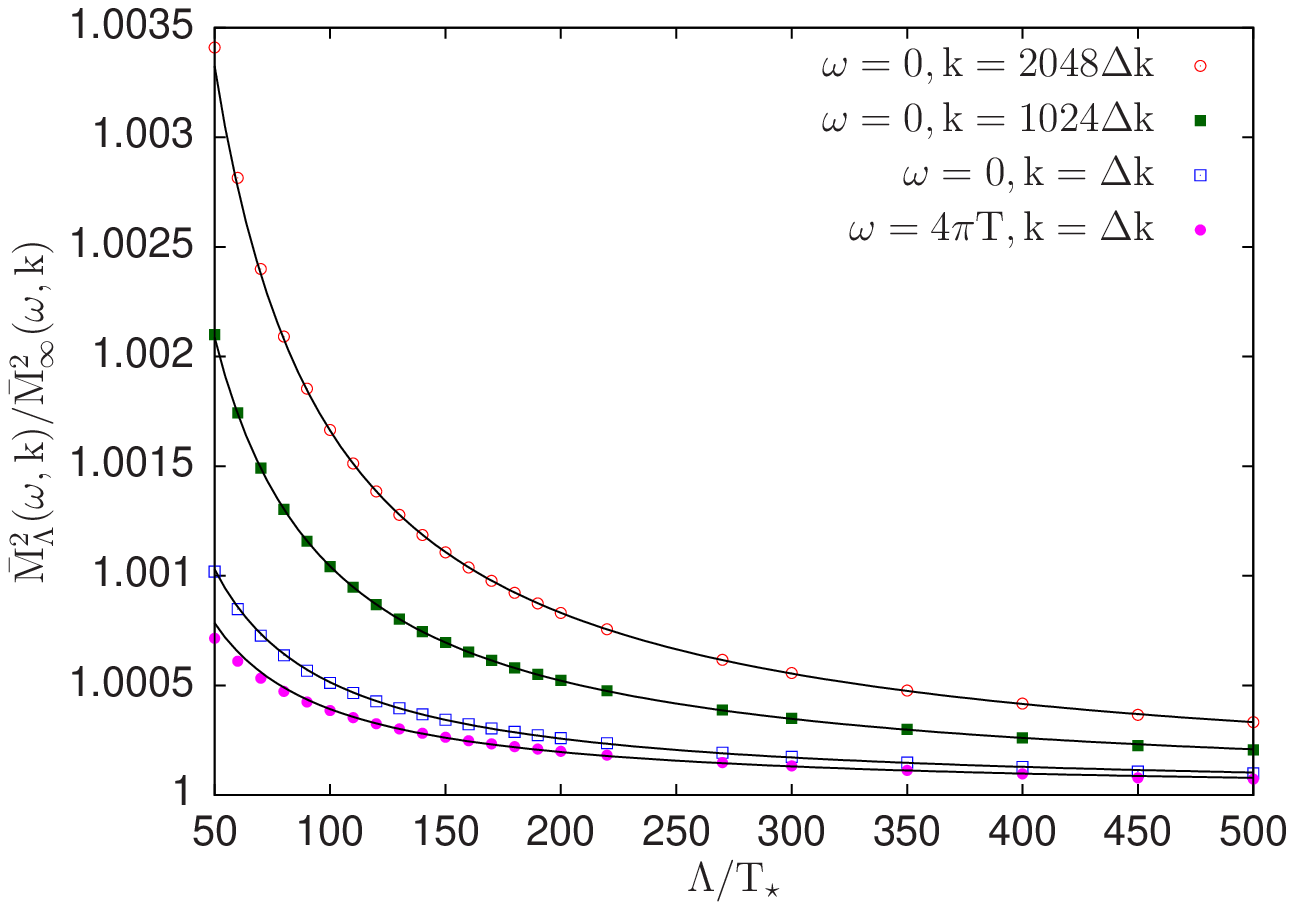}
\caption{Cutoff dependence of the solution to the coupled set of gap and field equations \eqref{eq:gap_num} and \eqref{eq:field_num} at $T/T_\star=0.8$ (points), scaled by the corresponding asymptotic value at $\Lambda\to\infty$ obtained by fitting $\bar\phi_\infty+c/\Lambda$ and $\bar M^2_\infty(k,\omega)+c_{k,\omega}/\Lambda$ (lines) to the corresponding set of points. The convergence of $\bar M^2(k,\omega)$ is slower for higher momenta (right panel). The parameters are $m_\star^2/T_\star^2=0.04,$ $\lambda_\star=3.$ The discretization is characterized by $\Delta k=\Lambda/N_s$ kept fixed at the value $10/2^{10}$ and $N_\tau=512,$ with the exceptions of the points at $\Lambda/T_\star>200,$ for which, in order to achieve accurate results, $N_\tau$ was increased by a factor of $2.$ \label{Fig:CO_dep}}
\end{center}
\end{figure}

The variation with the temperature of the order parameter and of the first bin of the self-energy obtained with the fully improved code shows (see Fig.~\ref{Fig:phi-T}) that the discretization effects are under control in the fully improved code for fixed value of the cutoff. In Fig.\ref{Fig:CO_dep} we show the cutoff dependence of the order parameter and the self-energy at different values of momenta at a given temperature. The quantities seems to converge as $1/\Lambda$, and practically one could regard them as cutoff insensitive, to a good accuracy. As already mentionned, a good description of the results at large cutoff values requires huge values of $N_s$. Moreover, in order to see the scaling behavior with $\Lambda,$ $N_\tau$ had to be increased for large values of the cutoff ($\Lambda/T_\star>200$) as well. This is because the error made by cutting the Matsubara sums depends on $\Lambda$. In the case of the unsubtracted tadpole for instance, the error is $\sim\Lambda^3/N_\tau T$, see App.~\ref{app:Matsubara}. Increasing $\Lambda$ without increasing $N_\tau$ would produce a cubic divergent which is not the correct UV behavior of the unsubtracted tadpole. 

All these tests convincingly show that the subtraction method described in details in Sec.~\ref{sec:numerics} accelerates the Matsubara sums and renders more efficient the evaluation of convolution using fast Fourier transformations. Therefore, it represents a reliable numerical method capable of providing accurate results.

\section{Proof of renormalizability}\label{sec:renorm_proof}
In this section, we show that the gap and field equations, and the effective potential, are rendered finite by the bare parameters given in  Eqs.~(\ref{eq:m0}), (\ref{eq:m2}), (\ref{eq:l0}), (\ref{eq:l2nl}), (\ref{eq:l2l}) and (\ref{eq:l4}). This result is nontrivial because the bare parameters do not depend on $T$ or $\phi$, whereas the gap and field equations, or the effective potential, do. The proof is also the opportunity to illustrate some useful techniques which allow to discuss the ultraviolet behavior of Matsubara sum-integrals.

\subsection{Renormalization of the gap equation}\label{sec:gap_ren}
We discuss the renormalization of the gap equation first. Using the expression (\ref{eq:m0}) for $m^2_0$, the gap equation (\ref{eq:gap2}) becomes
\beq\label{eq:gap_mixed}
\bar M^2(K)=m^2_\star+\frac{\lambda_2}{2}\phi^2+\frac{\lambda_0}{2}\big[{\cal T}[\bar G]-{\cal T}_\star[G_\star]\big]-\frac{\lambda_\star^2}{2}\phi^2{\cal B}[\bar G](K)\,.
\eeq
To proceed, it is convenient to decompose the momentum dependent mass $\bar M^2(K)$ into a local and a nonlocal part, that is $\bar M^2(K)=\bar M^2_{\rm l}+\bar M^2_{\rm nl}(K)$ with
\beq
\bar M^2_{\rm l} & \equiv & m^2_\star+\frac{\lambda_{2{\rm l}}}{2}\phi^2+\frac{\lambda_0}{2}\big[{\cal T}[\bar G]-{\cal T}_\star[G_\star]\big],\label{eq:Ml}\\
\bar M^2_{\rm nl}(K) & \equiv & -\frac{\lambda_\star^2}{2}\phi^2\big[{\cal B}[\bar G](K)-{\cal B}_\star[G_\star](0)\big]\,,\label{eq:Mnl}
\eeq
where we have used the decomposition $\lambda_2=\lambda_{2{\rm l}}+\delta\lambda_{2{\rm nl}}$ and the expression (\ref{eq:l2nl}) for $\delta\lambda_{2{\rm nl}}$. Let us now discuss the local and nonlocal parts separately and show that they are both convergent. Using the results of App.~\ref{app:th_exp}, the difference of tadpole sum-integrals appearing in $\bar M^2_{\rm l}$ can be written as
\beq\label{eq:TT}
{\cal T}[\bar G]-{\cal T}_\star[G_\star]=\int_{Q_\star}^{T_\star}\delta\bar G(Q_\star)+\int_{\tilde Q}\delta\sigma(\tilde Q)\,,
\eeq
where $\delta\bar G(Q_\star)\equiv \bar G(Q_\star)-G_\star(Q_\star)$. Here $\bar G(Q_\star)$ is the analytic continuation of the Matsubara propagator $\bar G(Q)$ to Matsubara frequencies at temperature $T_\star$, where $\bar G(Q)$ is originally defined for Matsubara frequencies at temperature $T$. The second integral is a Minkowski-type integral over $\tilde Q\equiv (q_0,q)$ with $\delta\sigma(\tilde Q)\equiv\rho(q_0,q)\varepsilon(q_0)(n_{|q_0|}-n^\star_{|q_0|})$ and $\rho(q_0,q)$ the spectral density which enters the spectral representation (\ref{eq:spectral}) of $\bar G(Q)$. As explained in App.~\ref{app:th_exp}, this formula is useful to discuss the ultraviolet behavior of ${\cal T}[\bar G]-{\cal T}_\star[G_\star]$. Indeed, if we write $\delta\bar G(Q_\star)=-(\bar M^2(Q_\star)-m^2_\star)G^2_\star(Q_\star)+\bar G_{\rm r}(Q_\star)$ with $\bar G_{\rm r}(Q_\star)\equiv (\bar M^2(Q_\star)-m^2_\star)^2G_\star^2(Q_\star)\bar G(Q_\star)$, we obtain
\beq\label{eq:al}
{\cal T}[\bar G]-{\cal T}_\star[G_\star]=-\int_{Q_\star}^{T_\star}(\bar M^2(Q_\star)-m^2_\star)G^2_\star(Q_\star)+\int_{Q_\star}^{T_\star}\bar G_{\rm r}(Q_\star)+\int_{\tilde Q}\delta\sigma(\tilde Q)\,,
\eeq
where the first term is the only one that can generate divergences in the gap equation. The second term is a sum-integral at temperature $T_\star$ whose integrand $\bar G_{\rm r}(Q_\star)$ decreases fast enough at large $Q_\star$, see App.~\ref{app:th_exp}. Moreover, the third term is convergent due to the presence of $\delta\sigma(\tilde Q)$, see App.~\ref{app:th_exp}. Note that the quantity $\bar M^2(Q_\star)$ which appears in the decomposition (\ref{eq:al}) needs to be regarded as the analytic continuation of $\bar M^2(Q)$ to Matsubara frequencies at temperature $T_\star$. Plugging this decomposition into the expression (\ref{eq:Ml}) for $\bar M^2_{\rm l}$, we arrive at
\beq
\bar M^2_{\rm l}= m^2_\star & \!\!-\!\! & \frac{\lambda_0}{2}(\bar M^2_{\rm l}-m^2_\star)\int_{Q_\star}^{T_\star} G_\star^2(Q_\star)+\frac{\lambda_0}{2}\left[\int_{Q_\star}^{T_\star}\bar G_{\rm r}(Q_\star)+\int_{\tilde Q}\delta\sigma(\tilde Q)\right]\nonumber\\
& \!\!+\!\! & \frac{\lambda_{2{\rm l}}}{2}\phi^2-\frac{\lambda_0}{2}\int_{Q_\star}^{T_\star} \bar M^2_{\rm nl}(Q_\star)G_\star^2(Q_\star)\,,
\eeq
where we have used the separation of $\bar M^2(Q_\star)$ into a local and a nonlocal part. The first line is very similar to what appears when one considers the Hartree approximation and can be treated along the same lines: dividing the equation by $\lambda_0$, gathering the contributions proportional to $\bar M^2_{\rm l}-m^2_\star$, using Eq.~(\ref{eq:l0}) and multiplying back the equation by $\lambda_\star$, we obtain
\beq\label{eq:final}
\bar M^2_{\rm l}= m^2_\star & + & \frac{\lambda_\star}{2}\left[\int_{Q_\star}^{T_\star}\bar G_{\rm r}(Q_\star)+\int_{\tilde Q}\delta\sigma(\tilde Q)\right]\nonumber\\
& + & \frac{\lambda_\star}{2}\left[\frac{\lambda_{2{\rm l}}}{\lambda_0}\phi^2-\int_{Q_\star}^{T_\star} \bar M^2_{\rm nl}(Q_\star)G_\star^2(Q_\star)\right].
\eeq
The first line is finite for both integrals are convergent, but the integral in the second line is still divergent. In order to treat this last integral, we need to discuss the nonlocal part $\bar M^2_{\rm nl}(Q)$ first and then its analytic continuation $\bar M^2_{\rm nl}(Q_\star)$ to Matsubara frequencies at temperature $T_\star$. According to Eq.~(\ref{eq:Mnl}), the nonlocal part $\bar M^2_{\rm nl}(Q)$ involves a difference of bubble sum-integrals which is shown to be convergent in App.~\ref{app:th_exp}. We also obtain a formula for the analytical continuation $\bar M^2_{\rm nl}(K_\star)$  which is needed to complete the discussion of Eq.~(\ref{eq:final}):
\begin{eqnarray}\label{eq:39}
{\cal B}[\bar G](K_\star)-{\cal B}_\star[G_\star](L_\star) & = & \int^{T_\star}_{Q_\star} G_\star(Q_\star)\big[G_\star(Q_\star+K_\star)-G_\star(Q_\star+L_\star)\big]\nonumber\\
& + & \int^{T_\star}_{Q_\star} \delta \bar G(Q_\star)\big[2G_\star(Q_\star+K_\star)+\delta\bar G(Q_\star+K_\star)\big]\nonumber\\
& + & 2\int_{\tilde Q}\delta\sigma(\tilde Q)\bar G(\tilde Q+K_\star)\,,
\end{eqnarray}
where we have introduced a general subtraction point $L_\star$ for later purpose. This formula is used in App.~\ref{app:th_exp} to show not only that the analytically continued difference of bubble sum-integrals, and in turn $\bar M^2_{\rm nl}(K_\star)$, converges but also that  it grows logarithmically at large $K_\star$, this logarithmic behavior being completely accounted for by the first term of Eq.~(\ref{eq:39}) with $L_\star=0$. But it is precisely this contribution which generates the remaining divergence in Eq.~(\ref{eq:final}). This divergence is then $T$-independent and proportional to $\phi^2$, as it should if it is to be absorbed by $\lambda_{2{\rm l}}$. In fact, after plugging Eq.~(\ref{eq:39}) with $L_\star=0$ into Eq.~(\ref{eq:final}), we obtain
\beq\label{eq:M2l_finite}
\bar M^2_{\rm l}= m^2_\star & + & \frac{\lambda_\star}{2}\left[\int_{Q_\star}^{T_\star}\bar G_{\rm r}(Q_\star)+\int_{\tilde Q}\delta\sigma(\tilde Q)\right]+\frac{\lambda_\star}{2}\phi^2\left[\frac{\lambda_{2{\rm l}}}{\lambda_0}+\frac{\lambda^2_\star}{2}\int_{Q_\star}^{T_\star} G_\star^2(Q_\star)\big[{\cal B}_\star[G_\star](Q_\star)-{\cal B}_\star[G_\star](0)\big]\right]\nonumber\\
& + & \frac{\lambda^2_\star}{4}\phi^3\int_{Q_\star}^{T_\star} G_\star^2(Q_\star)\left[\int^{T_\star}_{R_\star} \delta \bar G(R_\star)\big[2G_\star(R_\star+Q_\star)+\delta\bar G(R_\star+Q_\star)\big]+2\int_{\tilde R}\delta\sigma(\tilde R)\bar G(\tilde R+Q_\star)\right],
\eeq
where the only divergent contribution, that is the sum-integral in the second bracket, combines with $\lambda_{2{\rm l}}/\lambda_0$ to yield $1$, according to Eq.~(\ref{eq:l2l}). This completes the proof of renormalization of the gap equation.

\subsection{Renormalization of the field equation}\label{sec:eff_pot_ren}
In order to prove that the field equation is renormalized by the bare parameters $m_2$, $\lambda_2$ and $\lambda_4$, we prove a stronger result first, namely that the first derivative of the effective potential
\beq\label{eq:dg_dPhi}
\frac{\delta\gamma}{\delta\phi}=\phi\left(m^2_2+\frac{\lambda_4}{6}\phi^2+\frac{\lambda_2}{2}\,{\cal T}[\bar G]-\frac{\lambda^2_\star}{6}\,{\cal S}[\bar G]\right)
\eeq
can be put in an explicitly renormalized form. Using the expression (\ref{eq:m2}) for $m^2_2$, we obtain
\beq
\frac{\delta\gamma}{\delta\phi}=\phi\left[m^2_\star+\frac{\lambda_4}{6}\phi^2+\frac{\lambda_2}{2}\,\Big[{\cal T}[\bar G]-{\cal T}_\star[G_\star]\Big]-\frac{\lambda^2_\star}{6}\Big[{\cal S}[\bar G]-{\cal S}_\star[G_\star]\Big]\right].
\eeq
In App.~\ref{app:th_exp}, we show that Eq.~(see \eqref{eq:DfS})
\beq
{\cal S}[\bar G]-{\cal S}_\star[G_\star]=3\int_{Q_\star}^{T_\star}\delta\bar G(Q_\star)\,{\cal B}_\star[G_\star](Q_\star)+3\int_{\tilde Q}\delta\sigma(\tilde Q)\,{\cal B}^R_\star[G_\star](\tilde Q)+\Delta_{\rm f}{\cal S}\,,
\eeq
where $\delta\bar G(Q_\star)$ and $\delta\sigma(\tilde Q)$ have been defined in the previous section, ${\cal B}^R_\star[G_\star](\tilde Q)$ is the retarded bubble contribution, obtained from ${\cal B}_\star[G_\star](Q_\star)$ after analytic continuation and $\Delta_{\rm f}{\cal S}$ is a convergent quantity. From this decomposition, used together with $\lambda_2=\lambda_{2\rm{l}}+\lambda_{2\rm{nl}}$ and Eq.~(\ref{eq:TT}), it follows that
\beq\label{eq:vv}
\frac{\delta\gamma}{\delta\phi}=\phi\left(m^2_\star-\frac{\lambda^2_\star}{6}\,\Delta_{\rm f}{\cal S}+\frac{\lambda_4}{6}\phi^2+\frac{1}{2}\int_{Q_\star}^{T_\star}\delta \bar G(Q_\star)\,\Lambda_\star(Q_\star)+\frac{1}{2}\int_{\tilde Q} \delta\sigma(\tilde Q)\,\Lambda^R_\star(\tilde Q)\right)\,,
\eeq
where $\Lambda_\star(Q_\star)$ was defined in Eq.~(\ref{eq:Vs}) and $\Lambda^R_\star(\tilde Q)$ is the corresponding retarded contribution. It is natural to replace $\Lambda_\star(Q_\star)$ by $V_\star(Q_\star)$ because $V_\star(Q_\star)$ is renormalized, as it is clear from Eq.~(\ref{eq:Vren}). Similarly, it is convenient to replace $\Lambda^R_\star(\tilde Q)$ by $V^R_\star(\tilde Q)$. From the definition of $V_{\phi=0}$ in Eq.~(\ref{eq:V}) we can write
\beq
V_\star(Q_\star)-\Lambda_\star(Q_\star)=V^R_\star(\tilde Q)-\Lambda^R_\star(\tilde Q)=-\frac{\lambda_0}{2}\int_{Q_\star}^{T_\star} G^2_\star(Q_\star)\,V_\star(Q_\star)\,.
\eeq
Plugging these formulae into Eq.~(\ref{eq:vv}), we obtain
\beq\label{eq:abc}
\frac{\delta\gamma}{\delta\phi}=\phi\Bigg[m^2_\star-\frac{\lambda^2_\star}{6}\,\Delta_{\rm f}{\cal S}+\frac{\lambda_4}{6}\phi^2 & \!\!+\!\! & \frac{1}{2}\int_{Q_\star}^{T_\star}\delta \bar G(Q_\star)V_\star(Q_\star)+\frac{1}{2}\int_{\tilde Q} \delta\sigma(\tilde Q)V^R_\star(\tilde Q)\nonumber\\
& \!\!+\!\! & \left.\frac{\lambda_0}{4}\left(\int_{R_\star}^{T_\star}\delta\bar G(R_\star)+\int_{\tilde R}\delta\sigma(\tilde R)\right)\int_{Q_\star}^{T_\star} G^2_\star(Q_\star)\,V_\star(Q_\star)\right].
\eeq
Using Eq.~(\ref{eq:TT}) in the last term of this equation, we recognize ${\cal T}[\bar G]-{\cal T}_\star[G_\star]$ which we can rewrite using the gap equation (\ref{eq:gap_mixed}), analytically continued from $K$ to $Q_\star$, as
\beq\label{eq:the_trick}
\frac{\lambda_0}{2}\left(\int_{R_\star}^{T_\star}\delta\bar G(R_\star)+\int_{\tilde R}\delta\sigma(\tilde R)\right)=\frac{\lambda_0}{2}\big[{\cal T}[\bar G]-{\cal T}_\star[G_\star]\big]=\bar M^2(Q_\star)-m^2_\star-\frac{\phi^2}{2}\big[\lambda_2-\lambda^2_\star {\cal B}[\bar G](Q_\star)\big],
\eeq
where we note that the right-hand-side does not really depend on $Q_\star$ and can therefore be brought under the integral sign in the last line of Eq.~(\ref{eq:abc}). We obtain then
\beq\label{eq:abc2}
& & \frac{\delta\gamma}{\delta\phi}=\phi\left[m^2_\star-\frac{\lambda^2_\star}{6}\,\Delta_{\rm f}{\cal S}+\frac{1}{2}\int_{Q_\star}^{T_\star}\bar G_{\rm r}(Q_\star)V_\star(Q_\star)+\frac{1}{2}\int_{\tilde Q} \delta\sigma(\tilde Q)V^R_\star(\tilde Q)\right.\nonumber\\
& & \hspace{3.5cm}\left.+\,\frac{\phi^2}{6}\left(\lambda_4-\frac{3}{2}\int_{Q_\star}^{T_\star}\big[\lambda_2-\lambda^2_\star {\cal B}[\bar G](Q_\star)\big] G^2_\star(Q_\star)\,V_\star(Q_\star)\right)\right],
\eeq
where we have used that $\delta\bar G(Q_\star)+(\bar M^2(Q_\star)-m^2_\star)G^2_\star(Q_\star)=\bar G_{\rm r}(Q_\star)$. Using Eq.~(\ref{eq:39}) with $K_\star=L_\star=Q_\star$ and Eq.~(\ref{eq:Vs}) we arrive finally at
\beq\label{eq:jsp}
& & \frac{\delta\gamma}{\delta\phi}=\phi\left[m^2_\star-\frac{\lambda^2_\star}{6}\,\Delta_{\rm f}{\cal S}+\frac{1}{2}\int_{Q_\star}^{T_\star}\bar G_{\rm r}(Q_\star)V_\star(Q_\star)+\frac{1}{2}\int_{\tilde Q} \delta\sigma(\tilde Q)V^R_\star(\tilde Q)+\frac{\phi^2}{6}\left(\lambda_4-\frac{3}{2}\int_{Q_\star}^{T_\star}\Lambda_\star(Q_\star)G^2_\star(Q_\star)V_\star(Q_\star)\right)\right.\nonumber\\
& & \hspace{1.4cm}\left.+\,\frac{\lambda^2_\star}{4}\phi^2\int_{Q_\star}^{T_\star}V_\star(Q_\star)G^2_\star(Q_\star)\left(\int^{T_\star}_{R_\star} \delta \bar G(R_\star)\big[2G_\star(R_\star+Q_\star)+\delta\bar G(R_\star+Q_\star)\big]+2\int_{\tilde R}\delta\sigma(\tilde R)\bar G(\tilde R+Q_\star)\right)\right].\nonumber\\
\eeq
According to Eq.~(\ref{eq:l42}), the only integral which is still not convergent, that is the last sum integral in the first line of Eq.~(\ref{eq:jsp}), combines with $\lambda_4$ to yield the renormalized result $\lambda_\star$. This completes the proof of renormalization of the first derivative of the effective potential and in turn of the field equation.

\subsection{Renormalization of the subtracted effective potential}
Let us first give a simple argument valid as long as $T>\bar T_{\rm c}$, which includes the symmetric phase and part of the broken phase since $\bar T_{\rm c}\leq T_{\rm c}$, as discussed in Sec.~\ref{sec:transition}. In this range of temperatures the subtracted potential $\Delta\gamma(\phi)$ is defined down to $\phi=0$ and we can thus write
\beq
\Delta\gamma(\phi)=\Delta\gamma(0)+\int_0^\phi d\tilde\phi\,\,\frac{\delta\gamma}{\delta\tilde\phi}\,.
\eeq
The second term in the right hand side of this equation is convergent from the discussion of the previous subsection. The first term is the contribution to $\Delta\gamma(\phi)$ at vanishing field which coincides with that in the Hartree approximation and which is easily renormalized, see for instance \cite{Reinosa:2011ut}.\\

A more direct way, which is valid both in the symmetric and in the broken phase,\footnote{Of course the potential is defined only for those values of the fields where the gap equation admits a solution.} closely follows the derivation used for the field equation. From Eq.~\eqref{eq:the_trick} we obtain
\beq
\frac{\lambda_0}{8}\big[{\cal T}[\bar G]-{\cal T}_\star[\bar G_\star]\big]^2&=&\frac{1}{4}\left(\int_{Q_\star}^{T_\star}\delta\bar G(Q_\star)+\int_{\tilde Q}\delta\sigma(\tilde Q)\right)\, \frac{\lambda_0}{2}\big[{\cal T}[\bar G]-{\cal T}_\star[G_\star]\big]\nonumber\\
&=&\frac{1}{4} \int_{Q_\star}^{T_\star}\delta\bar G(Q_\star) \left[\bar M^2(Q_\star)-m^2_\star -\frac{\phi^2}{2}\big[\lambda_2-\lambda^2_\star {\cal B}[\bar G](Q_\star)\big]\right]\nonumber\\
&+& \frac{1}{4}\int_{\tilde Q}\delta\sigma(\tilde Q) \left[\bar M^2_R(\tilde Q)-m^2_\star -\frac{\phi^2}{2}\big[\lambda_2-\lambda^2_\star {\cal B}_R[\bar G](\tilde Q)\big]\right], 
\eeq
where $\bar M^2_R(\tilde Q)$ and ${\cal B}_R[\bar G](\tilde Q)$ are retarded functions obtained after analytical continuation. Using the equation above together with Eq.~(\ref{eq:39}) with $K_\star=L_\star=Q_\star$ and its analytical continuation in the subtracted effective potential \eqref{eq:pot_prat} one obtains
\beq
\Delta\gamma(\phi)&=&\gamma_0(m_\star,\Lambda)-\gamma_0^\star(m_\star,\Lambda) + 
\frac{1}{2}\int_Q^T\Big[\ln \big[1+\bar\Sigma(Q) G_\star(Q)\big]-\bar\Sigma(Q)\bar G(Q)\Big]+\frac{1}{4} \int_{Q_\star}^{T_\star}\delta\bar G(Q_\star)\bar\Sigma(Q_\star)\nonumber\\
&&-\frac{\phi^2}{4}\left(\frac{\lambda_4}{6}\phi^2+\frac{1}{2}\int_{Q_\star}^{T_\star}\delta \bar G(Q_\star)\,\Lambda_\star(Q_\star)+\frac{1}{2}\int_{\tilde Q} \delta\sigma(\tilde Q)\,\Lambda^R_\star(\tilde Q)\right)+\frac{1}{2}\phi\,\frac{\delta\gamma}{\delta\phi}+\frac{1}{4}\int_{\tilde Q}\delta\sigma(\tilde Q) \bar\Sigma_R(\tilde Q)\nonumber\\
&&-\frac{\phi^2}{8}\left[\int_{Q_\star}^{T_\star}\delta \bar G(Q_\star)\,\big[{\cal B}[\bar G](Q_\star)-{\cal B}_\star[G_\star](Q_\star)\big]+\int_{\tilde Q} \delta\sigma(\tilde Q)\,\big[{\cal B}_R[\bar G](\tilde Q)-{\cal B}^R_\star[G_\star](\tilde Q)\big] \right]\,,
\eeq
where we introduced the shorthand notation $\bar\Sigma(Q)=\bar M^2(Q)-m^2_\star.$ In the square brackets the first integral is finite because the difference of bubble sum-integrals decreases at least as $1/Q_\star$, while the second integral appears in the finite part of Eq.~\eqref{eq:DfS}. The combination in the round bracket is finite because it appeared in Eq.~\eqref{eq:vv}, which was proven finite. It remains to prove that the sum of the first two integrals is finite. To this purpose, note that since the divergences of these two integrals are overall divergences (this has to do with the fact that $\bar\Sigma(Q)$ grows at most logarithmically at large $Q$), they do not depend on the temperature. Moreover, since
\beq
\frac{1}{2}\Big[\ln \big[1+\bar\Sigma(Q) G_\star(Q)\big]-\bar\Sigma(Q)\bar G(Q)\Big] & = & \frac{1}{4}\,\bar\Sigma^2(Q)G^2_\star(Q)+{\cal O}(\bar\Sigma^3(Q)G^3_\star(Q))\,,\\
\frac{1}{4}\delta\bar G(Q_\star)\bar\Sigma(Q_\star) & = & -\frac{1}{4}\,\bar\Sigma^2(Q_\star)G^2_\star(Q_\star)+{\cal O}(\bar\Sigma^3(Q_\star)G^3_\star(Q_\star))\,,
\eeq
the two divergences are opposite to each other and thus cancel in the sum.

\pagebreak

\section{Conclusions}\label{sec:conclusion}
We have studied numerically the temperature phase transition of the real $\varphi^4$ model from the two-loop $\Phi$-derivable approximation. Our analysis reveals that the inclusion of the setting-sun diagram in the 2PI effective action turns the phase transition into a second order type, which is believed to be the true nature of the transition in the model. The correct description of the order of the phase transition, which is also reflected in the behavior of some thermodynamical quantities, like the heat capacity, speed of sound and trace anomaly, represents an improvement as compared to the Hartree approximation, where the transition is known analytically to be of the first order type. With this investigation we confirm (with a higher accuracy) and complete former numerical results obtained in Minkowski space within the same approximation~\cite{Arrizabalaga:2006hj}. 

Leaving the framework of strict $\Phi$-derivable approximations, we have checked that the phase transition remains of the second-order type even if the contribution of the setting-sun diagram is included only at the level of the 2PI effective action, whereas the gap equation is solved at a lower (Hartree) truncation, considerably simplifying in this way the numerical solution of the model. We will report on the renormalization of this ``hybrid'' $\Phi$-derivable approximation and some of its applications in a forthcoming publication \cite{wip}.

In the present two-loop $\Phi$-derivable approximation, using a combination of analytic and numerical methods, we have also determined the static critical exponents which turned out to be of the mean-field type. Implementing some ideas borrowed from the renormalization group approach, we found that if we let the mass and coupling run with the temperature the critical exponents depart from their mean-field values (also they remain of the integer or rational type). Some of them can be determined analytically. In particular, the critical exponent $\delta$ which characterizes the ``magnetization'' on the critical isotherm is found to be equal to $5$, closer to its expected value in three dimensions. Diagrams with higher number of loops have to be included at the level of the effective action in order to have wave function renormalization and in turn a nonzero anomalous dimension. It is an interesting question to know what type of resummation will eventually produce critical exponents which are not integer or rational numbers. The 2PI-$1/N$ provides one example of such a truncation, see \cite{Alford:2004jj,Saito:2011xq}.

Finally, the present work provides a concrete illustration, at finite temperature, of the general approach to renormalization in the 2PI formalism developed in \cite{Berges:2005hc}. Following this approach, we obtained expressions for the bare parameters which, in the present approximation, are given in terms of perturbative sum-integrals. Since we were interested in the effective potential and in thermodynamical quantities, we solved the approximation in the imaginary time formalism, which avoids the discretisation of sharply peaked functions such as the spectral density. The shortcoming of this approach is that we needed to determine the (slowly convergent) Matsubara sums numerically. However, owing to the simple asymptotic behavior of the propagator, we could accelerate the convergence of the Matsusbara sums. The same property allowed us to increase the accuracy of the momentum integrals which were computed using fast Fourier transform algorithms. The tests performed on different physical quantities concerning various discretization effects confirmed the gain in accuracy. It remains to be seen to what extent the numerical methods developed here can be applied to more complicated truncations in the 2PI formalism, like the next-to-leading order $1/N$ truncation in the $O(N)$ model, the difficulty being that beyond the present two-loop approximation the asymptotic properties of the propagator are changed in a nontrivial way.

\acknowledgments{We thank A. Arrizabalaga, Sz. Bors\'anyi and J. Serreau for valuable discussions on related work. G.\ M. and Zs.\ Sz. would like to thank the CPHT at Ecole Polytechnique for its hospitality and visitor support. U.\ R. would like to thank Andr\'as Patk\'os for his hospitality and visitor support from the E\"otv\"os Lor\'and University during part of the late stages of this work. G.\ M. and Zs.\ Sz. were supported by the Hungarian Scientific Research Fund (OTKA) under Contract Nos. K77534 and T068108.}

\appendix

\section{Thermal expansions}\label{app:th_exp}
When discussing the renormalization of the gap and field equations, we have to deal with differences of sum-integrals such as the following difference of two tadpole sum-integrals:
\beq\label{eq:taddiff}
{\cal T}[\bar G]-{\cal T}_\star[G_\star]=\int_Q^T \bar G(Q)-\int_{Q_\star}^{T_\star}G_\star(Q_\star)\,,
\eeq
where $\bar G(Q)$ is the Matsubara propagator at temperature $T$ and $G_\star(Q_\star)$ is the free-type propagator $1/(Q^2_\star+m^2_\star)$ that we introduced in Sec.~\ref{subsec:exp_bare_param}. The notations ${\cal T}_\star$ and $Q_\star$ are used to emphasize that the subtracted contribution ${\cal T}_\star[G_\star]$ differs from ${\cal T}[\bar G]$ not only by its propagator but also by the different temperature entering the Matsubara frequencies: $Q_\star=(i\omega_n^\star,q)$ with $\omega^\star_n=2\pi nT_\star$. In what follows, we bring Eq.~(\ref{eq:taddiff}) and similar differences for the bubble and setting-sun sum-integrals to a form which is convenient for discussing their ultraviolet behavior. To this purpose, we make extensive use of the ``analytic'' propagator
\beq\label{eq:spectral}
\bar G(Z)\equiv\int_{q_0}\frac{\rho(q_0,q)}{q_0-z}\equiv\int_{-\infty}^{+\infty}\frac{dq_0}{2\pi}\frac{\rho(q_0,q)}{q_0-z}\,,
\eeq
where $Z\equiv (z,q)$ and $z$ belongs to the complex plane minus some possible points and segments of the real axis where the spectral density $\rho$ is non zero. When evaluated for $Z=Q$, that is for $z=i\omega_n$ with $\omega_n=2\pi nT$ a Matsubara frequency at the same temperature $T$ than the spectral density $\rho$, we obtain the Matsubara propagator $\bar G(Q)$ which appears for instance in the first sum-integral ${\cal T}[\bar G]$. But we can also consider a ``hybrid'' propagator $\bar G(Q_\star)$ by evaluating the analytic propagator for $Z=Q_\star$, that is for $z=i\omega^\star_n$ with $\omega^\star_n=2\pi nT_\star$ a Matsubara frequency at a temperature $T_\star$ different from that of $\rho$, and use this hybrid propagator to compute a hybrid sum-integral such as ${\cal T}_\star[\bar G]\equiv\int_{Q_\star}^{T_\star}\bar G(Q_\star)$. This type of hybrid sum-integrals will be useful in what follows.

\subsection{Tadpole sum-integrals}\label{app:tad_diff}
Using the spectral representation (\ref{eq:spectral}) for the Matsubara propagator $\bar G(Q)$ in ${\cal T}[\bar G]$, and performing the Matsubara sum, it is a simple exercise to arrive at
\beq\label{eq:A3}
{\cal T}[\bar G]=\int_{q_0}\int_q\rho(q_0,q)n_{q_0}\,.
\eeq
Writing $n_{q_0}=n^\star_{q_0}+\delta n_{q_0}$, we obtain
\beq\label{eq:A4}
{\cal T}[\bar G]=\int_{q_0}\int_q\rho(q_0,q)n^\star_{q_0}+\int_{q_0}\int_q \rho(q_0,q)\delta n_{q_0}\,.
\eeq
By repeating the calculation that leads to Eq.~(\ref{eq:A3}), it is easily checked that the hybrid sum-integral ${\cal T}_\star[\bar G]$ is nothing but the first term of Eq.~(\ref{eq:A4}). Then, if we introduce the notations $\tilde Q\equiv (q_0,q)$ and $\delta\sigma(\tilde Q)\equiv \rho(q_0,q)\delta n_{q_0}$, Eq.~(\ref{eq:A4}) can be written finally as
\beq\label{eq:dtkw}
{\cal T}[\bar G]-{\cal T}_\star[\bar G]=\int_{\tilde Q}\delta\sigma(\tilde Q)\,,
\eeq
where the right-hand-side is a Minkowski-type integral, in the sense that it involves an integral over a real frequency $q_0$. We can use this formula to express the original difference of tadpole sum-integrals (\ref{eq:taddiff}) as
\beq\label{eq:taddiffres}
{\cal T}[\bar G]-{\cal T}_\star[G_\star] & \!\!=\!\! & \big[{\cal T}[\bar G]-{\cal T}_\star[\bar G]\big]+\big[{\cal T}_\star[\bar G]-{\cal T}_\star[G_\star]\big]\nonumber\\
& \!\!=\!\! & \int_{\tilde Q}\delta\sigma(\tilde Q)+\int_{Q_\star}^{T_\star}\delta \bar G(Q_\star)\,,
\eeq
where we have introduced $\delta\bar G(Q_\star)\equiv \bar G(Q_\star)-G_\star(Q_\star)$ which involves the hybrid propagator $\bar G(Q_\star)$. As we now show, this formula facilitates the discussion of the ultraviolet behavior. Notice first that $\delta n_{q_0}=\varepsilon(q_0)\delta n_{|q_0|}$ with $\delta n_{|q_0|}\equiv n_{|q_0|}-n^\star_{|q_0|}$, and thus $\delta\sigma(\tilde Q)=\rho(q_0,q)\varepsilon(q_0)\delta n_{|q_0|},$ where $\varepsilon(q_0)$ denotes the sign function. It follows that the Minkowski integral in Eq.~(\ref{eq:taddiffres}) converges. Indeed, the integral over $q_0$ is cut off by $\delta n_{|q_0|}$ both for positive and negative values of $q_0$. Moreover, although we cannot really prove this fact, the spectral density is expected to decrease fast enough at large $q$ and fixed $q_0$: a perturbative estimate in the present approximation shows that the spectral density decreases like $1/q^8$ at large $q$ and fixed $q_0$. As for the second term in Eq.~(\ref{eq:taddiffres}), it involves a sum-integral at temperature $T_\star$. If we write
\beq\label{eq:exp}
\delta \bar G(Q_\star)=-(\bar M^2(Q_\star)-m^2_\star)G^2_\star(Q_\star)+\bar G_{\rm r}(Q_\star)\,,
\eeq
with $\bar G_{\rm r}(Q_\star)\equiv(\bar M^2(Q_\star)-m^2_\star)^2G^2_\star(Q_\star)\bar G(Q_\star)$ and use the fact that, in the present approximation, $\bar M^2(Q_\star)$ is expected to grow at most logarithmically at large $Q_\star$, we see that the second term of Eq.~(\ref{eq:taddiffres}) diverges logarithmically and that the divergence is generated entirely by the first term of Eq.~(\ref{eq:exp}). For the purpose of renormalization, it is then convenient to rewrite Eq.~(\ref{eq:taddiffres}) as
\beq
{\cal T}[\bar G]-{\cal T}_\star[G_\star]=-\int_{Q_\star}^{T_\star}(\bar M^2(Q_\star)-m^2_\star)G_\star^2(Q_\star)+\int_{Q_\star}^{T_\star}\bar G_{\rm r}(Q_\star)+\int_{\tilde Q}\delta\sigma(\tilde Q)\,.
\eeq
We mention that the quantity $\bar M^2(Q_\star)$ needs to be regarded as the analytic continuation of $\bar M^2(Q)$, which is initially defined for Matsubara frequencies at temperature $T$, to Matsubara frequencies at temperature $T_\star$, just as $\bar G(Q_\star)$ which enters $\delta\bar G(Q_\star)$ or $\bar G_{\rm r}(Q_\star)$ is the analytic continuation of $\bar G(Q)$.

\subsection{Bubble sum-integrals}\label{app:bub_diff}
A similar analysis can be done for the difference of bubble sum-integrals:
\beq\label{eq:bubdiff}
{\cal B}[\bar G](K)-{\cal B}_\star[G_\star](L_\star)=\int^T_Q \bar G(Q)\bar G(Q+K)-\int^{T_\star}_{Q_\star} G_\star(Q_\star)G_\star(Q_\star+L_\star)\,.
\eeq
Introducing an additional momentum integral by means of a $\delta$-function in order to symmetrize the role of each propagator, using the spectral representation (\ref{eq:spectral}) and performing the Matsubara sum, we obtain\footnote{It is convenient to use the parity of the propagator to replace $\bar G(Q+K)$ by $\bar G(-Q-K)$ at the beginning of the calculation.}
\beq\label{eq:A9}
{\cal B}[\bar G](K) & = & \int_{q_0}\int_{p_0}\int_q\int_p(2\pi)^3\delta^{(3)}(\p+\q+\k)\,\rho(q_0,q)\rho(p_0,p)\,\frac{1+n_{q_0}+n_{p_0}}{q_0+p_0+i\omega}\nonumber\\
& = & \int_{q_0}\int_{p_0}\int_q\int_p(2\pi)^3\delta^{(3)}(\p+\q+\k)\,\rho(q_0,q)\varepsilon(q_0)\rho(p_0,p)\,\frac{1+2n_{|q_0|}}{q_0+p_0+i\omega}\,,
\eeq
where, in going from the first to the second line, we have used the symmetry between $(p_0,p)$ and $(q_0,q)$ as well as the formula $1+2n_{q_0}=\varepsilon(q_0)(1+2n_{|q_0|})$. Note that, in contrast to the case of ${\cal T}[\bar G]$, it is not straightforward here to compare ${\cal B}[\bar G](K)$ to the hybrid sum-integral ${\cal B}_\star[\bar G](K_\star)$ because the external frequency has changed. We compare it instead to ${\cal B}^{(0)}[\bar G](K)$ where the superscript $(0)$ refers to the zero temperature Euclidean-type integral
\beq
{\cal B}^{(0)}[\bar G](K)\equiv\int_Q \bar G(Q)\bar G(Q+K)\equiv\int\frac{d^4Q}{(2\pi)^4}\,\bar G(Q)\bar G(Q+K)\,.
\eeq
Since the frequencies are not constrainted in the zero temperature integral, we can consider the same external frequency as in ${\cal B}[\bar G](K)$. We can now check that the contribution involving ``$1$'' in the second line of Eq.~(\ref{eq:A9}) is nothing but the one we would obtain by computing ${\cal B}^{(0)}[\bar G](K).$ As for the contribution involving $n_{|q_0|}$, one can use the spectral representation (\ref{eq:spectral}) to perform the integral over $p_0$. After performing the trivial integral over $p$, one obtains finally
\beq\label{eq:bubble}
{\cal B}[\bar G](K)={\cal B}^{(0)}[\bar G](K)+2\int_{\tilde Q}\sigma(\tilde Q)\bar G(\tilde Q+K)\,,
\eeq
where we have used the notation $\tilde Q$ that we introduced in the previous section as well as $\sigma(\tilde Q)\equiv\rho(q_0,q)\varepsilon(q_0)n_{|q_0|}$. Note that the propagator $\bar G(\tilde Q+K)$ needs to be understood as the analytic propagator defined in Eq.~(\ref{eq:spectral}). Back to our original calculation (\ref{eq:bubdiff}), we write as before 
\beq\label{eq:decomp2}
{\cal B}[\bar G](K)-{\cal B}_\star[G_\star](L_\star)=\big[{\cal B}[\bar G](K)-{\cal B}_\star[\bar G](H_\star)\big]+\big[{\cal B}_\star[\bar G](H_\star)-{\cal B}_\star[G_\star](L_\star)\big]\,,
\eeq
where we have introduced an arbitrary external momentum $H_\star$ for later convenience. The first bracket can be treated using Eq.~(\ref{eq:bubble}) and the second bracket can be put in the form of a single sum-integral at temperature $T_\star$. We finally arrive at
\begin{eqnarray}\label{eq:bubble2}
{\cal B}[\bar G](K)-{\cal B}_\star[G_\star](L_\star) & = &  \int^{T_\star}_{Q_\star} \big[\bar G(Q_\star)\bar G(Q_\star+H_\star)-G_\star(Q_\star)G_\star(Q_\star+L_\star)\big]\nonumber\\
& + & \int_Q \bar G(Q)\big[\bar G(Q+K)-\bar G(Q+H_\star)\big]\nonumber\\
& + & 2\int_{\tilde Q}\sigma(\tilde Q)\bar G(\tilde Q+K)-2\int_{\tilde Q}\sigma^\star(\tilde Q)\bar G(\tilde Q+H_\star)\,,
\end{eqnarray}
where $\sigma^\star(\tilde Q)\equiv\rho(q_0,q)\varepsilon(q_0)n^\star_{|q_0|}$. This formula is the generalization of Eq.~(\ref{eq:taddiffres}) to the case of a difference of bubble sum-integrals with non-zero external frequency and momentum. We will also need its analytic continuation from the Matsubara frequencies at temperature $T$ in $K$ to the Matsubara frequencies at temperature $T_\star$ of some vector $K_\star$. This continuation can be done readily on Eq.~(\ref{eq:bubble2}) by replacing $K$ by $K_\star$. The decomposition (\ref{eq:bubble2}) and its analytic continuation from $K$ to $K_\star$ are particularly useful to discuss the ultraviolet behavior. Using $\bar G(Q_\star)=G_\star(Q_\star)+\delta \bar G(Q_\star)$ and choosing $H_\star=L_\star$, Eq.~(\ref{eq:bubble2}) becomes
\begin{eqnarray}\label{eq:A15}
{\cal B}[\bar G](K)-{\cal B}_\star[G_\star](L_\star) & = & \int_Q \bar G(Q)\big[\bar G(Q+K)-\bar G(Q+L_\star)\big]\nonumber\\
& + & \int^{T_\star}_{Q_\star} \delta\bar G(Q_\star)\big[2G_\star(Q_\star+L_\star)+\delta\bar G(Q_\star+L_\star)\big]\nonumber\\
& + & 2\int_{\tilde Q}\sigma(\tilde Q)\bar G(\tilde Q+K)-2\int_{\tilde Q}\sigma^\star(\tilde Q)\bar G(\tilde Q+L_\star)\,.
\end{eqnarray}
Similarly, making the choice $K=K_\star=H_\star$, we obtain
\begin{eqnarray}\label{eq:A16}
{\cal B}[\bar G](K_\star)-{\cal B}_\star[G_\star](L_\star) & = & \int^{T_\star}_{Q_\star} G_\star(Q_\star)\big[G_\star(Q_\star+K_\star)-G_\star(Q_\star+L_\star)\big]\nonumber\\
& + & \int^{T_\star}_{Q_\star} \delta \bar G(Q_\star)\big[2G_\star(Q_\star+K_\star)+\delta\bar G(Q_\star+K_\star)\big]\nonumber\\
& + & 2\int_{\tilde Q}\delta\sigma(\tilde Q)\bar G(\tilde Q+K_\star)\,.
\end{eqnarray}
Using similar arguments as those presented in the previous section, it is easily checked that all the contributions in Eqs.~(\ref{eq:A15}) and (\ref{eq:A16}) are convergent. In particular the Minkowski integrals are convergent due to the presence of the functions $\sigma(\tilde Q)$, $\sigma^\star(\tilde Q)$ or $\delta\sigma(\tilde Q)$. The sum-integrals at temperature $T_\star$ which appear in the second lines of Eqs.~(\ref{eq:A15}) and (\ref{eq:A16}) are convergent because they involve either $\delta\bar G(Q_\star) G_\star(Q_\star+K_\star)$ or $\delta\bar G(Q_\star)\delta\bar G(Q_\star+K)$ which decrease fast enough at large $Q_\star$. As for the first lines of Eqs.~(\ref{eq:A15}) and (\ref{eq:A16}) they differ only by the temperature at which the sum-integral is computed and by the propagator which is used ($T=0$ and $\bar G(Q)$ for Eq.~(\ref{eq:A15}) and $T_\star$ and $G_\star(Q_\star)$ for (\ref{eq:A16})). Each of the integrands appears as a difference of two terms which decrease both exactly as $1/(Q^2)^2$ in Eq.~(\ref{eq:A15}) or as $1/(Q_\star^2)^2$ in (\ref{eq:A16}). It follows that the integrands decrease strictly faster than $1/(Q^2)^2$ and $1/(Q_\star^2)^2$, and the corresponding sum-integrals are convergent. We will also need to know the asymptotic behavior of Eq.~(\ref{eq:A16}) as $K_\star$ becomes large. The Minkowski integral is subleading for it behaves as 
\beq
\int_{\tilde Q}\delta\sigma(\tilde Q)\bar G(\tilde Q+K_\star)\sim\frac{1}{K^2_\star}\int_{\tilde Q}\delta\sigma(\tilde Q)\,.
\eeq
Some analysis that we shall not reproduce here allows us to argue that the second line of Eq.~(\ref{eq:A16}) behaves also like $1/K^2_\star$, up to logarithmic corrections, and the first one behaves logarithmically. Thus, the dominant contribution at large $K_\star$ is encoded in the first line of Eq.~(\ref{eq:A16}).

\subsection{Setting-sun sum-integrals}\label{app:sun_diff}
Let us finally consider the difference of setting-sun sum-integrals at zero momentum
\beq\label{eq:Sdiff}
{\cal S}[\bar G]-{\cal S}_\star[G_\star]=\int_Q^T\int_R^T\,\bar G(Q)\bar G(R)\bar G(R+Q)-\int_{Q_\star}^{T_\star}\int_{R_\star}^{T_\star}\,G_\star(Q_\star)G_\star(R_\star)G_\star(R_\star+Q_\star)\,.
\eeq
Introducing an additional momentum integral by means of a $\delta$-function in order to symmetrize the role of each propagator, using the spectral representation (\ref{eq:spectral}) and performing the Matsubara sums, we obtain
\beq\label{eq:tt}
{\cal S}[\bar G]=\int_{r_0}\int_{q_0}\int_{p_0}\int_r\int_q\int_p(2\pi)^3\delta^{(3)}(\p+\q+\rb)\, \rho(r_0,r)\rho(q_0,q)\rho(p_0,p)\,\frac{n_{p_0}n_{q_0}-n_{p_0}n_{-r_0}+n_{-q_0}n_{-r_0}}{r_0+q_0+p_0}\,.
\eeq
Note that there is no singularity as the denominator $r_0+q_0+p_0$ approaches $0$ due to the particular combination of statistical factors in the numerator. For later convenience, we shall then replace $r_0+q_0+p_0$ by $r_0+q_0+p_0+i\alpha$. This does not bring any imaginary part because the sign of $\alpha$ can be chosen arbitrarily. Generalizing the approach of \cite{Blaizot:2004bg}, we now write $n_{q_0}=n^\star_{q_0}+\delta n_{q_0}$ and $-n_{-q_0}=-n^*_{-q_0}+\delta n_{q_0}$, and obtain
\beq\label{eq:tt2}
{\cal S}[\bar G] & = & \int_{r_0}\int_{q_0}\int_{p_0}\int_r\int_q\int_p(2\pi)^3\delta^{(3)}(\p+\q+\rb)\, \rho(r_0,r)\rho(q_0,q)\rho(p_0,p)\,\frac{n^\star_{p_0}n^\star_{q_0}-n^\star_{p_0}n^\star_{-r_0}+n^\star_{-q_0}n^\star_{-r_0}}{r_0+q_0+p_0+i\alpha}\nonumber\\
& + & 3\int_{r_0}\int_{q_0}\int_{p_0}\int_r\int_q\int_p(2\pi)^3\delta^{(3)}(\p+\q+\rb)\, \rho(r_0,r)\rho(q_0,q)\rho(p_0,p)\,\frac{\delta n_{q_0}(n^\star_{p_0}-n^\star_{-r_0})}{r_0+q_0+p_0+i\alpha}\nonumber\\
& + & 3\int_{r_0}\int_{q_0}\int_{p_0}\int_r\int_q\int_p(2\pi)^3\delta^{(3)}(\p+\q+\rb)\, \rho(r_0,r)\rho(q_0,q)\rho(p_0,p)\,\frac{\delta n_{q_0}\delta n_{r_0}}{r_0+q_0+p_0+i\alpha}\,,
\eeq
where we have exploited the permutation symmetry between the $(p_0,p)$, $(q_0,q)$ and $(r_0,r)$ pairs. Applying the same steps that lead to Eq.~(\ref{eq:tt}) to the hybrid sum-integral ${\cal S}_\star[\bar G]$, it is easily checked that the first line of Eq.~(\ref{eq:tt2}) is nothing but ${\cal S}_\star[\bar G]$. Comparing the second line to Eq.~(\ref{eq:A9}), we see that it can be written as $\int_{\tilde Q}\delta\sigma(\tilde Q){\cal B}_\star[\bar G](\tilde Q+i\alpha)$ where $\tilde Q+i\alpha\equiv (q_0+i\alpha,q)$ and ${\cal B}_\star[\bar G](\tilde Q+i\alpha)\equiv {\cal B}^R_\star[\bar G](\tilde Q)$ is the retarded (if we choose $\alpha>0$) contribution obtained after analytically continuing ${\cal B}_\star[\bar G](Q_\star)$. Finally, in the third line of Eq.~(\ref{eq:tt2}), we can perform the integral over $p_0$ using the spectral representation which leads to the retarded propagator $\bar G(\tilde R+\tilde Q+i\alpha)$, as well as the integral over $p$. We obtain finally
\beq\label{eq:dtkw2}
{\cal S}[\bar G]={\cal S}_\star[\bar G]+3\int_{\tilde Q}\,\delta\sigma(\tilde Q)\,{\cal B}_\star[\bar G](\tilde Q+i\alpha)+3\int_{\tilde Q}\int_{\tilde R}\delta\sigma(\tilde Q)\delta\sigma(\tilde R)\bar G(\tilde R+\tilde Q+i\alpha)\,.
\eeq
Once again, we can use this formula to compute the original difference (\ref{eq:Sdiff}) of setting-sun sum-integrals for vanishing external frequency and momentum. We write
\beq
{\cal S}[\bar G]-{\cal S}_\star[G_\star] & = & \big[{\cal S}[\bar G]-{\cal S}_\star[\bar G]\big]+\big[{\cal S}_\star[\bar G]-{\cal S}_\star[G_\star]\big]\nonumber\\
& = & 3\int_{\tilde Q}\,\delta\sigma(\tilde Q)\,{\cal B}_\star[\bar G](\tilde Q+i\alpha)+3\int_{\tilde Q}\int_{\tilde R}\delta\sigma(\tilde Q)\delta\sigma(\tilde R)\,\bar G(\tilde R+\tilde Q+i\alpha)\nonumber\\
& + & \int_{Q_\star}^{T_\star}\int_{R_\star}^{T_\star}\big[\bar G(Q_\star)\bar G(R_\star)\bar G(R_\star+Q_\star)-G_\star(Q_\star)G_\star(R_\star)G_\star(R_\star+Q_\star)\big].
\eeq
This formula is the generalization of Eq.~(\ref{eq:taddiffres}) to the case of the difference of two setting-sun sum-integrals for zero external frequency and momentum. To obtain a formula suited for the discussion of ultraviolet divergences, we use $\bar G(Q_\star)=G_\star(Q_\star)+\delta\bar G(Q_\star)$. We obtain
\beq
{\cal S}[\bar G]-{\cal S}_\star[G_\star] & = & 3\int_{Q_\star}^{T_\star}\delta\bar G(Q_\star){\cal B}_\star[G_\star](Q_\star)+\int_{Q_\star}^{T_\star}\int_{R_\star}^{T_\star}\delta \bar G(Q_\star)\delta \bar G(R_\star)\big[3G_\star(R_\star+Q_\star)+\delta \bar G(R_\star+Q_\star)\big]\nonumber\\
& + & 3\int_{\tilde Q}\,\delta\sigma(\tilde Q)\,{\cal B}_\star[\bar G](\tilde Q+i\alpha)+3\int_{\tilde Q}\int_{\tilde R}\delta\sigma(\tilde Q)\delta\sigma(\tilde R)\,\bar G(\tilde R+\tilde Q+i\alpha)\,.
\eeq
It is also convenient to use Eq.~(\ref{eq:bubble}) to rewrite
\beq
{\cal B}_\star[\bar G](\tilde Q+i\alpha)={\cal B}_\star[G_\star](\tilde Q+i\alpha) & \!\!+\!\! & \big[{\cal B}_\star[\bar G](\tilde Q+i\alpha)-{\cal B}_\star[G_\star](\tilde Q+i\alpha)\big]\nonumber\\
 ={\cal B}_\star[G_\star](\tilde Q+i\alpha) & \!\!+\!\! & \int_R \big[\bar G(R)\bar G(R+\tilde Q+i\alpha)-G_\star(R)G_\star(R+\tilde Q+i\alpha)\big]\nonumber\\
& \!\!+\!\! & 2\int_{\tilde R}\sigma_\star(\tilde R)\bar G(\tilde R+\tilde Q+i\alpha)-2\int_{\tilde R}\sigma^\star_\star(\tilde R) G_\star(\tilde R+\tilde Q+i\alpha)\,,
\eeq
where $\sigma^\star_\star(\tilde Q)\equiv\rho_\star(q_0,q)\varepsilon(q_0)n^\star_{|q_0|}$ and $\rho_\star(q_0,q)=(2\pi)\varepsilon(q_0)\delta(q^2_0-q^2-m^2_\star)$ is the spectral density corresponding to $G_\star(Q_\star)$. We can write finally
\beq
{\cal S}[\bar G]-{\cal S}_\star[G_\star] & = & 3\int_{Q_\star}^{T_\star}\delta\bar G(Q_\star){\cal B}_\star[G_\star](Q_\star)+3\int_{\tilde Q}\,\delta\sigma(\tilde Q)\,{\cal B}_\star[G_\star](\tilde Q+i\alpha)\nonumber\\
& + & \int_{Q_\star}^{T_\star}\int_{R_\star}^{T_\star}\delta \bar G(Q_\star)\delta \bar G(R_\star)\big[3G_\star(R_\star+Q_\star)+\delta \bar G(R_\star+Q_\star)\big]\nonumber\\
& + & 3\int_{\tilde Q}\,\delta\sigma(\tilde Q)\,\int_R \big[\bar G(R)\bar G(R+\tilde Q+i\alpha)-G_\star(R)G_\star(R+\tilde Q+i\alpha)\big]\nonumber\\
& + & 6\int_{\tilde Q}\,\delta\sigma(\tilde Q)\int_{\tilde R}\sigma_\star(\tilde R)\bar G(\tilde R+\tilde Q+i\alpha)-6\int_{\tilde Q}\,\delta\sigma(\tilde Q)\int_{\tilde R}\sigma^\star_\star(\tilde R) G_\star(\tilde R+\tilde Q+i\alpha)\nonumber\\
& + & 3\int_{\tilde Q}\int_{\tilde R}\delta\sigma(\tilde Q)\delta\sigma(\tilde R)\,\bar G(\tilde R+\tilde Q+i\alpha)\,.
\label{eq:DfS}
\eeq
A similar discussion as the one presented in the previous sections shows that only the first two terms are divergent. The sum of the remaining terms will be denoted $\Delta_{\rm f}{\cal S}.$

\section{Perturbative sum-integrals}\label{app:pert}
 In this section, we give the explicit expressions for the perturbative tadpole, bubble and setting-sun sum-integrals at temperature $T$ involving a free-type propagator $G(Q)=1/(Q^2+M^2)$ of mass $M$, both in dimensional regularization in $d=4-2\epsilon$ dimensions and in the presence of a sharp 3d cutoff for each propagator. We also discuss the monotonous behavior of the perturbative bubble sum-integral in the presence of a sharp cutoff.
 
\subsection{Perturbative results using dimensional regularization \label{ss:dimreg}} 

From Eq.~(\ref{eq:dtkw}), in which we set $T_\star=0$ and $\delta\sigma(\tilde Q)=(2\pi)n_{|q_0|}\delta(q^2_0-q^2-M^2)$, we obtain the decomposition ${\cal T}_\epsilon[G]={\cal T}^{(0)}_\epsilon[G]+{\cal T}^{(1)}_\epsilon[G].$ ${\cal T}^{(0)}_\epsilon[G]$ is the tadpole at zero temperature, which is easily computed to be 
\beq
{\cal T}^{(0)}_\epsilon[G]=\frac{M^2}{16\pi^2}\left[-\frac{1}{\epsilon}+\ln\frac{M^2}{\bar\mu^2}-1\right]-\epsilon\,\frac{M^2}{32\pi^2}\left[\left(\ln\frac{M^2}{\bar\mu^2}-1\right)^2+\frac{\pi^2}{6}+1\right]+{\cal O}(\epsilon^2),
\eeq
where we introduced the standard notation $\bar\mu^2=4\pi\mu^2 e^{-\gamma_{\mbox{\tiny E}}}$ with $\gamma_{\mbox{\tiny E}}$ standing for Euler's constant, while ${\cal T}^{(1)}_\epsilon[G]$ the finite temperature part of the tadpole, given by
\beq
{\cal T}^{(1)}_\epsilon[G]=\mu^{2\epsilon}\int\frac{d^{d-1}q}{(2\pi)^{d-1}}\frac{n_{\varepsilon_q}}{\varepsilon_q}=\frac{2\mu^{2\epsilon}}{(4\pi)^{\frac{d-1}{2}}\Gamma(\frac{d-1}{2})}\int_0^\infty dq\,q^{d-2}\,\frac{n_{\varepsilon_q}}{\varepsilon_q}\,,
\eeq
with $\varepsilon_q=\sqrt{q^2+M^2}$. Taking a derivative with respect to $M^2$ in the previous expressions, we obtain a decomposition of the bubble sum-integral at zero external momentum ${\cal B}_\epsilon[G](0)={\cal B}^{(0)}_\epsilon[G](0)+{\cal B}^{(1)}_\epsilon[G](0)$ with
\beq
{\cal B}^{(0)}_\epsilon[G](0)=\frac{1}{16\pi^2}\left[\frac{1}{\epsilon}-\ln\frac{M^2}{\bar\mu^2}\right]+\epsilon\,\frac{1}{32\pi^2}\left[\ln^2\frac{M^2}{\bar\mu^2}+\frac{\pi^2}{6}\right]+{\cal O}(\epsilon^2)
\label{eq:bubble_dimreg}
\eeq
and
\beq
{\cal B}_\epsilon^{(1)}[G](0)=-\mu^{2\epsilon}\int\frac{d^{d-1}q}{(2\pi)^{d-1}}\frac{d}{dM^2}\frac{n_{\varepsilon_q}}{\varepsilon_q}=\frac{2\mu^{2\epsilon}}{(4\pi)^{\frac{d-1}{2}}\Gamma(\frac{d-1}{2})}\int_0^\infty dq\,q^{d-4}\,\frac{n_{\varepsilon_q}}{\varepsilon_q}\,,
\eeq
where, in this last integral, we have used the fact that $d/dM^2$ can be replaced by $d/dq^2$ in the integrand and we have integrated by parts assuming $d\geq 2$. We have expanded the vacuum pieces ${\cal T}^{(0)}_\epsilon[G]$ and ${\cal B}^{(0)}_\epsilon[G](0)$ up to and including order $\epsilon$ for later convenience. The thermal parts ${\cal T}^{(1)}_\epsilon[G]$ and ${\cal B}^{(1)}_\epsilon[G]$ will be needed only to order $\epsilon^0$ and we can thus take the limit $\epsilon\to 0$ in those contributions.\\

We proceed similarly for the setting-sun sum-integral. From Eq.~(\ref{eq:dtkw2}), in which we set $T_\star=0$ and also $\varepsilon_r\equiv\sqrt{(\k-\q)^2+M^2},$ we obtain the following decomposition:
\beq\label{eq:dcom}
{\cal S}_\epsilon[G]&=&\mu^{4\epsilon}\int\frac{d^{d-1}k}{(2\pi)^{d-1}}\int\frac{d^{d-1}q}{(2\pi)^{d-1}}\,\frac{1}{4\varepsilon_k\varepsilon_q\varepsilon_r}\,\frac{1}{\varepsilon_k+\varepsilon_q+\varepsilon_r} + 3\mu^{4\epsilon}\int\frac{d^{d-1}k}{(2\pi)^d}\,\frac{n_{\varepsilon_k}}{2\varepsilon_k}\,\Big[{\cal B}_\epsilon^{(0)}(\varepsilon_k+i\alpha,{\bf k})+{\cal B}_\epsilon^{(0)}(-\varepsilon_k+i\alpha,{\bf k})\Big]\nonumber\\
& + & 3\mu^{4\epsilon}\int\frac{d^{d-1}k}{(2\pi)^{d-1}}\int\frac{d^{d-1}q}{(2\pi)^{d-1}}\,\frac{n_{\varepsilon_k}}{2\varepsilon_k}\frac{n_{\varepsilon_q}}{2\varepsilon_q}\left[\frac{1}{(\varepsilon_r+i\alpha)^2-(\varepsilon_k+\varepsilon_q)^2}+\frac{1}{(\varepsilon_r+i\alpha)^2-(\varepsilon_k-\varepsilon_q)^2}\right].
\eeq
The terms of the sum above contain in order zero, one and two statistical factors with positive argument and will be denoted respectively as ${\cal S}^{(0)}_\epsilon[G]$, ${\cal S}^{(1)}_\epsilon[G],$ and ${\cal S}^{(2)}_\epsilon[G]$. Note that in the present perturbative calculation, the regulator $\alpha$ plays no role if we assume $M^2>0$. This is always true for ${\cal S}^{(0)}[G]$ for it is the zero temperature limit of a diagram which does not depend on $\alpha$. For ${\cal S}_\epsilon^{(1)}[G],$ the analytically continued bubble contribution is evaluated on the mass shell and therefore does not generate any imaginary part. Similarly, the denominators in ${\cal S}_\epsilon^{(2)}[G]$ never vanish since the equation $0=\varepsilon_r^2-(\varepsilon_k\pm\varepsilon_q)^2$ implies $4(k^2q^2-(\k\cdot\q)^2)+4(k^2-kq+q^2)M^2+3M^4=0,$ which has no solution if $M^2>0$ because the three terms are positive and one of them is strictly positive.

We can also write the zero temperature contribution ${\cal S}_\epsilon^{(0)}[G]$ in a covariant form as:
\beq\label{eq:SS0}
{\cal S}_\epsilon^{(0)}[G]=\mu^{4\epsilon}\int\frac{d^d Q}{(2\pi)^d} \int\frac{d^d K}{(2\pi)^d} G(Q) G(K) G(K-Q)\,.
\eeq
This integral can be evaluated with the method given in Sec.~11.5 of Ref.~\cite{Zinn-Justin:2002}. We only have to obtain the ${\cal O}(\epsilon)$ contribution of the integral $J$ defined in (11.53) of this reference, which using our relation between $d$ and $\epsilon$ is $J=3/\epsilon+3+\epsilon\left(3-11\pi^2/6+2\Psi_1(2/3)\right)+{\cal O}(\epsilon^2),$ with $\Psi_1(x)=d^2 \Gamma(x)/d x^2$ being the trigamma function. Expanding in series of $\epsilon$ one obtains
\beq
{\cal S}_{\epsilon}^{(0)}[G]=\frac{M^2}{(16\pi^2)^2}\left[-\frac{3}{2\epsilon^2}+\frac{3}{\epsilon}\left(\ln\frac{M^2}{\bar\mu^2}-\frac{3}{2}\right)-3\left(\left(\ln\frac{M^2}{\bar\mu^2}-\frac{3}{2}\right)^2+\frac{5}{4}-\frac{5}{36}\pi^2\right)-\Psi_1\left(\frac{2}{3}\right) \right]+\mathcal{O}(\epsilon),
\label{Eq:SS-0stat}
\eeq
This expression agrees with the form given in \cite{Davydychev:1992mt}, upon exploiting a relation between specific values of the trigamma function and the Clausen function $\textnormal{Cl}_2(x)=-\int\limits_0^x d\theta\,\ln (2\sin\left(\theta/2\right))$, namely
\[
\Psi_1\left(\frac{2}{3}\right)=\frac{2}{3}\pi^2-2\sqrt{3}\textnormal{Cl}_2\left(\frac{\pi}{3}\right).
\]
Since ${\cal B}^{(0)}_\epsilon(k_0,\k)$ is Lorentz covariant (in dimensional regularization), ${\cal B}^{(0)}_\epsilon(\pm\varepsilon_k,\k)$ does not depend on ${\bf k}$ and can be pulled out of the integral in ${\cal S}^{(1)}_\epsilon[G]$. In fact this constant contribution can be computed analytically. We obtain finally 
\beq
{\cal S}_\epsilon^{(1)}[G]&=&\frac{3}{16\pi^2}\left(\frac{1}{\epsilon}-\ln\frac{M^2}{\bar\mu^2}+2-\frac{\pi}{\sqrt{3}}+\mathcal{O}(\epsilon)\right) {\cal T}_\epsilon^{(1)}[G],
\label{Eq:SS-1stat}
\eeq
where, for later purpose, it is enough to expand up to order $\epsilon^0$ the prefactor of ${\cal T}^{(1)}_\epsilon[G]$. Finally, the contribution ${\cal S}^{(2)}_\epsilon[G]$ will only be needed in the limit $\epsilon\to 0$ where it yields a finite result due to the presence of the two thermal factors. After integrating over the angles, we obtain
\beq
{\cal S}^{(2)}[G]&=&\frac{3}{32\pi^4}\int_0^\infty dk\,k\,\frac{n_{\varepsilon_k}}{\varepsilon_k} \int_0^\infty dq\,q\, \frac{n_{\varepsilon_q}}{\varepsilon_q} \ln\frac{4(k^2+kq+q^2)+3 M^2}{4(k^2-kq+q^2)+3 M^2}\,.
\label{Eq:SS-2stat}
\eeq
\vglue6mm

Using these results, we can now check that the combination
\beq\label{eq:comb_C}
{\cal C}(M,m_\star)\equiv\Big[{\cal T}[G]-{\cal T}_\star[G_\star]+(M^2-m^2_\star){\cal B}_\star[G_\star](0)\Big] {\cal B}_\star[G_\star](0)
-\frac{1}{3}\left[{\cal S}[G]-S_\star[G_\star] -(M^2-m^2_\star)\frac{d {\cal S}_\star[G_\star]}{d m^2_\star}\right],\ \ \ 
\eeq
which appears in Eq.~\eqref{eq:def_Tc} leads to a finite expression from which the scale $\mu$ drops out. To see this, we use that the expression inside the first pair of brackets is finite and that in terms of the form {\it finite} $\times$ {\it divergent} we have to expand the {\it finite} part to $\mathcal{O}(\epsilon^a),$ where the integer $a$ is the power in the most divergent, $\mathcal{O}(\epsilon^{-a})$ piece of the divergent part. Note however that we do not need to consider the order $\epsilon$ term originating from ${\cal B}^{(1)}_{\star,\epsilon}[G_\star](0)$ in this first bracket because it is identically canceled by the contribution originating from ${\cal T}^{(1)}_{\star,\epsilon}[G]$ in the one thermal factor contribution to $d{\cal S}_{\star,\epsilon}[G_\star]/dm^2_\star$. In the limit $\epsilon\to0$ the expression reads
\beq
\nonumber
{\cal C}(M,m_\star)&=&\frac{1}{(16\pi^2)^2}\left[\frac{M^2}{2}\left(\ln\frac{M^2}{m^2_\star}-2\right)^2 - 2 m^2_\star\right] + \frac{1}{16\pi^2}\left[\ln\frac{M^2}{m^2_\star}+\frac{\pi}{\sqrt{3}}-2\right]{\cal T}^{(1)}[G]\\
&-&\frac{1}{16\pi^2}\left[\frac{M^2}{m^2_\star} + \frac{\pi}{\sqrt{3}}-3\right]{\cal T}^{(1)}_\star[G_\star] + \frac{1}{16\pi^2}\left[M^2\ln\frac{M^2}{m^2_\star}+(M^2-m^2_\star)\left(\frac{\pi}{\sqrt{3}}-3\right)\right]{\cal B}_{\star}^{(1)}[G_\star](0)\nonumber\\
&+&\Big[{\cal T}^{(1)}[G]-{\cal T}_{\star}^{(1)}[G_\star]+(M^2-m^2_\star){\cal B}_{\star}^{(1)}[G_\star](0)\Big]{\cal B}_{\star}^{(1)}[G_\star](0)\nonumber\\
&-&\frac{1}{3}\left[{\cal S}^{(2)}[G]-{\cal S}_\star^{(2)}[G_\star]-(M^2-m^2_\star)\frac{d{\cal S}_\star^{(2)}[G_\star]}{d m^2_\star}\right].
\label{eq:comb_C2}
\eeq
Similarly, for the combination appearing in Eq.~\eqref{eq:V_phi0_conv}
\beq\label{eq:comb_D}
{\cal D}(M, m_\star)\equiv\Big[{\cal B}_\star[G_\star](0) - {\cal B}[G](0)\Big]{\cal B}_\star[G_\star](0)+\frac{1}{3}\left[\frac{d {\cal S}_\star[G_\star]}{d m^2_\star} - \frac{d {\cal S}[G]}{d M^2}\right]\, ,
\eeq  
the following explicitly finite expression can be obtained: 
\beq
\nonumber
{\cal D}(M,m_\star)&=& \frac{1}{128\pi^4}\left(\ln^2\frac{m_\star}{M}+\ln\frac{m_\star}{M}\right)+\frac{1}{16\pi^2}\left[\frac{{\cal T}^{(1)}[G]}{M^2}-\frac{{\cal T}^{(1)}_\star[G_\star]}{m_\star^2}\right]+\frac{1}{3}\left[\frac{d {\cal S}^{(2)}_\star[G_\star]}{dm_\star^2}-\frac{d{\cal S}^{(2)}[G]}{d M^2}\right]\nonumber\\
&+& \left[\frac{1}{16\pi^2}\left(\ln\frac{M^2}{m^2_\star}-2+\frac{\pi}{\sqrt{3}}\right)+{\cal B}^{(1)}_\star[G_\star](0)\right]\left[{\cal B}^{(1)}_\star[G_\star](0)-{\cal B}^{(1)}[G](0)\right].\ \ \ \ 
\label{eq:comb_D2}
\eeq

\subsection{Perturbative results using a 3d cutoff\label{ss:cutoff}}

\begin{figure}[t]                                                               
\begin{center}
\includegraphics[width=0.4\textwidth]{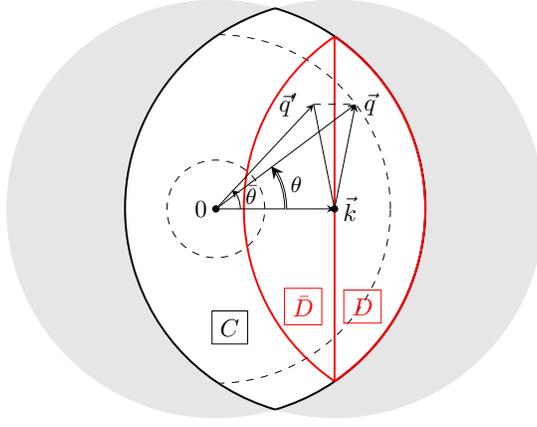}
\caption{Region of integration for the bubble integral.\label{Fig:spheres}}                                                            
\end{center}                                                                    
\end{figure}

 The techniques of App.~\ref{app:th_exp} can also be applied in the presence of a 3d cutoff. From Eq.~(\ref{eq:A3}) with $\rho(q_0,q)=(2\pi)\varepsilon(q_0)\delta(q^2_0-q^2-m^2)$, we obtain
\beq
{\cal T}[G]=\frac{1}{2\pi^2}\int_0^\Lambda dq \,q^2\,\frac{1+2n_{\varepsilon_{q}}}{\varepsilon_q}=\frac{1}{8\pi^2}\left[\Lambda\varepsilon_\Lambda-M^2\,{\rm arcsinh}\left(\frac{\Lambda}{M}\right)\right]+\frac{1}{2\pi^2}\int_0^\Lambda dq \,q^2\,\frac{n_{\varepsilon_{q}}}{\varepsilon_q},
\label{Eq:TAD}
\eeq
where $\varepsilon_q=\sqrt{q^2+M^2}.$ In the second line of Eq.~\eqref{eq:A9}, we can perform the integral over $p$ trivially, as well as the integral over $p_0$ using Eq.~(\ref{eq:spectral}). After some obvious shifts of the integration variables, we obtain
\beq\label{eq:new}
{\cal B}[G](K) & = & \int_{q_0}\mathop{\int_{|\q|<\Lambda}}_{|\k-\q|<\Lambda}\,\rho(q_0,q)\varepsilon(q_0)(1+2n_{|q_0|})G(i\omega-q_0,{\bf k}-{\bf q})\nonumber\\
& = & \mathop{\int_{|\q|<\Lambda}}_{|\k-\q|<\Lambda}\frac{1+2n_{\varepsilon_q}}{2\varepsilon_q}\Big[G(i\omega-\varepsilon_q,{\bf k}-{\bf q})+G(i\omega+\varepsilon_q,{\bf k}-{\bf q})\Big]\,,
\eeq
where in the second line we have performed the frequency integral. The angular integrals are computed using the geometrical picture shown in Fig.~\ref{Fig:spheres}.  With a 3d spherical coordinate system we integrate over the common region of two spheres of radius $\Lambda$ whose centers are separated by a distance $k\equiv|\k|$. The contribution is clearly $0$ if $k>2\Lambda$. There is a non zero contribution if $\Lambda<k<2\Lambda$ but we will be mainly interested in the case $k<\Lambda$. The integral over one angle gives $2\pi,$ while the remaining angle $\theta$ goes from $0$ to $\pi$ if $0\leq q\equiv|\q|\leq \Lambda-k$ and from $0$ to $\arccos\,\alpha<\pi$ if $\Lambda-k\leq q\leq \Lambda$,  with $\alpha\equiv (k^2+q^2-\Lambda^2)/(2k q)$ determined by the intersection ``point'' of the two spheres.  Using the formula
\beq
\int_\alpha^1 d(\cos\theta) G(i\omega\pm\varepsilon_q,\k-\q)=\int_\alpha^1 \frac{d(\cos\theta)}{\omega^2+k^2\mp 2i\omega\varepsilon_q-2kq\cos\theta}=\frac{1}{2kq}\ln\frac{k^2-2kq\alpha+\omega^2\mp 2i\omega\varepsilon_q}{k^2-2kq+\omega^2\mp 2i\omega\varepsilon_q}\,,
\eeq
its limit for $\alpha\to-1$, as well as $\ln z+\ln\bar z=\ln |z|^2$, the final form of the bubble integral reads (if $k<\Lambda$)
\beq
\nonumber
{\cal B}[G](K)&=&\frac{1}{16 \pi^2 k}\left[\int_0^{\Lambda-k} dq\,q\,\frac{1+2n_{\varepsilon_q}}{\varepsilon_q}\ln\frac{(k^2+2 k q +\omega^2)^2 + 4\omega^2 \varepsilon_q^2}{(k^2-2 k q +\omega^2)^2 + 4\omega^2 \varepsilon_q^2}\right.\nonumber\\
&&\left.\hspace*{1.0cm}+\int_{\Lambda-k}^\Lambda dq\,q\,\frac{1+2n_{\varepsilon_q}}{\varepsilon_q}\ln\frac{(\Lambda^2-q^2 +\omega^2)^2 + 4\omega^2 \varepsilon_q^2}{(k^2-2 k q +\omega^2)^2 + 4\omega^2 \varepsilon_q^2}\right].
\label{Eq:bubble_final}
\eeq
Setting $\omega=0$ and taking the limit $k\to0,$ one easily obtains the expression of the bubble integral at vanishing frequency and momentum
\beq
{\cal B}[G](0)=\frac{1}{8\pi^2}\left[-\Lambda\,\frac{1+2n_{\varepsilon_\Lambda}}{\varepsilon_\Lambda}+\int_0^\Lambda dq\,\frac{1+2n_{\varepsilon_q}}{\varepsilon_q}\right].
\label{Eq:BUB}
\eeq
which can also be obtained from \eqref{Eq:TAD} by taking a derivative with respect to $M^2$ and using an integration by parts.

For the setting-sun sum-integral we start from Eq.~(\ref{eq:tt2}) with $T_\star=0$. In this limit $n^\star_{p_0}$ becomes $-\theta(-p_0)$ and $\delta n_{p_0}$ becomes $\varepsilon(p_0)n_{|p_0|}$. The first line of Eq.~(\ref{eq:tt2}) gives just the zero temperature contribution ${\cal S}^{(0)}[G]$. After performing the frequency integrals and the trivial integral over $\rb$, we obtain
\beq
{\cal S}^{(0)}[G]=\int_{|\k|<\Lambda}\int_{|\q|<\Lambda}\frac{1}{4 \varepsilon_k \varepsilon_q \varepsilon_r}\frac{\theta(\Lambda-|\k-\q|)}{\varepsilon_k+\varepsilon_q+\varepsilon_r},
\eeq
where we have renamed $\p$ as $\k$ and introduced $\varepsilon_r=\sqrt{(\k-\q)^2+M^2}.$ The angular integrals can be performed with the method given in \cite{Reinosa:2011cs} and one obtains
\beq
{\cal S}^{(0)}[G]=\frac{1}{32 \pi^4} \int_0^\Lambda dk \frac{k}{\varepsilon_k} \left[\int_0^{\Lambda-k} d q \frac{q}{\varepsilon_q} \ln \frac{\varepsilon_k+\varepsilon_q+\varepsilon_+}{\varepsilon_k+\varepsilon_q+\varepsilon_-}+\int_{\Lambda-k}^\Lambda d q \frac{q}{\varepsilon_q} \ln \frac{\varepsilon_k+\varepsilon_q+\varepsilon_\Lambda}{\varepsilon_k+\varepsilon_q+\varepsilon_-}
\right],
\eeq
where $\varepsilon_\pm=\sqrt{(k\pm q)^2+M^2}.$ The second and third lines of Eq.~(\ref{eq:tt2}) can be treated simultaneously. We note first that in the second line the factor $-\theta_{-p_0}+\theta_{r_0}=-1+\theta_{p_0}+\theta_{r_0}$ can be replaced by $-1+2\theta_{r_0}=\varepsilon(r_0)$, owing to the symmetry under $q_0\leftrightarrow r_0$ of the integrand. Combining this contribution with the third line of Eq.~(\ref{eq:tt2}) and performing the frequency integrals, we arrive at
\beq
{\cal S}[G]-{\cal S}^{(0)}[G]=3 \int_{|\k|<\Lambda}\int_{|\q|<\Lambda} \theta(\Lambda-|\k-\q|)\frac{n_{\varepsilon_k} (1+n_{\varepsilon_q})}{2 \varepsilon_k \varepsilon_q} \left[\frac{1}{\varepsilon_r^2-(\varepsilon_k+\varepsilon_q)^2}+\frac{1}{\varepsilon_r^2-(\varepsilon_k-\varepsilon_q)^2}
\right],
\eeq
Doing the frequency integrals, one obtains finally
\beq
\nonumber
{\cal S}[G]-{\cal S}^{(0)}[G] &=&\frac{3}{32 \pi^4} \int_0^\Lambda dk\,k\, \frac{n_{\varepsilon_k}}{\varepsilon_k} \left[\int_0^{\Lambda-k} dq\,q\,\frac{1+n_{\varepsilon_q}}{\varepsilon_q} \ln\frac{4(k^2+kq+q^2)+ 3 M^2}{4(k^2-kq+q^2) + 3 M^2} \right.
\\
&&\left.\qquad\qquad\qquad\qquad+ 
\int_{\Lambda-k}^\Lambda dq\,q\,\frac{1+n_{\varepsilon_q}}{\varepsilon_q}\ln\frac{4 (\varepsilon_q \varepsilon_k)^2 - (\Lambda^2 - q^2 -k^2 -M^2)^2}{M^2\big(4(k^2-kq+q^2) + 3 M^2\big)}
\right].
\label{eq:SS_1}
\eeq

\subsection{Monotonous behavior of the perturbative bubble sum-integral} 
Let us prove that the perturbative bubble sum-integral at finite temperature
\beq
{\cal B}[G](K)\equiv T\sum_n\int\frac{d^3q}{(2\pi)^3}\frac{\Theta(\Lambda^2-q^2)}{\omega^2_n+q^2+M^2}\frac{\Theta(\Lambda^2-(\q-\k)^2)}{(\omega_n-\omega)^2+(\q-\k)^2+M^2}\,,
\eeq
decreases as one increases $\omega$ or $k\equiv |\k|$. This is clear for the frequency dependence. To prove it for the momentum dependence, we consider
\beq
\frac{\partial}{\partial k}{\cal B}[G](K)\equiv 2\,{\cal I}_1(K)+2\,{\cal I}_2(K)
\eeq
with (we set $\hat\k=\k/k$)
\beq
{\cal I}_1(K) & \equiv & T\sum_n\int\frac{d^3q}{(2\pi)^3}\frac{\Theta(\Lambda^2-q^2)}{\omega^2_n+q^2+M^2}\frac{\Theta(\Lambda^2-(\q-\k)^2)}{((\omega_n-\omega)^2+(\q-\k)^2+M^2)^2}\,(\q-\k)\cdot\hat{\k}\,,\nonumber\\
{\cal I}_2(K) & \equiv & T\sum_n\int\frac{d^3q}{(2\pi)^3}\frac{\Theta(\Lambda^2-q^2)}{\omega^2_n+q^2+M^2}\frac{\delta(\Lambda^2-(\q-\k)^2)}{(\omega_n-\omega)^2+(\q-\k)^2+M^2}\,(\q-\k)\cdot\hat{\k}
\eeq
and prove that both ${\cal I}_1(K)$ and ${\cal I}_2(K)$ are negative.  Let us treat the contribution ${\cal I}_1(K)$ first. It is convenient to decompose the integration domain, which is defined by the $\Theta$ functions, into three different regions $C$, $D$ and $\bar D$, see Fig.~\ref{Fig:spheres}. The region $D$ corresponds to $\{q<\Lambda\}\cap\{|\q-\k|<\Lambda\}\cap\{(\q-\k)\cdot\hat{\k} > 0\}$. The region $\bar D$ is the mirror symmetric of $D$ with respect to the axis $(\q-\k)\cdot\hat{\k}=0$. The region $C$ is $\{q<\Lambda\}\cap\{|\q-\k|<\Lambda\}\backslash (D\cup\bar D)$. One has ${\cal I}_1={\cal I}_1^C+{\cal I}_1^D+{\cal I}_1^{\bar D}$. In region $C$ (and also in region  $\bar D$), one has $(\q-\k)\cdot\hat{\k}<0$, from which it follows that ${\cal I}_1^C<0$.  In order to treat the remaining contributions, for each point $\q$ in region $D$, we introduce its mirror symmetrized  $\q'=\q'(\q)$. More precisely, we have the formula
\beq
\q'(\q)=\q-2\big((\q-\k)\cdot\hat{\k}\big)\hat{\k}\,.
\eeq
From this formula or from the geometrical interpretation given in Fig.~\ref{Fig:spheres}, it is easily checked that $|\q'-\k|=|\q-\k|$, $(\q'-\k)\cdot\hat{\k}=-(\q-\k)\cdot\hat{\k}$ and $|\q'|>|\q|$. Moreover the previous transformation has the Jacobian $J_{ij}=\delta_{ij}-2k_ik_j/k^2$ which is such that $J^2=1$ and thus $|{\rm det}\,J|=1$. We can now use this transformation to express the contribution from region $\bar D$ as an integral over region $D$. It follows that
\beq
{\cal I}_1^D(K)+{\cal I}_1^{\bar D}(K)=T\sum_n\int_D\frac{d^3q}{(2\pi)^3}\frac{(\q-\k)\cdot\hat{\k}}{((\omega_n-\omega)^2+(\q-\k)^2+M^2)^2}\left[\frac{1}{\omega^2_n+q^2+M^2}-\frac{1}{\omega^2_n+{q'}^2(\q)+M^2}\right]\,,
\eeq
which is negative because on region $D$, $|\q|>|\q'|$ and $(\q-\k)\cdot\hat{\k}>0$. 

In order to treat the contribution ${\cal I}_2(K)$ we can use the following geometrical argument. Noting that ${\cal I}_2(K)$ represents the contribution to the variation of ${\cal B}[G](K)$ which comes from the modification of the integration region as $k$ increases, that the two spheres separate apart in this cases and that the integrand of ${\cal B}[G](K)$ is positive, it follows that ${\cal I}_2(K)$ is negative. Alternatively, we can also perform the angular integral using the $\delta$-function. We obtain
\beq
{\cal I}_2(K)=\frac{T}{16\pi^2k^2}\sum_n\frac{1}{(\omega_n-\omega)^2+\Lambda^2+M^2}\int^\Lambda_{\Lambda-k} dq\,q\,\frac{q^2-k^2-\Lambda^2}{\omega^2_n+q^2+M^2}\,,
\eeq
which is indeed negative since $q^2-k^2-\Lambda^2$ increases on the interval $[\Lambda-k,\Lambda]$ and is equal to $-k^2<0$ for $q=\Lambda$.

\section{Rate of convergence of Matsubara sums}\label{app:Matsubara}

In this section we study the rate of convergence of Matsubara sums and relate it to the asymptotic behavior of the summand at large Matsubara frequencies. Consider first the perturbative tadpole sum-integral ${\cal T}[G]$, which we approximate by
\beq\label{eq:approx2}
{\cal V}_{N_\tau}[G]\equiv T\!\!\sum_{n=-N_\tau}^{N_\tau}\int\frac{d^3q}{(2\pi)^3}\frac{1}{\omega_n^2+q^2+M^2}\,.
\eeq
In order to study how this finite sum converges to its limit ${\cal T}[G]$, we introduce the error
\beq\label{eq:error}
{\cal E}_{N_\tau}[G]\equiv {\cal T}[G]-{\cal V}_{N_\tau}[G]=T\!\!\sum_{|n|>N_\tau}\int\frac{d^3q}{(2\pi)^3}\frac{1}{\omega_n^2+q^2+M^2}\,.
\eeq
Note first that a very simple bound of the error is obtained by setting $q^2+M^2$ to $0$ in the previous expression. We obtain $|{\cal E}_{N_\tau}[G]|\leq c_0\varphi_0(N_\tau)$ with $c_0\equiv\Lambda^3/(24\pi^4 T)$ and $\varphi_0(N_\tau)\equiv \sum_{|n|>N_\tau}1/n^2$. From
\beq\label{eq:bound}
\sum_{|n|>N_\tau} \frac{1}{n^2}\leq 2\sum_{n>N_\tau} \frac{1}{n(n-1)}=\frac{2}{N_\tau}\,,
\eeq
we obtain an even simpler bound $|{\cal E}_{N_\tau}[G]|\leq 2c_0/N_\tau$. A numerical investigation reveals that the bounds are saturated already for values of $N_\tau<\Lambda/T$, which shows that the bounds provide a good description of the error in this range of $N_\tau$ and in turn that the convergence of the Matsubara sum is slow. 

The functions $c_0\varphi_0(N_\tau)$ or $2c_0/N_\tau$ are in fact the first terms of asymptotic expansions of the error at large $N_\tau$ in some appropriate asymptotic scales which we now discuss.\footnote{An asymptotic scale is any collection of functions $(\psi_k(N_\tau))_k$ such that each $\psi_k(N_\tau)$ does not vanish above some value of $N_\tau$ and that for $k<k'$, $\psi_{k'}(N_\tau)/\psi_k(N_\tau)\rightarrow 0$ as $N_\tau\rightarrow\infty$. This last condition is also written $\psi_{k'}(N_\tau)=o(\psi_k(N_\tau))$.} We start from the following identity:
\beq
\frac{1}{a+b}=\sum_{k=0}^K(-1)^k\frac{b^k}{a^{k+1}}+r_k \quad {\rm with} \quad r_k\equiv\frac{(-1)^{K+1}}{a+b}\frac{b^{K+1}}{a^{K+1}}\,.
\eeq
Note that, when $a$ and $b$ are both positive, $|r_k|<{\rm Min}\,(b^{K+1}/a^{K+2},b^K/a^{K+1})$, that is the rest of the geometric series is bounded both by the last term kept in the sum and by the first term neglected. Plugging this identity into Eq.~(\ref{eq:error}), we arrive at
\beq
{\cal E}_{N_\tau}[G]=\sum_{k=0}^K c_k[G]\,\varphi_k(N_\tau)+R^{(K)}_{N_\tau}[G]\,,
\eeq
with
\beq
c_k[G]\equiv \frac{2}{(4\pi^2)^{k+2}T^{2k+1}}\int_0^\Lambda dq\,q^2\, (q^2+M^2)^k \quad {\rm and} \quad \varphi_k(N_\tau)\equiv \sum_{|n|>N_\tau}\frac{1}{n^{2k+2}}\,.
\eeq
It can be shown that he functions $\varphi_k(N_\tau)$ form an asymptotic scale at large $N_\tau$, in the sense that, $\varphi_k(N_\tau)/\varphi_{k'}(N_\tau)\rightarrow 0$ if $k'>k$. Now, from the inequality obeyed by $r_k$, it is easy to show that
\beq
|R_K|\leq c_{K+1}[G]\varphi_{K+1}(N_\tau)\,.
\eeq
In other words
\beq\label{eq:C8}
{\cal E}_{N_\tau}[G]=\sum_{k=0}^K c_k[G]\,\varphi_k(N_\tau)+O(\varphi_{K+1}(N_\tau))\,,
\eeq
for any value of $K\geq -1$, if we decide conventionally that the sum is empty when $K=-1$. This shows precisely that the error admits an asymptotic expansion in the scale of functions $\varphi_k(N_\tau)$. The functions $\varphi_k(N_\tau)$ admit themselves an asymptotic expansion in a scale of inverse powers of $N_\tau$ from which we could also deduce an expansion of the error in such a simpler scale. The advantage of (\ref{eq:C8}) is that the corresponding series converges as $K\rightarrow\infty$ for fixed $N_\tau$. We get then a better estimate of the error than with an expansion in a scale of inverse powers of $N_\tau$ for which the corresponding series turns out to be divergent.\\

Let us now see how to use this results to accelerate the Matsubara sums. Note first that the coefficient $c_0[G]$ does not depend on $M$. The simple bound in \eqref{eq:bound} can be used to accelerate the convergence of the Matsubara sum by writing 
\beq
{\cal T}[G]={\cal V}_{N_\tau}[G]+{\cal E}_{N_\tau}[G]=\Big[{\cal V}_{N_\tau}[G]+c_0\varphi_0(N_\tau)\Big]+\Big[{\cal E}_{N_\tau}[G]-c_0\varphi_0(N_\tau)\Big],
\eeq
and computing the first bracket instead of ${\cal V}_{N_\tau}[G].$ The new error is bounded by $c_1[G]\varphi_1(N_\tau)$ which is parametrically smaller than $c_0\varphi_0(N_\tau)$. Using that $\varphi_k(N_\tau)\leq 2/N_\tau^{2k+1}$ and taking the dominant contribution of $c_k[G]$, that is $c_k[G]\sim 2\Lambda^{2k+3}/((2k+3)(4\pi^2)^{k+2}T^{2k+1})$, we obtain 
\beq
\frac{c_1[G]\varphi_1(N_\tau)}{c_0[G]\varphi_0(N_\tau)}\lesssim\frac{3}{20\pi^2}\left(\frac{\Lambda}{N_\tau T}\right)^2\,.
\eeq
It is easy to see that in combination with the trapezoidal rule \eqref{eq:trapezoid} the improvement amounts to using the following approximate evaluation of the tadpole sum-integral:
\beq\label{eq:Tad_improved} 
{\cal V'}_{N_\tau,N_s}[G]={\cal V}_{N_\tau,N_s}[G]+\frac{(\Delta k)^3}{24\pi^4}\frac{N_s^2+2 N_s^3}{N_\tau T}\,,
\eeq
where the notation ${\cal V}_{N_\tau,N_s}[f]$ was introduced in Eq.~\eqref{eq:vol_d}. Similar considerations for the bubble sum-integral implies the following improved evaluation:
\beq\label{eq:bub_improved}
{\cal V'}_{N_\tau,N_s}[G^2]={\cal V}_{N_\tau,N_s}[G^2] +\frac{(\Delta k)^3}{288\pi^4}\frac{N_s^2+2 N_s^3}{N_\tau^3T^3}\,,
\eeq
where we used that the analogous simple bound here is $\sum_{|n|>N_\tau}1/n^4<2/(3 N_\tau^3)$. We can do even better by noticing that the leading part of the coefficient $c_k[G]$ is of the order or $\Lambda^{2k+3}$ and it is thus independent of $M$. We consider then
\beq\label{eq:approx3}
{\cal V}''_{N_\tau}[G]={\cal V}_{N_\tau}[G]-{\cal V}_{N_\tau}[G_\star]+{\cal T}[G_\star]\,.
\eeq
where $G_\star$ is a free-type propagator of reference for which we assume that ${\cal T}[G_\star]$ is known exactly. The error becomes now
\beq
{\cal E}''_{N_\tau}[G]=\sum_{k=1}^K (c_k[G]-c_k[G_0])\,\varphi_k(N_\tau)+O(\varphi_{K+1}(N_\tau))\,.
\eeq
Note first that, because $c_0[G]$ does not depend on $M$, the leading contribution of the error (k=0) has dropped. Moreover the leading contribution to the term of order $k$, which was previously of the order of $\Lambda^{2k+3}$ is now of the order $\Lambda^{2k+1}M^2$.

Of course the previous discussion is not very usefull in the case where the mass $M$ is momentum independent because our acceleration method requires that we are able to evaluate $T[G_\star]$ very accurately which is the same as computing the perturbative $T[G]$ very accurately. The approach becomes interesting when we evaluate the tadpole sum-integral in the presence of a propagator $\bar G(Q)=1/(Q^2+\bar M^2(Q))$ with a momentum dependent mass. If the latter remains positive and grows only logarithmically at large $Q$, which is precisely what happens in the two-loop $\Phi$-derivable approximation, then it makes a difference to evaluate $T[\bar G]$ from (\ref{eq:approx2}) which has an error bounded by $|{\cal E}_{N_\tau}[\bar G]|\leq c_0\varphi_0(N)\leq 2c_0/N_\tau$, or from (\ref{eq:approx3}) which yields an error bounded by
\beq
|{\cal E}_{N_\tau}[\bar G]-{\cal E}_{N_\tau}[G_\star]|\leq T\sum_{|n|>N_\tau}\int\frac{d^3q}{(2\pi)^3}\frac{|\bar M^2(Q)-m^2_\star|}{\omega^4_n}\,.
\eeq
which is expected to be suppressed with respect to $c_0\varphi_0(N_\tau)$ if $\bar M^2(Q)$ grows only logarithmically at large $Q$, as we verified numerically.

\end{document}